\documentclass[10pt,journal,compsoc]{IEEEtran}

\usepackage{graphicx}
\usepackage{subfigure}

\usepackage{gensymb}
\usepackage{amsmath}

\usepackage{cases}
\usepackage{float}
\usepackage{amssymb}
\usepackage{multirow}
\usepackage{mathrsfs}
\usepackage{algorithmic}
\usepackage[ruled]{algorithm2e}
\usepackage{rotating}
\usepackage{balance}
\usepackage{threeparttable}
\usepackage{color}
\usepackage{colortbl}
\definecolor{mygray}{gray}{.9}
\usepackage{makecell}
\usepackage{pdflscape}
\usepackage{siunitx}
\usepackage[pagewise]{lineno}

\usepackage{stfloats}
\usepackage{verbatim}
\usepackage{setspace}
\usepackage{threeparttable}
\usepackage{supertabular,booktabs}
\usepackage{rotating}

\usepackage{supertabular}
\usepackage{pifont}

\usepackage{xcolor,pict2e}

\usepackage{tikz}

\begin{document}

\title{Identification of Failure Regions for Programs with Numeric Inputs}

\author{Rubing Huang,~\IEEEmembership{Senior Member,~IEEE,}
        Weifeng Sun,
        Tsong Yueh Chen,~\IEEEmembership{Senior Member,~IEEE,}
        Sebastian Ng,
        Jinfu Chen,~\IEEEmembership{Member,~IEEE}

\IEEEcompsocitemizethanks{
\IEEEcompsocthanksitem R. Huang is with the School of Computer Science and Communication Engineering, and Jiangsu Key Laboratory of Security Technology for Industrial Cyberspace, Jiangsu University, Zhenjiang, Jiangsu 212013, China.\protect\\
E-mail: rbhuang@ujs.edu.cn

\IEEEcompsocthanksitem W. Sun and J. Chen are with the School of Computer Science and Communication Engineering, Jiangsu University, Zhenjiang, Jiangsu 212013, China.\protect\\
E-mail: 2211808031@stmail.ujs.edu.cn, jinfuchen@ujs.edu.cn.

\IEEEcompsocthanksitem T. Y. Chen and S. Ng are with the
Department of Computer Science and Software Engineering, Swinburne University of Technology, Hawthorn, VIC 3122, Australia.\protect\\
E-mail: tychen@swin.edu.au, sng@swin.edu.au.

}
}

\markboth{Identification of Failure Regions for Programs with Numeric Inputs}
{Shell \MakeLowercase{\textit{et al.}}: Bare Demo of IEEEtran.cls for Computer Society Journals}
\IEEEtitleabstractindextext{
\begin{abstract}
Failure region, where failure-causing inputs reside, has provided many insights to enhance testing effectiveness of many testing methods. Failure region may also provide some important information to support other processes such as software debugging. When a testing method detects a software failure, indicating that a failure-causing input is identified, the next important question is about how to identify the failure region based on this failure-causing input, i.e., \textit{Identification of Failure Regions} (IFR). In this paper, we introduce a new IFR strategy, namely \textit{Search for Boundary} (SB), to identify an approximate failure region of a numeric input domain. SB attempts to identify additional failure-causing inputs {that are as close to the boundary of the failure region as possible.} To support SB, we provide a basic procedure, and then propose two methods, namely \textit{Fixed-orientation Search for Boundary (FSB)} and \textit{Diverse-orientation Search for Boundary} (DSB). In addition, we implemented an automated experimentation platform to integrate these methods. In the experiments, we evaluated the proposed SB methods using a series of simulation studies andempirical studies with different types of failure regions. The results show that our methods can effectively identify a failure region, within the limited testing resources.
\end{abstract}

}

\maketitle

\IEEEdisplaynontitleabstractindextext
\IEEEpeerreviewmaketitle

\section{Introduction}
\label{introduction}

\IEEEPARstart{A}{ccording} to the IEEE standard~\cite{STD2010}, a software developer makes a \textit{mistake}, which may introduce a \textit{fault} (\textit{defect} or \textit{bug}) in the software. When a fault is encountered, a \textit{failure} may be produced, i.e., the software behaves unexpectedly. \textit{Software testing} plays an important role in ensuring the quality of software, which generally contains four steps: 1) Determining test objectives; 2) Selecting some inputs from the input domain as test cases; 3) Executing these test cases to test the software; and 4) Observing and comparing the execution outputs against the expected results (i.e., \textit{test oracles}).

\begin{figure*}[!b]
\centering
    \subfigure[Block pattern]
    {
        \includegraphics[width=0.25\textwidth]{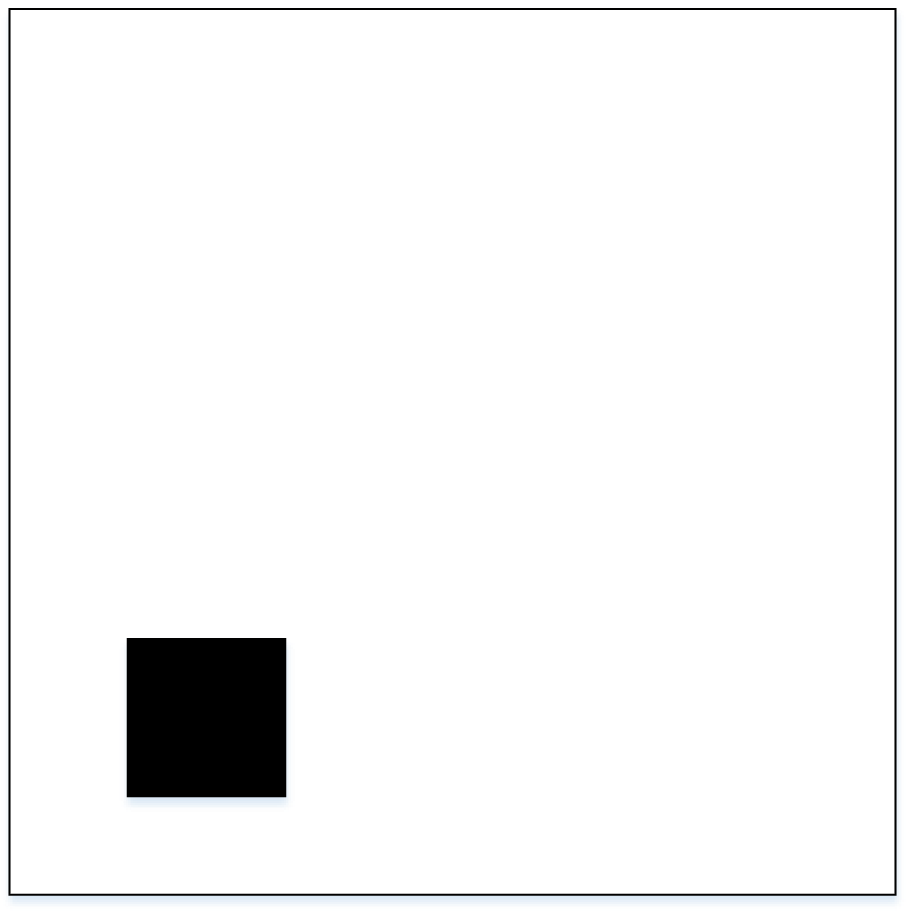}
        \label{FP:bp}
    }
    \hspace{10mm}
    \subfigure[Strip pattern]
    {
        \includegraphics[width=0.25\textwidth]{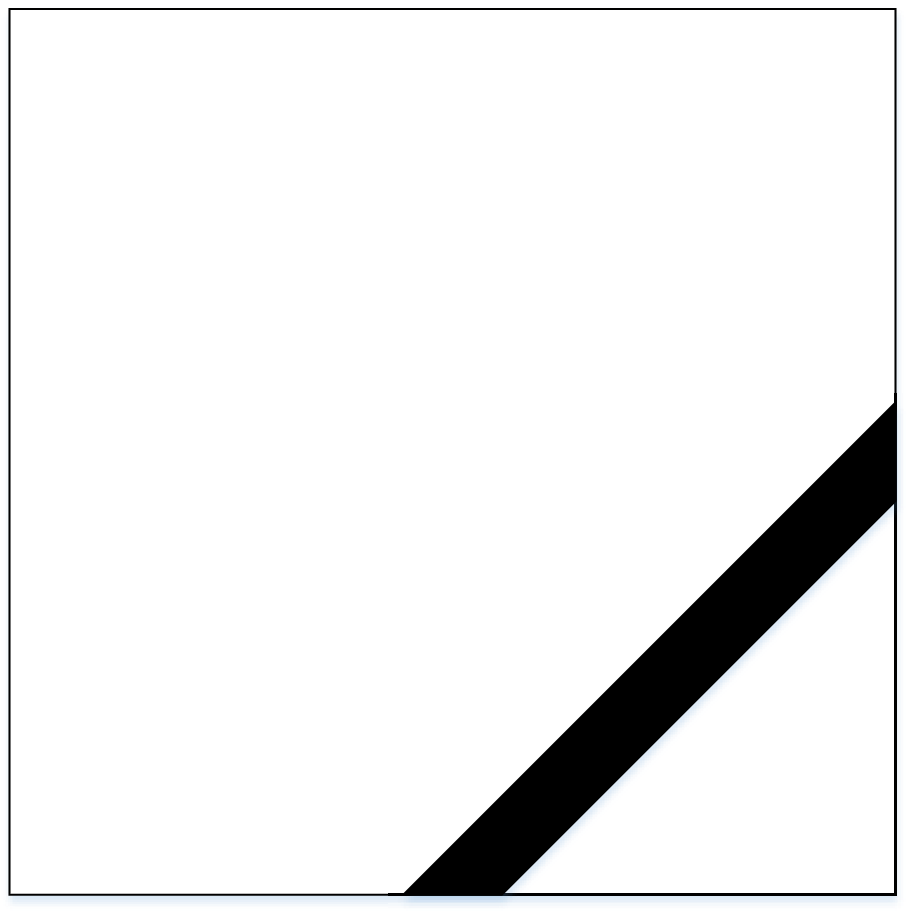}
        \label{FIG:sp}
    }
    \hspace{10mm}
    \subfigure[Point pattern]
    {
        \includegraphics[width=0.25\textwidth]{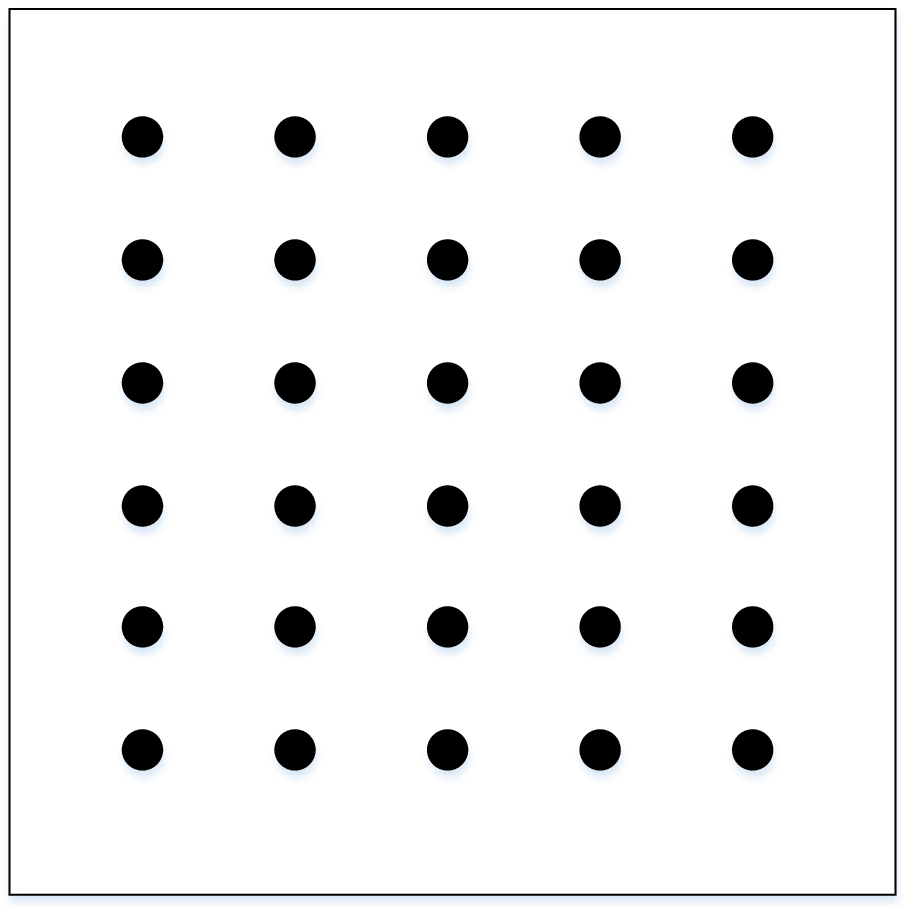}
        \label{FIG:pp}
    }

    \caption{Three failure patterns in a two-dimension input domain.}
    \label{FIG:FP}
\end{figure*}

A test input that causes a software failure is generally called a \textit{failure-causing input}. Intuitively speaking, the information about failure-causing inputs is fixed after coding but unknown before testing, such as the shape, size, and location of \textit{failure regions} (i.e., regions where failure-causing inputs reside)\footnote{The shape of failure region together with its distribution over the input domain is also called \textit{failure pattern}; while its size is also called \textit{failure rate}, defined as the ratio of the number of failure-causing inputs to the number of all possible inputs from the input domain.}.
The information of failure regions has provided many insights for enhancing software testing techniques. For example, previous studies have independently observed that failure regions are likely to be contiguous over the input domain~\cite{White1980,Ammann1988,Finelli1991,Bishop1993,Schneckenburger2007}. In other words, if a test case is a failure-causing input, its neighbours are most likely also failure-causing inputs; similarly, if a test case is a successful input (i.e., no failures is triggered by this test case), its neighbours are likely to be successful. Based on this observation, Chen et al.~\cite{Chen2010} proposed an enhancement of \textit{Random Testing} (RT)~\cite{Orso2014}, namely \textit{Adaptive Random Testing} (ART), which aims at generating new test cases that are as far away from the successful inputs as possible~\cite{Zhou2013,Huang2015,Liu2015,Barus2016,Huang2019,Chen2004c}. In addition, Cai et al.~\cite{Cai2009} proposed another enhancement of RT through the integration of \textit{Partition Testing}~(PT)~\cite{Chen1994}, namely \textit{Dynamic Random Testing} (DRT), which aims at dynamically updating the selection probability of the partition of the input domain: If a new test case is failure-causing input, the selection probability of its related partition will be increased; otherwise decreased. Sun et al.~\cite{Sun2019} proposed an enhanced version of DRT, called \textit{Adaptive Partition Testing} (APT), by adaptively adjusting the selection probability of each partition.

Furthermore, failure regions provide other valuable information such as sizes, shapes, locations, and numbers, which may help us to develop new and effective strategies of test case generation. Such family of test case generation is referred to as the approach of {\textit{Failure-Based Testing} (FBT)}{~\cite{Chen2010,Selay2018}.} In fact, ART is a special instance of failure-based testing, and it has been well received by the software testing community~\cite{Huang2019}. Therefore, the knowledge about failure regions may promote the development of FBT. In addition, failure regions may also provide some insights for different application scenarios such as \textit{fault localization}~\cite{Wong2016} and \emph{program repair}~\cite{Monperrus2018}. In this paper, therefore, we attempt to identify failure regions after identifying a single failure-causing input. More specifically, a testing tool is built to generate test cases with the goal of revealing the failure region. During the test case execution, when a software failure is triggered by a single failure-causing test case, we turn to the question about how to identify its own failure region, i.e., \textit{Identification of Failure Regions} (IFR).

Previous studies~\cite{Ahmad2013,Ahmad2014} have proposed some methods to address IFR with numeric inputs. More specifically, Ahmad and Oriol~\cite{Ahmad2013,Ahmad2014} proposed a IFR method, namely \textit{Automated Discovery of Failure Domain} (ADFD), which attempts to identify the failure region by exhaustively executing neighbours around a given failure-causing input within an inclusion region. However, ADFD suffers from the following drawbacks: 1) It can be adopted for input domains with only integer inputs rather than floating inputs, because it is impossible to exhaustively execute floating inputs; and 2) The size of the inclusion region is a parameter of ADFD, which means that different parameter values may produce highly different failure regions.

\begin{figure*}[!b]
\centering
    \subfigure[A circular  region]
    {
        \includegraphics[width=0.25\textwidth]{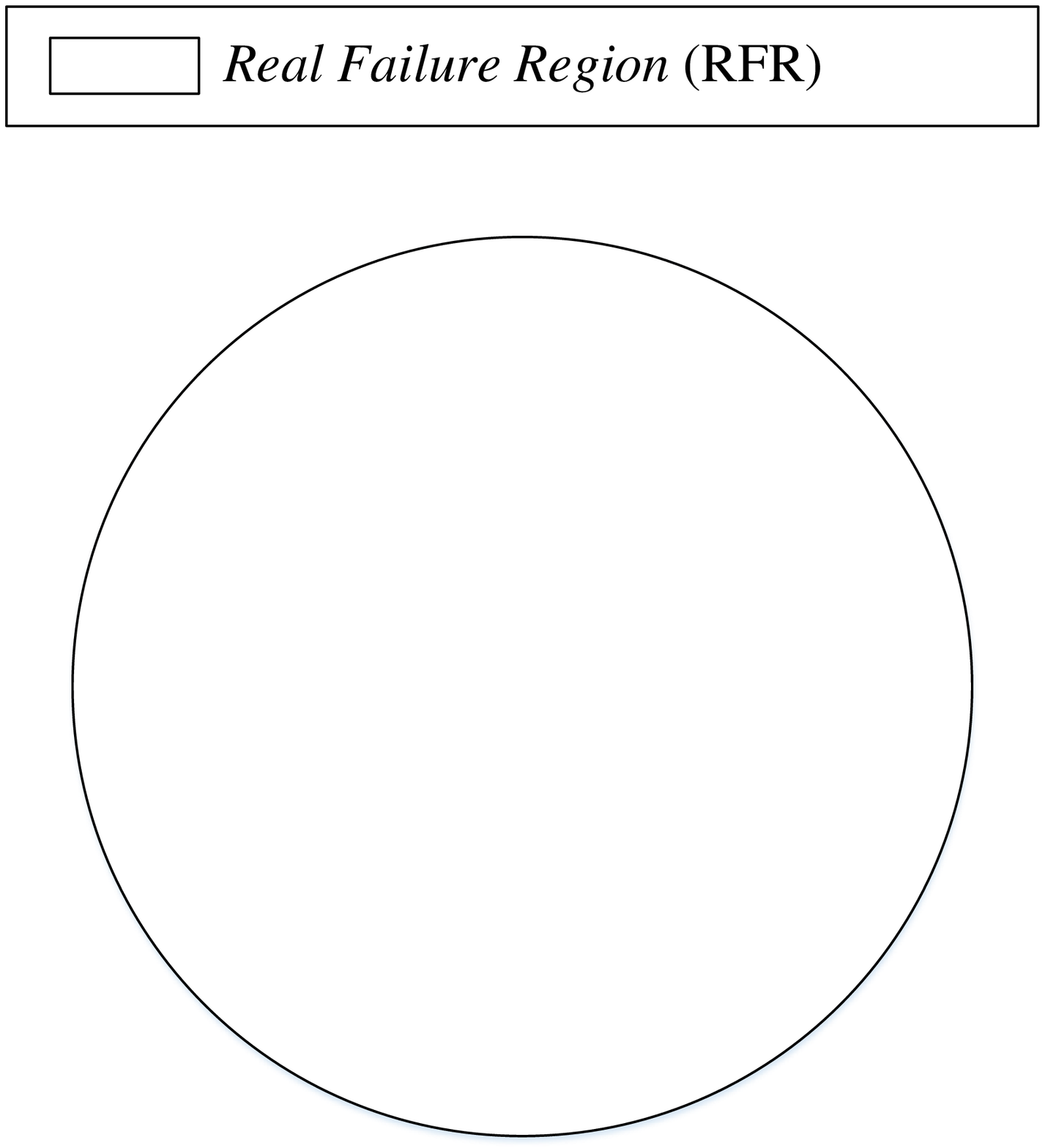}
        \label{FIG:circle}
    }
    \hspace{10mm}
    \subfigure[Approximate solution 1]
    {
        \includegraphics[width=0.25\textwidth]{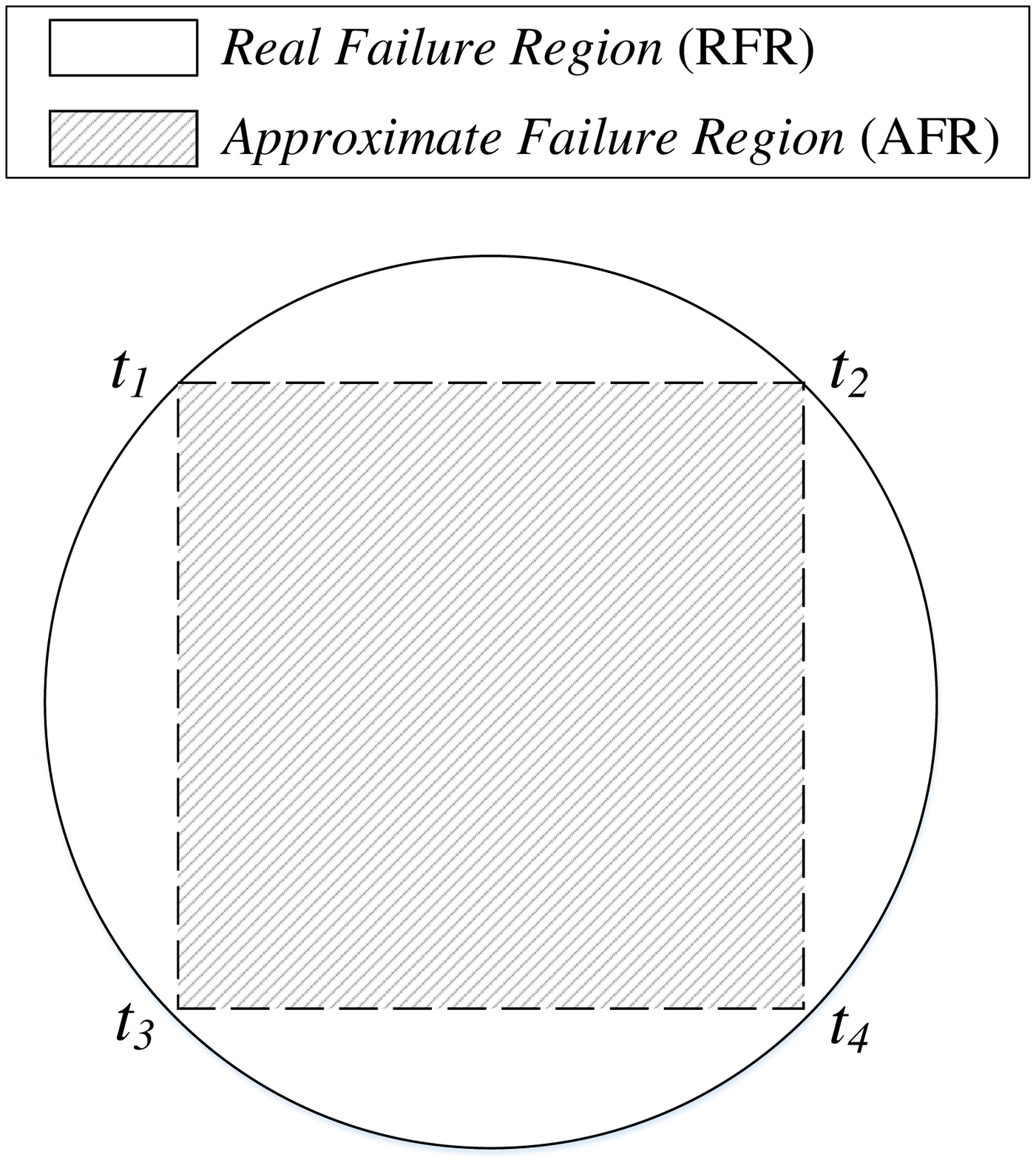}
        \label{FIG:circle1}
    }
    \hspace{10mm}
    \subfigure[Approximate solution 2]
    {
        \includegraphics[width=0.25\textwidth]{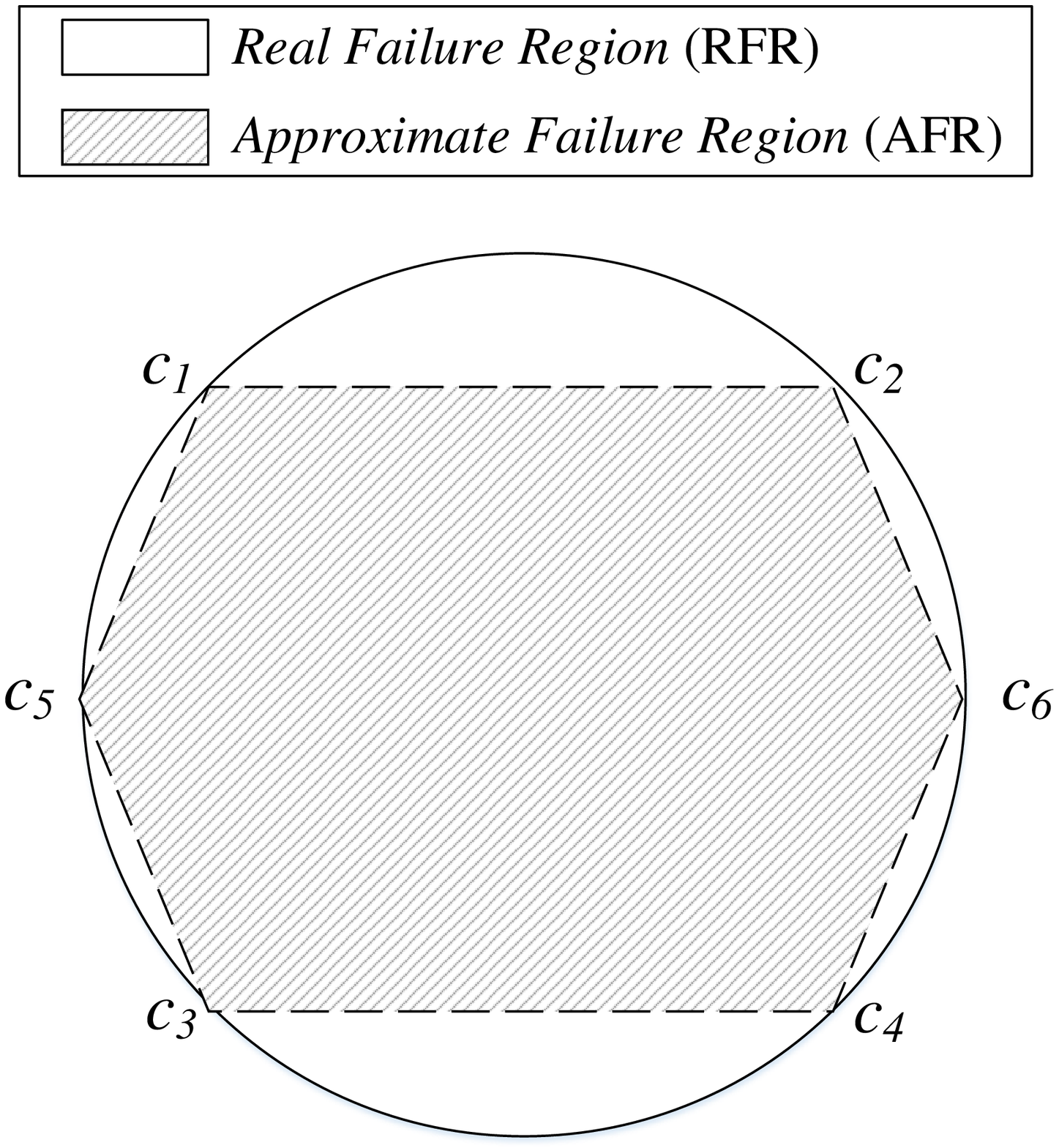}
        \label{FIG:circle2}
    }

    \caption{Approximate failure regions for a circular failure region.}
    \label{circle_solution}
\end{figure*}

To overcome the above drawbacks, in this paper we propose a new IFR strategy, namely \textit{Search for Boundary} (SB), which attempts to identify the failure region of a numeric input domain\footnote{The numeric input domain includes all possible types of numeric inputs.} based on a single failure-causing test case. The key component of SB is to identify \emph{additional} failure-causing inputs as close to the boundary of the real failure region as possible. The main contributions of this paper are listed as follows:
\begin{itemize}
\item[1)] We proposed a new IFR strategy, i.e., \textit{Search for Boundary} (SB), and then built two methods to support SB. To the best of our knowledge, this is the first study that comprehensively addresses IFR for programs with numeric inputs.

\item[2)] We conducted a series of simulation studies and empirical studies with different failure regions to investigate the effectiveness and efficiency of our SB methods.

\item[3)] We developed an automated experimentation platform to identify and visualize the failure region.
\end{itemize}

The rest of this paper is organized as follows: Section \ref{SEC:Background} presents some background information; while Section \ref{SEC:ABM} describes the details about SB, including assumptions, a motivating example, and method. The simulation studies are reported in Section \ref{SEC:Simulations}. Section~\ref{SEC:results} provides and analyzes the experimental results aiming at answering the research questions. Section \ref{SEC:Weaken} discusses how to relax the assumptions; while Section~\ref{SEC:Tool} provides an automated experimentation platform for SB. Section \ref{SEC:RelatedWork} discusses some related work, and Section~\ref{SEC:Conclusions} concludes the paper, and discusses the future work.

\section{Background}
\label{SEC:Background}
In this section, some background information will be presented, including failure region, and failure-based testing.
\subsection{{Failure Region}}

{For a faulty program, a \emph{failure region} is a group of all test inputs that cluster together and cause the software failure (these test inputs are known as \textit{failure-causing inputs}). Generally speaking, a failure region has two basic features, namely \textit{failure rate} and \textit{failure pattern}. The failure rate is to measure the size of the failure region; while the failure pattern is to describe the geometrical distribution of the failure region including its shape and orientation.}

Many previous studies have independently found that failure-causing inputs tend to be clustered into contiguous failure regions over the input domain~\cite{White1980,Ammann1988,Finelli1991,Bishop1993,Schneckenburger2007}. Chan et al.~\cite{Chan1996} identified three major failure patterns in a general sense, i.e., \textit{block pattern}, \textit{strip pattern}, and \textit{point pattern}. Figure~\ref{FIG:FP} depicts these three failure patterns in a two-dimensional input domain, where the bounding box stands for the boundary of the input domain; and the black block, strip, or dots represents the failure-causing inputs. Chan et al.~\cite{Chan1996} also claimed that block and strip patterns are more commonly encountered than point patterns.

\subsection{Failure-Based Testing}
A \emph{Failure-Based Testing} (FBT)~\cite{Chen2010} method is defined as one that chooses test cases by leveraging the information about various aspects of failure patterns, such as shapes, numbers, and locations. To support FBT, many implementations have been proposed. For example, White and Cohen~\cite{White1980} proposed the \textit{domain testing strategy}, which can be considered as the first FBT technique (because the proposed method aims at identifying the domain faults, which result in a specific failure pattern in the input domain). Another example of FBT is \textit{Adaptive Random Testing} (ART)~\cite{Huang2019}, which attempts to achieve an even spread of test cases over the input domain, taking the advantage of the contiguity of failure patterns. ART uses less test cases than RT to detect the first failure, and also achieves a higher code coverage than that of RT when both use the same number of test cases~\cite{Chen2013}.

Many studies have investigated the impact of failure patterns on different software testing techniques. For example, Chen et al.~\cite{Chen2006a,Chen2007} conducted a series of experiments to analyze how different aspects of failure patterns influence testing effectiveness of ART.
Cai et al.~\cite{Cai2009} proposed an effective testing technique, namely dynamic random testing (DRT), which adjusts the selection probability of each partition of input domain to improve its effectiveness. In addition, Selay et al.~\cite{Selay2018} leveraged peculiar characteristics of failure patterns found in browser layout rendering, that is layout faults of web applications tending to propagate to the lower-right corner of the screen, to guide the testing process for more cost-effective results.
Chen and Merkel~\cite{Chen2008} analytically examined the possible effect of failure patterns on software testing, especially the degree to which testing effectiveness can be improved based on the knowledge about failure patterns. They also found that if all of the knowledge about the size, shape, or orientation are known, it is possible to develop a testing strategy that guarantees to detect a failure using not less than half of the expected number of test cases required by RT to detect a failure~\cite{Chen2008}.

\section{Search for Boundary}
\label{SEC:ABM}
In this section, we present some assumptions about the identification of the failure region, and then give an example to illustrate inspiration for the \textit{Search for Boundary} (SB). Finally, we provide a basic procedure to support SB, and then propose two methods.

\subsection{Assumptions}
\label{SEC:assumptions}
To simplify the problem, we make three assumptions to help us building a framework. However, these assumptions may seem to be somehow too strong to hold in many practical applications. In Section~\ref{SEC:Weaken}, we will discuss how to relax these assumptions, i.e., to narrow the gap between ideal and practical problem framework.

\textbf{\textit{Assumption 1: Only one failure-causing input is available to start and support the process of SB.}}
It is intuitive that more failure-causing inputs may provide more information to support SB. In this study, we assumed that only one failure-causing input is available.

\textbf{\textit{Assumption 2: There is only one failure region in the input domain.}}
In practice, it is common to have more than one failure region in faulty programs. In this paper, we assumed there is only one failure region within the input domain.

\textbf{\textit{Assumption 3: The failure region is convex.}} Generally speaking, the shape of failure region is fixed but unknown to testers before testing.
In addition, there are many types of the shape for the failure region. This study assumed that the shape of the failure region is of the {convex}\footnote{{Convex means bending outwards; conversely, it is denoted as concave.}} type.

\subsection{A Motivating Example}

Consider a two-dimensional input domain, the \textit{Real Failure Region} (RFR) is a circle (as shown in Figure~\ref{FIG:circle}). As we know, the RFR is unknown before identification. If we obtain some points (i.e., failure-causing inputs) that are close to the boundary, we would obtain an approximate region for the RFR, i.e., an \textit{Approximate Failure Region} (AFR). For example, Figure~\ref{FIG:circle1} provides an AFR with the square shape (the shaded region) by adopting four points ($t_1$, $t_2$, $t_3$, and $t_4$); while Figure~\ref{FIG:circle2} shows a hexagonal region as the AFR by using six failure-causing inputs ($c_1$, $c_2$, $c_3$, $c_4$, $c_5$, and $c_6$).

{We can then use mathematical inequalities to describe the AFR. Assuming that the center coordinate of the circle is equal to $(1,1)$, and let four points $t_1$, $t_2$, $t_3$, and $t_4$ be $(1-\frac{\sqrt{2}}{2},1+\frac{\sqrt{2}}{2})$, $(1+\frac{\sqrt{2}}{2},1+\frac{\sqrt{2}}{2})$, $(1-\frac{\sqrt{2}}{2},1-\frac{\sqrt{2}}{2})$, and $(1+\frac{\sqrt{2}}{2},1-\frac{\sqrt{2}}{2})$, respectively. Therefore, the approximate failure region in Figure~\ref{FIG:circle1} could be described as follows:}
\begin{numcases}{}
1-\frac{\sqrt{2}}{2} \leq x \leq 1+\frac{\sqrt{2}}{2} \\
1-\frac{\sqrt{2}}{2} \leq y \leq 1+\frac{\sqrt{2}}{2}
\end{numcases}

Based on this example, it can be observed that more failure-causing inputs that are closer to the boundary may provide more precise AFR, although it is difficult to identify the RFR. Consequently, it is essential to identify more failure-causing inputs around the boundary of the RFR.

\subsection{Strategy}
\label{SEC:method}
In this section, we first present the basic procedure of SB, and then provide two methods to achieve such procedure.

\subsubsection{Basic Procedure}
\label{frame}

\begin{figure}[!t]
    \centering
    \includegraphics[width=0.48\textwidth]{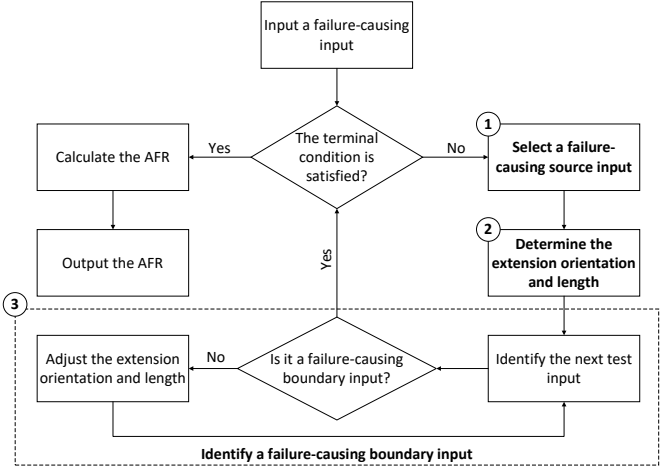}
    \caption{The basic procedure of SB.}
    \label{FIG:Procedure}
\end{figure}

The focus of SB is to find additional and diverse failure-causing inputs, which are as close to the boundary of the RFR as possible. Therefore, the core of SB is to identify a new failure-causing input near the boundary of the RFR (namely a \textit{failure-causing boundary input}), based on a previous failure-causing input (namely a \textit{failure-causing source input}). Figure~\ref{FIG:Procedure} shows the basic procedure of SB, which consists of three main steps.

\textbf{\textit{(1) Selection of a failure-causing source input:}} During the process of identifying failure-causing boundary inputs, it is necessary to choose a previous failure-causing input as the source input to be used for the next search. Obviously, when selecting the failure-causing source input, we consider the first failure-causing input (i.e., the parameter input of SB), and also adopt the following failure-causing test cases during the whole process.

\textbf{\textit{(2) Determination of the extension orientation and length:}} Once a failure-causing source input is selected, we need to determine the extension orientation and length that are used for finding more potential failure-causing inputs, and then identify the failure-causing boundary input.

\textbf{\textit{(3) Identification of a failure-causing boundary input:}} By adopting the failure-causing source input, the extension orientation $\overrightarrow{\alpha}$ and length $L$ (as discussed in the previous two steps), it is easy to identify the next test input $ti$. By executing $ti$ with the subject program, we can then determine whether it is failure-causing.
\begin{itemize}
\item If $ti$ is not a failure-causing input, it can be concluded that the extension length may be too long to exceed the boundary of the failure region. Therefore, we need to adjust the extension orientation and length to retract $ti$. Here, the extension orientation is set as $(-\overrightarrow{\alpha})$; while the length is assigned to $L/2$.
\item If $ti$ is a failure-causing input, there are the following two cases:
1) If $ti$ is not a retracted point, the extension orientation is still equal to $\overrightarrow{\alpha}$, while the length is also equal to $L$, and
2) If $ti$ is a retracted point, the extension orientation is changed from $(-\overrightarrow{\alpha})$ to $\overrightarrow{\alpha}$, while the length becomes half of the current value.
\end{itemize}

Algorithm \ref{ALG:find_boundary} provides detailed information to identify a failure-causing boundary input. As for the terminal condition to stop the algorithm, it is common to set a threshold $\lambda$, which is generally decided by the human tester in advance. In other words, the algorithm is repeated until the threshold is reached. The threshold $\lambda$ can have different definitions, according to different perspectives. For example, $\lambda$ can be a measure of the number of iterations, the number of failure-causing inputs, or the total required time for execution.

\begin{algorithm}[!t]
\footnotesize
\caption{Identify a failure-causing boundary input}
\label{ALG:find_boundary}
\begin{algorithmic}[1]
    \renewcommand{\algorithmicrequire}{\textbf{Input:}}
    \renewcommand{\algorithmicensure}{\textbf{Output:}}
    \renewcommand{\algorithmicelsif}{\algorithmicelse}
\REQUIRE
A failure-causing source input $tc$\\
~~~~~~An extension length $L$\\
~~~~~~An extension orientation $\overrightarrow{\alpha}$
\ENSURE A failure-causing boundary input $tb$
\STATE $flag \leftarrow false$\\
/* The $flag$ is to check whether $tc$ is a retracted point */
\WHILE {The terminal condition is not satisfied}
\STATE $tc \leftarrow tc + L \cdot \overrightarrow{\alpha}$
\IF {$tc$ is a not failure-causing input}
\STATE $L \leftarrow L/2$
\STATE $\overrightarrow{\alpha} \leftarrow -\overrightarrow{\alpha}$
\STATE $flag \leftarrow true$ /* Set $tc$ as a retracted point */
\ELSE
\STATE $tb \leftarrow tc$
    \IF {$flag == true$}
    \STATE $L \leftarrow L/2$
    \STATE $\overrightarrow{\alpha} \leftarrow -\overrightarrow{\alpha}$
    \STATE $flag \leftarrow false$ /* Set $tc$ as a non retracted point */
    \ENDIF
\ENDIF
\ENDWHILE
\STATE Return $tb$
\end{algorithmic}
\end{algorithm}

\begin{figure}[!b]
\centering
    \subfigure[]
    {
        \includegraphics[width=0.23\textwidth]{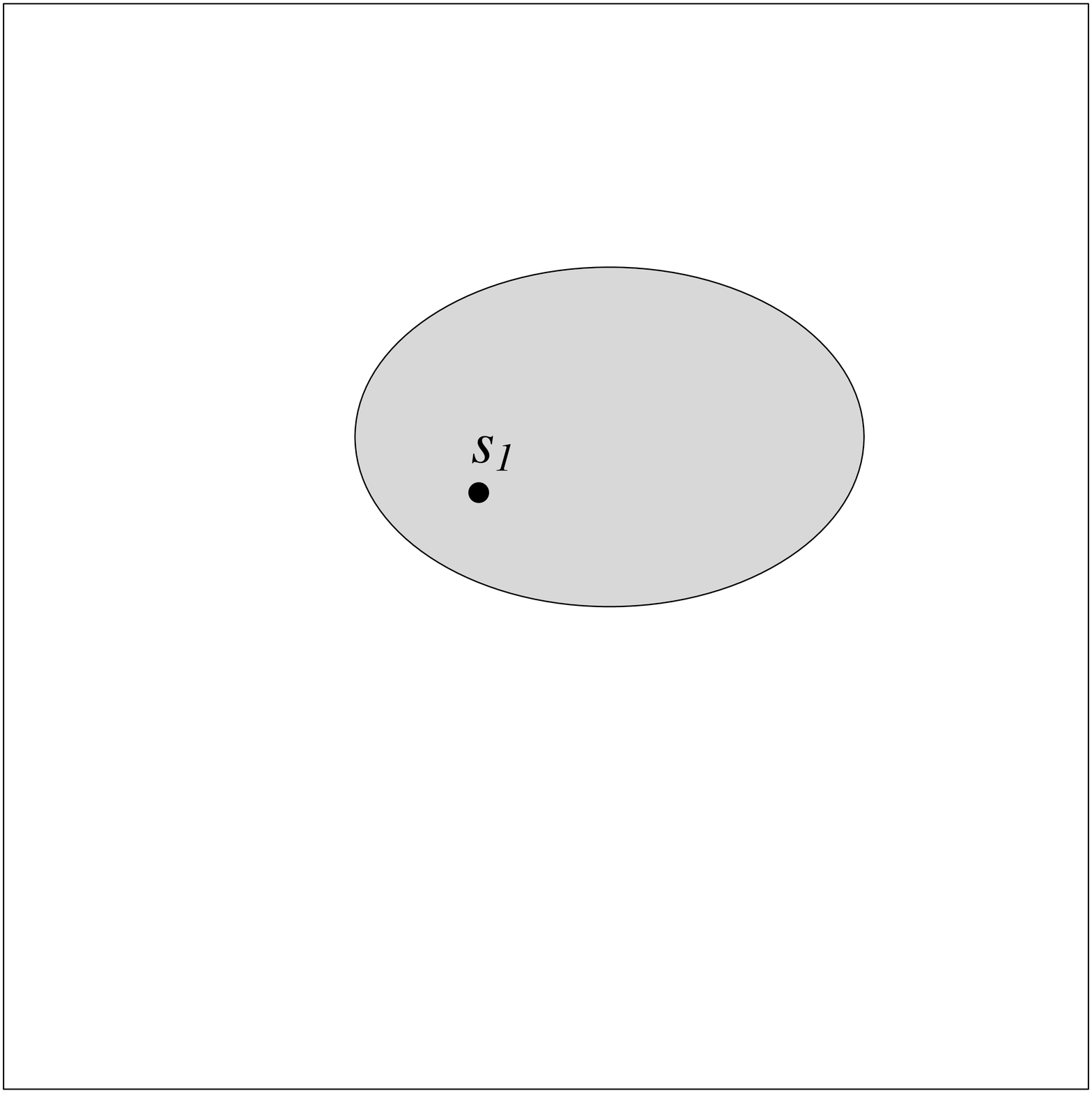}
        \label{boundary1}
    }
    \subfigure[]
    {
        \includegraphics[width=0.23\textwidth]{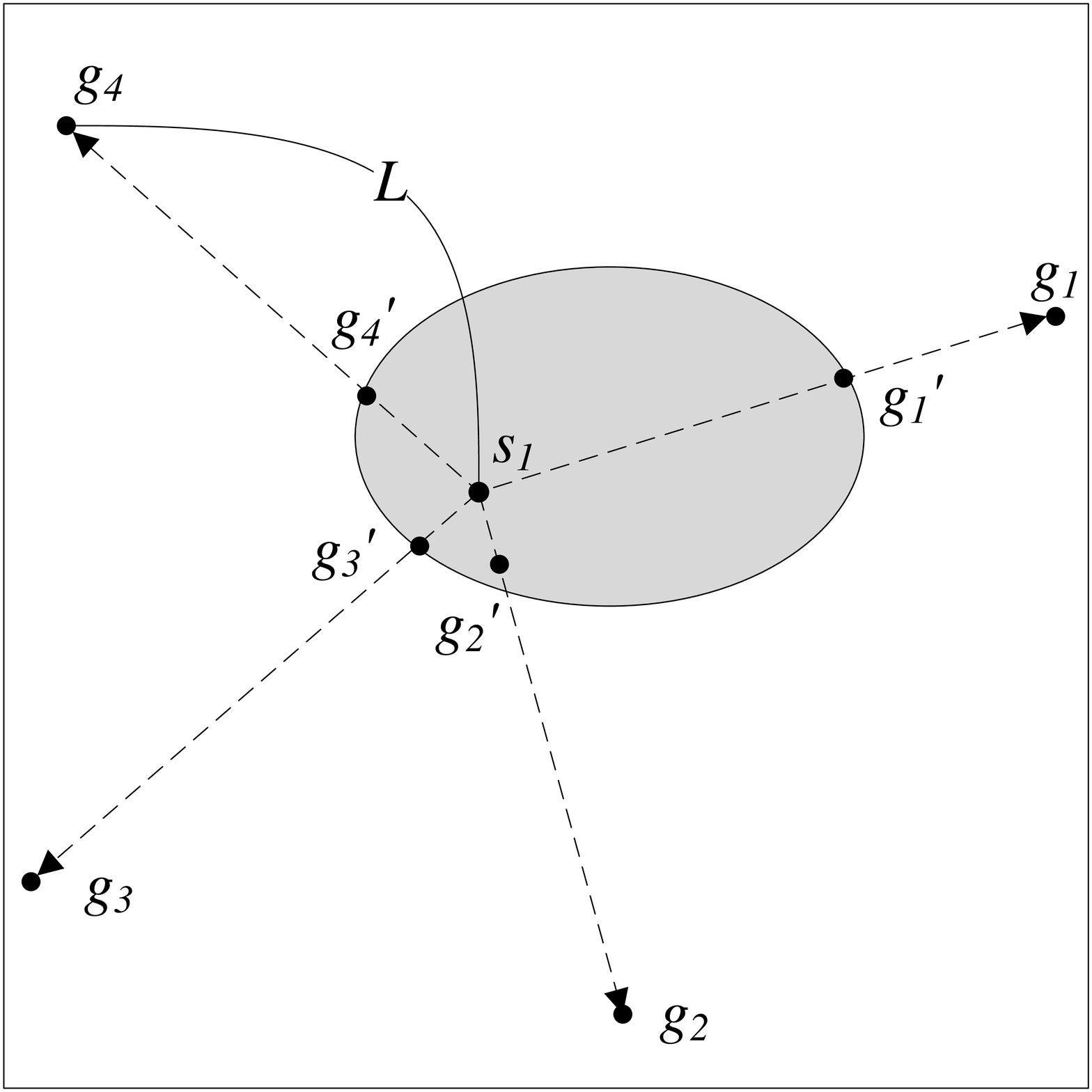}
        \label{boundary2}
    }

    \caption{An illustrative example of SB in a two-dimensional input domain.}
    \label{FIG:ABMexample}
\end{figure}

\begin{figure}[!t]
\centering
    \subfigure[FSB-1]
    {
        \includegraphics[width=0.23\textwidth]{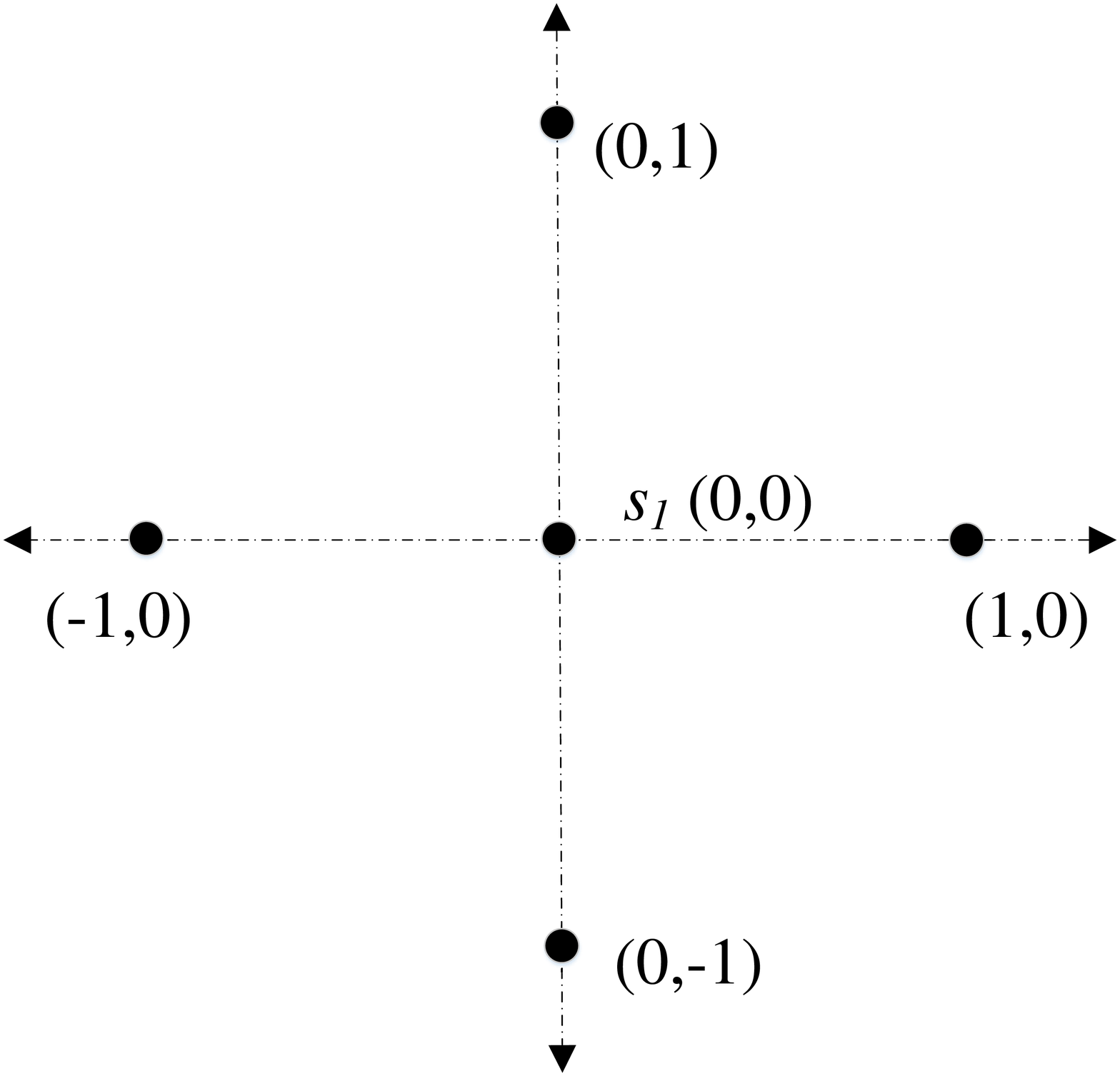}
        \label{FIG:direction1}
    }
    \subfigure[FSB-2]
    {
        \includegraphics[width=0.23\textwidth]{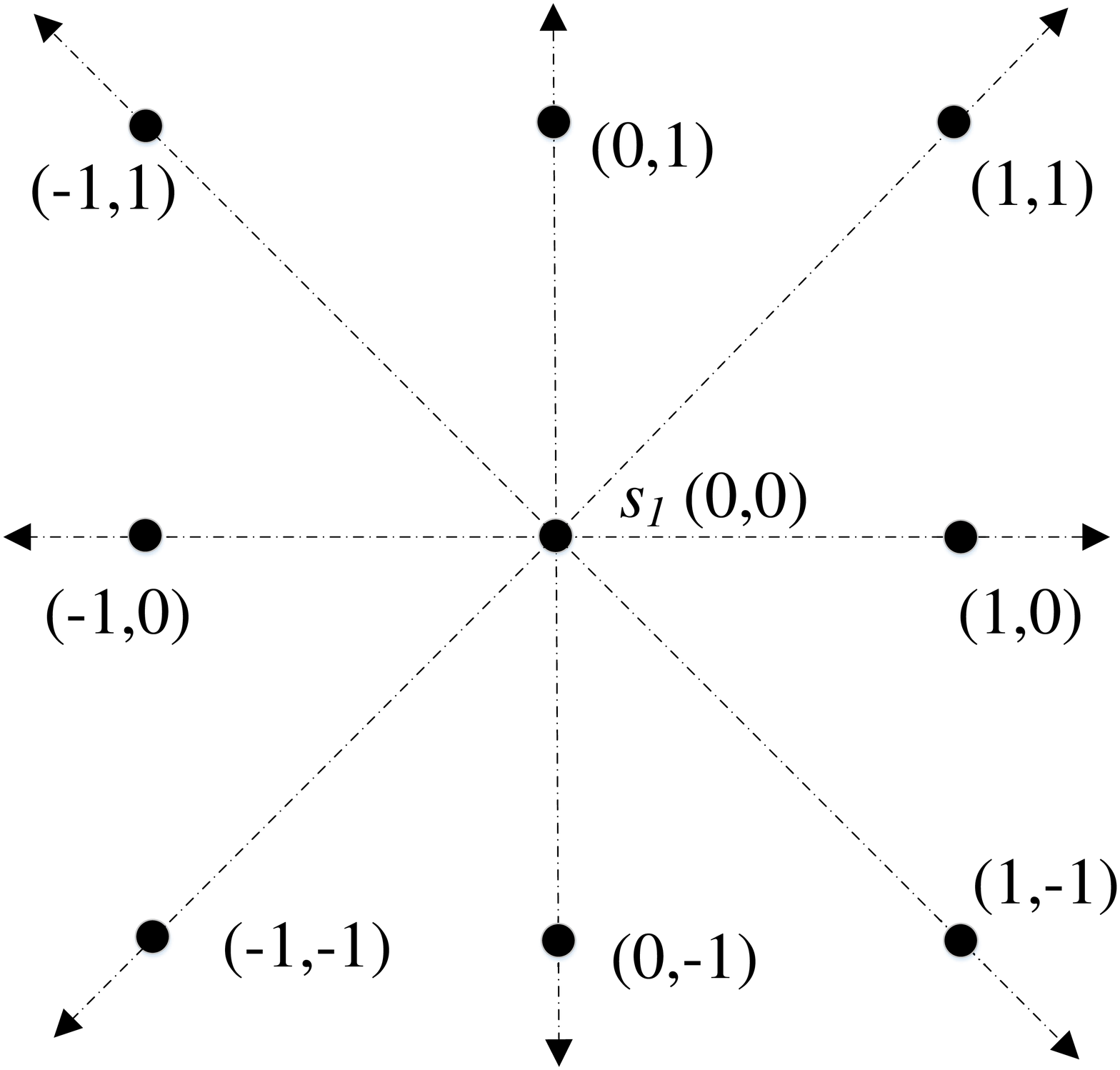}
        \label{FIG:direction2}
    }
    \caption{An illustrative example of extension orientations for FSB in a two-dimensional input domain.}
    \label{FIG:direction}
\end{figure}

\begin{figure*}[!b]
\centering
    \subfigure[]
    {
        \includegraphics[width=0.485\textwidth]{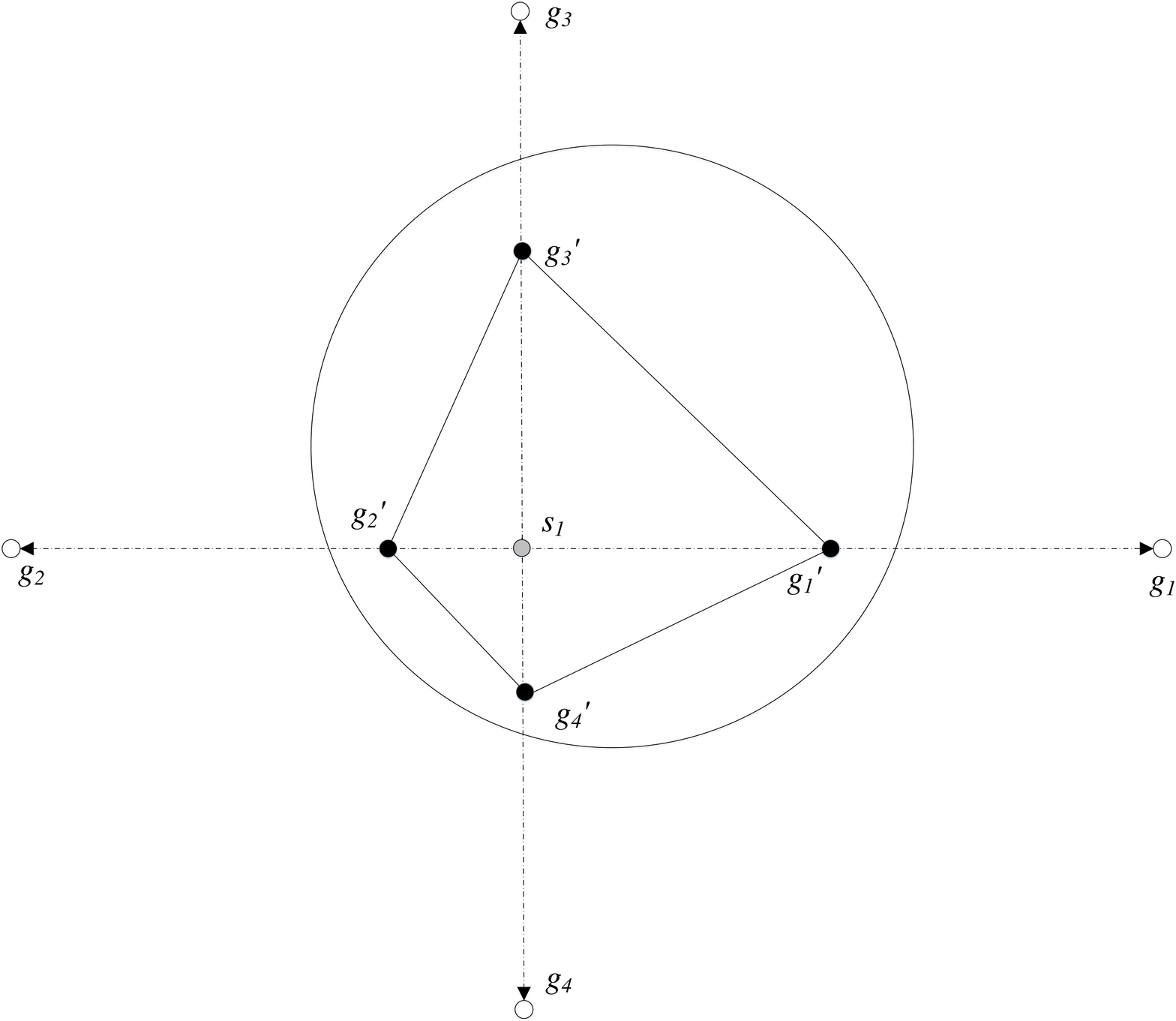}
        \label{iteration1}
    }
    \subfigure[]
    {
        \includegraphics[width=0.485\textwidth]{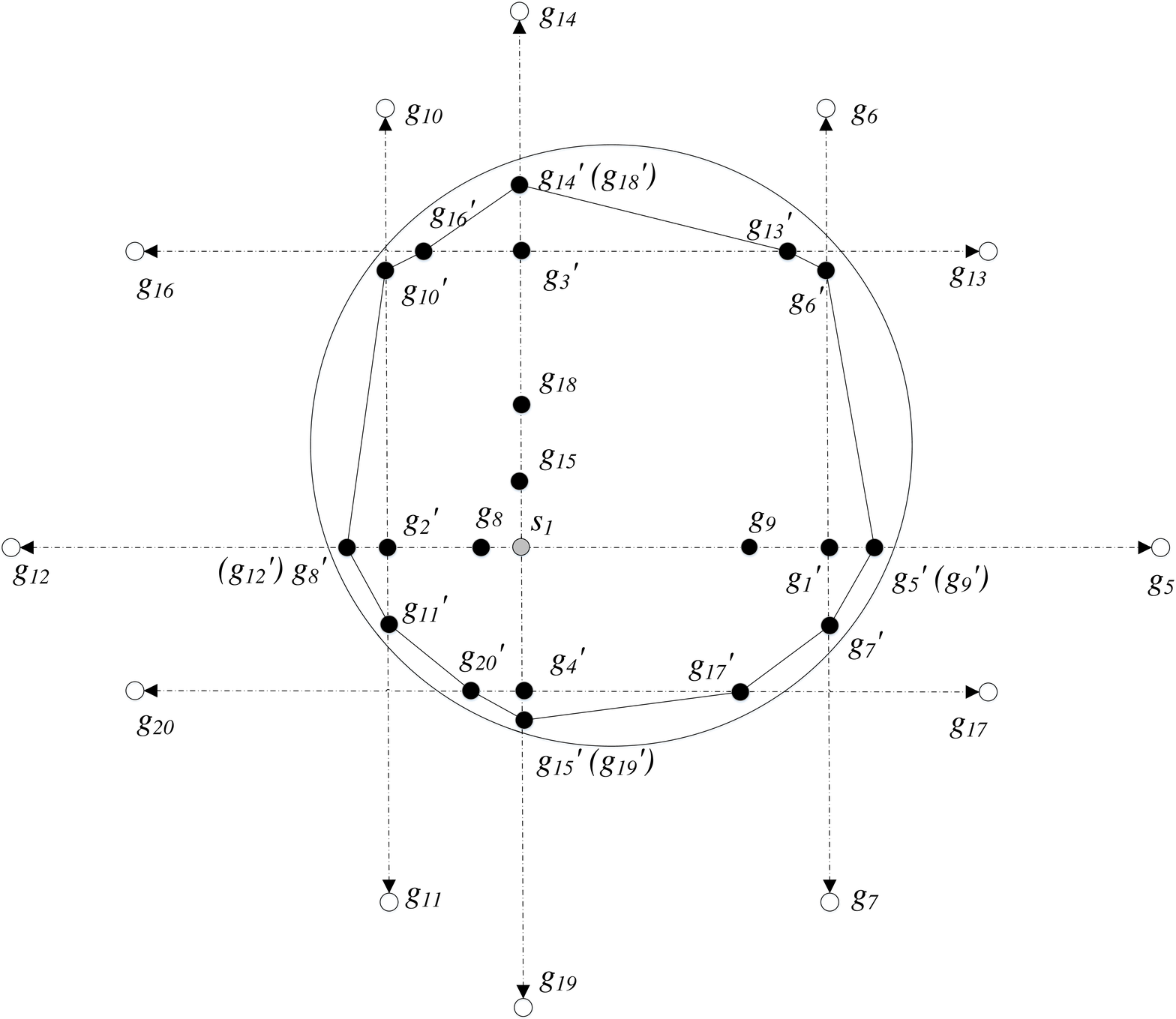}
        \label{iteration2}
    }
    \caption{An illustrative example of FSB-1.}
    \label{FIG:FABM1}
\end{figure*}

Figure~\ref{FIG:ABMexample} shows an example to illustrate SB applying in a two-dimensional input domain. As shown in Figure~\ref{boundary1}, the square is the input domain, and the shaded region is the failure region; while $s_1$ is a failure-causing input (i.e., an input of SB). Figure~\ref{boundary2} shows the SB process, which starts with four extension orientations. It can be seen that based on the failure-causing source input $s_1$, the first extension with the length $L$ could generate the next four test inputs (i.e., $g_1$, $g_2$, $g_3$, and $g_4$, respectively). After many iterations, SB obtains four failure-causing boundary inputs, i.e., $g_1'$, $g_2'$, $g_3'$, and $g_4'$, respectively.

\subsubsection{Method}
\label{SEC:algorithm}

In this study, we propose two main methods to support SB, i.e., \textit{Fixed-orientation Search for Boundary} (FSB) and \textit{Diverse-orientation Search for Boundary} (DSB). The former applies the extension with the fixed orientations to SB; while the latter makes use of diverse orientations for the extension process.

\begin{figure*}[!b]
\centering
    \subfigure[]
    {
        \includegraphics[width=0.38\textwidth]{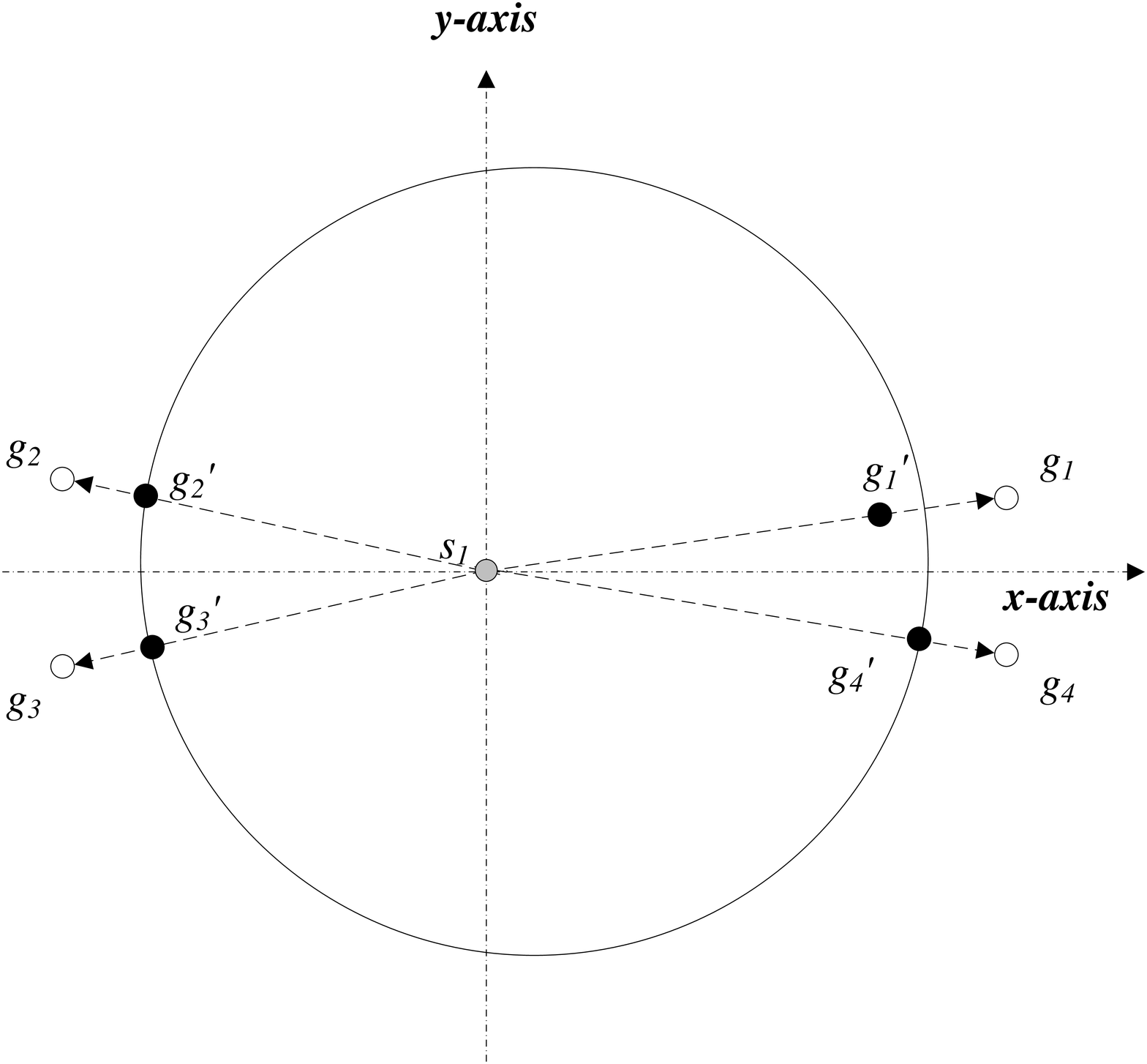}
        \label{DABM1}
    }
    \hspace{10mm}
    \subfigure[]
    {
        \includegraphics[width=0.38\textwidth]{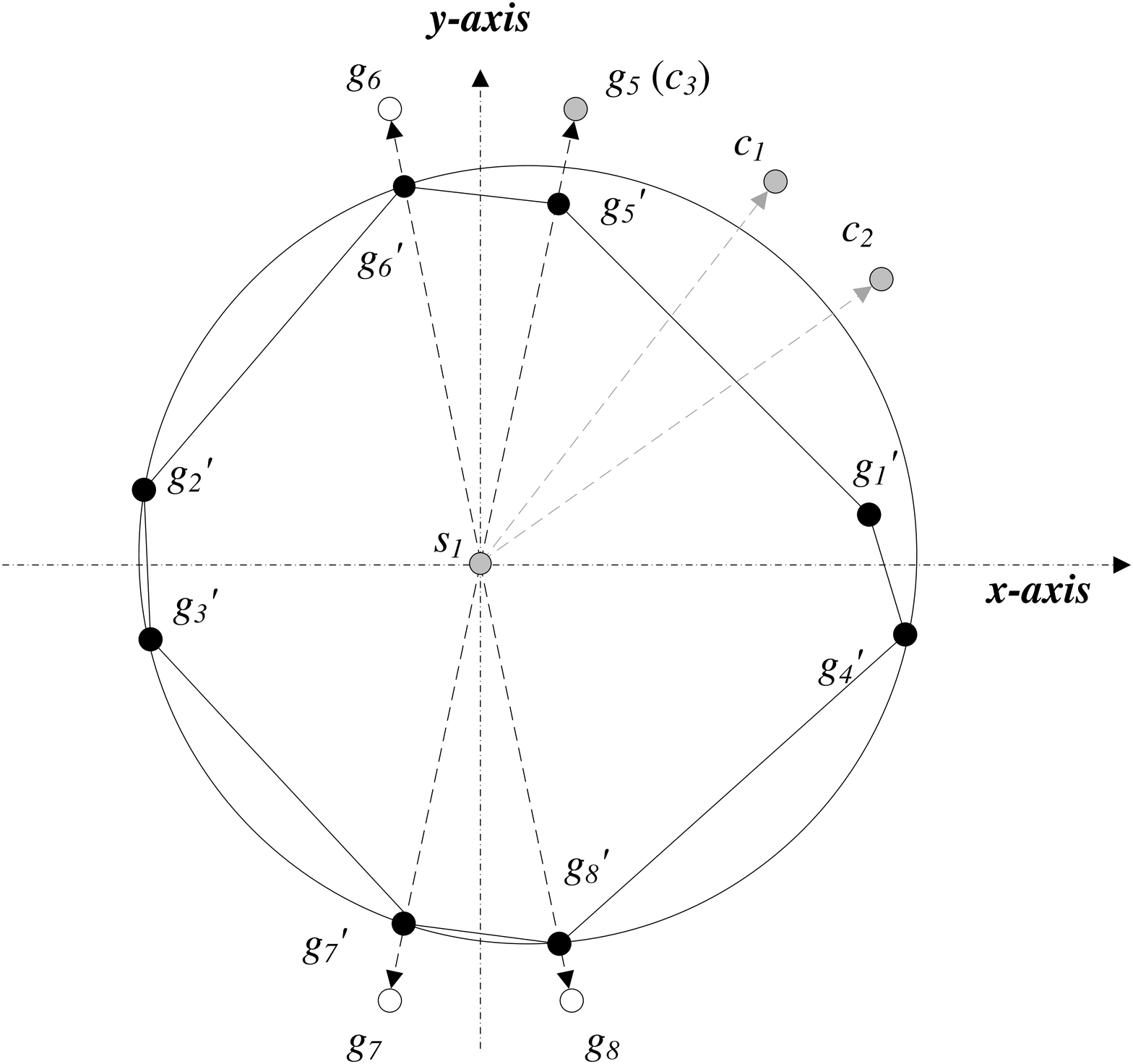}
        \label{DABM2}
    }
    \caption{An illustrative example of DSB.}
    \label{DABM}
\end{figure*}

\textbf{\textit{1) Fixed-orientation Search for Boundary (FSB):}} With a failure-causing source input, FSB makes use of the fixed extension orientations to implement the process of SB. Generally speaking, the number of extension orientations can be determined by the tester. In this study we mainly focus on two basic methods (namely FSB-1 and FSB-2) to guide the selection of extension orientations. These two methods attempt to use the information about coordinate axes or {orthants}\footnote{{The orthant means the analogue in a $n$-dimensional Euclidean space of a quadrant in a two-dimensional plan or an octant in a three-dimensional space.}} for determining the extension orientations, which means that the number of extension orientations depends on the dimension of the input domain. { FSB-1 only uses the information about coordinate axes; while FSB-2 uses the information about both coordinate axes and orthants. More specifically, FSB-1 adopts the positive and negative coordinate axis as two extension orientations for each dimension. However, in addition to extension orientations, FSB-2 makes use of new information from each orthant. The direction of expansion in each orthant is arbitrary for FSB-2. The testers can choose any angle (as an input parameter) before identifying a failure region and keep the value constant during the subsequent identification. In this paper, without loss of generality, we set the angle to $45\degree$ that guarantees to partition each orthant equally.} In other words, consider a $d$-dimensional input domain, FSB-1 could obtain $2d$ extension orientations; while FSB-2 could obtain $(2d+2^d)$ extension orientations, where $d$ is larger than one\footnote{When $d=1$, there is no orthant in fact, and hence FSB-2 becomes FSB-1.}. Figure~\ref{FIG:direction} provides an example to illustrate extension orientations for FSB-1 and FSB-2 in a two-dimensional input domain, where four and eight extension orientations are used for FSB-1 and FSB-2, respectively.

Since FSB adopts fixed extension orientations during the whole process, it should consider different failure-causing source inputs for further extension. The main reason for this is that each failure-causing source input can only be used for a fixed number of extensions to generate fixed failure-causing boundary inputs. Therefore, if a failure-causing input $tc$ is selected as the failure-causing source input for the next extension work, $tc$ will not be chosen again in FSB. In other words, each failure-causing input has only one chance to be used for identifying failure-causing boundary inputs. Moreover, with multiple failure-causing source inputs, FSB randomly chooses one among them for extension. Similarly, with more than one extension orientation, FSB randomly selects one orientation for the next extension.

It is clear that FSB-1 and FSB-2 are very similar, and the only difference between them is in selecting extension orientations. {In Figure~\ref{FIG:FABM1}, we presented an example to illustrate FSB-1 with white circles representing the successful inputs, a grey circle for failure-causing source input, and black circles for other failure-causing inputs.} Figure~\ref{iteration1} shows the first extension process to identify four failure-causing boundary inputs (i.e., $g_1'$, $g_2'$, $g_3'$, and $g_4$', from four extension orientations), based on a failure-causing source input $s_1$. After that, each failure-causing boundary input is considered as the next failure-causing source input, and the extension process will be repeated to obtain other new failure-causing boundary inputs, as shown in Figure~\ref{iteration2}.

\textbf{\textit{2) Diverse-orientation Search for Boundary (DSB):}}
Different to FSB, DSB adopts adaptive rather than fixed extension orientations, indicating that it may make use of different extension orientations during the SB process. Intuitively speaking, DSB may identify better failure regions than FSB, because DSB could identify failure-causing boundary inputs with a greater variety.

To generate diverse extension orientations, DSB applies the principle of one \textit{Adaptive Random Testing} (ART)~\cite{Huang2019} algorithm, i.e., \textit{Fixed-Size-Candidate-Set ART} (FSCS-ART)~\cite{Chen2010}. Initially, DSB selects an extension orientation in a random manner. When it is required to choose the next extension orientation, DSB randomly generates $k$ candidate extension orientations, and then selects a candidate as the next extension orientation such that it is farthest away from previously selected extension orientations. More specifically, suppose that $C=\{\overrightarrow{\mu_1},\overrightarrow{\mu_2},\cdots,\overrightarrow{\mu_k}\}$ is the set of candidate extension orientations; while $S=\{\overrightarrow{\nu_1},\overrightarrow{\nu_2},\cdots,\overrightarrow{\nu_n}\}$ is the set of already selected extension orientations. The criterion is to choose a candidate $\overrightarrow{\mu_h}~(1 \leq h \leq k)$ from $C$ as the next extension orientation, satisfying $\forall i \in \{1,2,\cdots,n\}$,
\begin{equation}
\label{EQ:fscs}
\min_{j=1}^{n} dist(\overrightarrow{\mu_h}, \overrightarrow{\nu_j}) \geq \min_{j=1}^{n} dist(\overrightarrow{\mu_i}, \overrightarrow{\nu_j})
\end{equation}
where $dist(\overrightarrow{\mu_i}, \overrightarrow{\nu_j})$ can be defined as the cosine distance, i.e.,
\begin{equation}
dist(\overrightarrow{\mu_i}, \overrightarrow{\nu_j}) = 1 - \frac{\overrightarrow{\mu_i} \cdot \overrightarrow{\nu_j}}{\big\|\overrightarrow{\mu_i}\big\|\big\|\overrightarrow{\nu_j}\big\|}
\end{equation}
To reduce the computational overhead, DSB follows the concept of \textit{mirroring}~\cite{Huang2015}. In detail, DSB applies the above process to a orthant for generating each extension orientation $\overrightarrow{\alpha}$, and then directly maps $\overrightarrow{\alpha}$ to each of remaining orthants for generating their own extension orientations. Obviously, each orthant could then have a wider range of orientations.

Since DSB makes use of diverse extension directions, it is generally recommended to use the same failure-causing input as the failure-causing source input. In this paper, we have therefore used the first failure-causing input (the parameter of SB) as the failure-causing source input, and kept it remain the same during the whole process.

Figure~\ref{DABM} provides an example to illustrate DSB in the two-dimensional input domain. DSB first generates a random extension orientation, and then extends a certain length from the failure-causing source input $s_1$ to $g_1$ (in fact this extension orientation is $\overrightarrow{\nu_1} = g_1-s_1$). By mapping $\overrightarrow{\nu_1}$ from the first quadrant to another three quadrants, we could obtain three extension orientations, i.e., $\overrightarrow{\nu_2} = g_2-s_1$, $\overrightarrow{\nu_3} = g_3-s_1$, and $\overrightarrow{\nu_4} = g_4-s_1$, respectively. DSB then applies Algorithm~\ref{ALG:find_boundary} to each quadrant, resulting in four failure-causing boundary inputs, i.e., $g_1'$, $g_2'$, $g_3'$, and $g_4'$, respectively (as shown in Figure~\ref{DABM1}). To generate the second extension orientation in the first quadrant, three candidates are assumed to be randomly selected ($\overrightarrow{\mu_1}=c_1-s_1$, $\overrightarrow{\mu_2}=c_2-s_1$, and $\overrightarrow{\mu_3}=c_3-s_1$). By applying the criterion (Eq.~(\ref{EQ:fscs})), the next extension orientation would be $\overrightarrow{\mu_3}$ (i.e., $c_3$ becomes $g_5$). Similar to the process in Figure~\ref{DABM1}, an extension orientation could be confirmed in each of the other three quadrants, resulting in identifying all four failure-causing boundary inputs (i.e., $g_5'$, $g_6'$, $g_7'$, and $g_8'$, respectively).

\section{Experimental Setup}
\label{SEC:Simulations}
We conducted a series of simulations and empirical studies to evaluate the performance of our proposed SB methods over different failure regions. In this section, we provide the details about our experimental settings and results.

\subsection{Research Questions}
\label{SEC:questions}
{SB attempts to identify the AFR rather than the RFR, such that the AFR is as close to the RFR as possible. In addition, we also need to check the identification time (i.e., the time taken to identify the failure region) of SB. Therefore, we aim at investigating identification effectiveness and efficiency of SB.}

\begin{itemize}
\item \emph{\textbf{RQ1:} How effective is SB in identifying failure regions?}
\item \emph{\textbf{RQ2:} What is the performance of SB in terms of the identification time?}
\end{itemize}

\begin{table*}[!t]
\footnotesize
\centering
 \caption{Six object programs used}
  \label{TAB:Program}

    \begin{tabular}{@{}c|c|c|c|c|c|c|c@{}}
     \hline

      \multirow{2}*{\textbf{Program}} &\multirow{2}*{\textbf{Dimension}} &\multicolumn{2}{c|}{\textbf{Input Domain}} &\textbf{{Source Lines}} &\multirow{2}*{\textbf{Fault Types}} &\textbf{Total} &\multirow{2}*{\textbf{Failure Pattern Description}}\\
      \cline{3-4}
       & &\textbf{From} &\textbf{To} &\textbf{{of Code}} & &\textbf{Faults} &\\ \hline

       \texttt{airy}   &1  &(-5000.0)   &(5000.0)   &43 &CR                &1 &A line segment\\ \hline

       \texttt{gammq}  &2   &(0, 0)     &(1700.0, 40.0)   &106 &ROR, CR      &5  &A strip with the rotation angle $45^\circ$  \\ \hline

        \texttt{expint}  &2   &(0, 0)     &(4095.0, 4095.0)   &87 &AOIS      &1  & A irregular failure region \\ \hline

        \texttt{bessj}  &2   &(2, 0)     &(300.0, 1500.0)   &99 &AOR, ROR, CR      &4  &A right-angled triangle \\ \hline

        \multirow{2}*{\texttt{triangle}}  &\multirow{2}*{3}   &(0, 0,    &(255.0, 255.0,    &\multirow{2}*{26} &\multirow{2}*{COI}      &\multirow{2}*{1}  &\multirow{2}*{A irregular failure region} \\

        & &0) &255.0) & & & &\\ \hline

       \multirow{2}*{\texttt{cel}} &\multirow{2}*{4} &(0.001, 0.001, &(1.0, 300.0, &\multirow{2}*{49} &\multirow{2}*{AOR, ROR, CR} &\multirow{2}*{3} &\multirow{2}*{A strip with the rotation angle $0^\circ$}\\
       & &0.001, 0.001) &1.0, 1.0) & & & &\\
       \hline

    \multicolumn{8}{l}{AOR: arithmetic operator replacement, CR: constant replacement, ROR: relational operator replacement, AOIS: arithmetic operator insertion } \\
       \multicolumn{8}{l} {short-cut, and COI: conditional operator insertion.}
  \end{tabular}

\end{table*}

\subsection{Variables and Measures}

\subsubsection{Independent Variables}

{Our SB methods proposed in this paper are independent variables in the experiment studies. The IFR methods described in ~\cite{Ahmad2013,Ahmad2014} are limited to only integer input domains, and hence are not compared in this study (please refer to Section~\ref{SEC:RelatedWork} for more detailed discussions). We conducted experiments to compare different versions of SB including FSB-1, FSB-2, and DSB by adopting the `float' and `double' input domains.}

\subsubsection{Dependent Variables}
In the simulations, we used a unit square/hypercube as the input domain $D$. \textit{Failure rate}, denoted by $\theta$, is required to be determined in advance. Let $\mathcal{S}_\textit{RFR}$ be the size of the RFR, which can be calculated by $\mathcal{S}_\textit{RFR}=\theta * |D|$; while $\mathcal{S}_\textit{AFR}$ be the size of the AFR, which can be calculated by the convex hull algorithm~\cite{Barber1996}. To answer RQ1, we used $\mathcal{S}_\textit{ratio}$ to measure the effectiveness of the proposed methods, i.e.,
\begin{equation}
\mathcal{S}_\textit{ratio}=\frac{\mathcal{S}_\textit{AFR}}{\mathcal{S}_\textit{RFR}}
\label{a_ratio}
\end{equation}
Intuitively, higher $\mathcal{S}_\textit{ratio}$ indicates better effectiveness for identifying the failure region. {It should be noted that the shape of RFR involved in the experiments is convex, which is after Assumption 3. Therefore, for the simulations and empirical studies, $S_\textit{AFR}$ is always less than or equal to $S_\textit{RFR}$. As a reminder, when the shape of RFR is concave, $S_\textit{ratio}$ may be greater than 1.0, because the identified failure region may contain some successful inputs. For such case, we will discuss more details in Section~\ref{SEC:Weaken}.}
In addition, to answer RQ2, we collected the identification time for obtaining a given number of failure-causing boundary inputs to measure the efficiency of the proposed methods. Obviously, shorter identification time implies better efficiency.

\subsection{Simulations and Empirical Studies}

\subsubsection{Simulations}
In our experiments, we attempted to simulate faulty programs under different situations.

{ Considering a $d$-dimensional input domain $D$, without loss of generality, we assume that $D = \{(x_1,x_2,\cdots,x_d)| 0 \leq x_i < 1.0, i =1,2,\cdots,d\}$, abbreviated as $[0,1.0)^d$. The failure rate of the RFR is denoted by $\theta$, where $0 < \theta \leq 1.0$. Obviously, the failure pattern of the RFR may be influenced by factors such as shapes, compactness, and orientation. Due to page limitation, this paper only focuses on two-dimensional, three-dimensional and four-dimensional input domains, i.e., $d=2$, $d=3$ and $d=4$. With the same reason, only rectangle (including cuboid, hypercube) and ellipse (including ellipsoid, hyperellipsoid) will be taken into consideration as RFR shapes.}

{ RFR compactness refers to the ratio among edge lengths for rectangle or cuboid RFR; and the ratio among axis lengths for ellipse or ellipsoid RFR. In this study, we used a parameter $\delta~(\delta \geq 1)$ to describe compactness of RFR. In other words, the ratio among edge or axis lengths could be described as $1:\delta$, $1:\delta:\delta$, and $1:\delta:\delta:\delta$ in 2D, 3D, and 4D space, respectively. Intuitively, the smaller $\delta$ is, the more compact the RFR. In addition, we adopted a parameter $\gamma$ to represent rotation angle of the RFR (i.e., its orientation); and another parameter $N$ to represent the number of failure-causing boundary inputs for stopping the SB process. As discussed in Algorithm~\ref{ALG:find_boundary}, there exists a parameter $L$ named \textit{extension length}, which is required to be set in advance. Moreover, a threshold parameter $\lambda$ is defined as the number of consecutive iterations without revealing any failure-causing inputs.}

{It is noted that the value of $\lambda$ has an impact on the performance of SB. Therefore, we conducted a preliminary study to investigate the effect of $\lambda$ on performance of our methods. The range of value of $\lambda$ was first arbitrarily set from $2$ to $10$ with an increment of $2$ and then from $10$ to $100$ with an increment of $10$. From the results, it is observed that in general, a larger $\lambda$ gives rise to a better estimated boundary, thus obtaining a more accurate AFR. On the other hand, a larger $\lambda$ may require longer identification time. Furthermore, for $\lambda \geq 20$, the difference in areas of identified failure regions is small in our simulations. Hence, we set $\lambda = 20$ in our experiment.}

With this in mind, all parameters in our simulation experiments were assigned as follows:
\begin{itemize}
  \item Dimension of the input domain $d$: 2, 3, and 4.
  \item Failure rate of the RFR $\theta$ : 0.001 and 0.005.
  \item Shape of the RFR: Rectangle and ellipse when $d=2$, cuboid and ellipsoid when $d=3$, while hypercube and hyperellipsoid when $d=4$.
   \item Ratio among edge or axis lengths of the RFR $\delta$: 1, 10, and 100.
  \item Rotation angle of the RFR $\gamma$: $0^\circ$, $30^\circ$, $60^\circ$, $90^\circ$, $120^\circ$, $150^\circ$, and $180^\circ$.
  \item Threshold $\lambda$ to stop Algorithm~\ref{ALG:find_boundary}: 20.
  \item Extension length $L$: Length of the input domain, i.e., $1.0$.
  \item Number of failure-causing boundary inputs $N$: 100 and 1000.
\end{itemize}
Therefore, the number of all possible settings would be $3\times 2 \times 2 \times 3 \times 7 \times 1 \times 1 \times 2 =504$.

\subsubsection{Empirical Studies}
Following the assumptions described in Section~\ref{SEC:assumptions}, we selected six representative object programs, \texttt{airy}, \texttt{gammq}, \texttt{cel}, \texttt{expint}, \texttt{bessj}, and \texttt{triangle} from Numerical Recipes~\cite{Press1986} and ACM Collected Algorithms~\cite{ACM1980}, among which \texttt{airy}, \texttt{gammq}, and \texttt{cel} were converted into C or C++; while \texttt{expint}, \texttt{bessj}, and \texttt{triangle} were written in or translated into Java. These object programs were popularly used in other similar studies~\cite{Chen2010,Huang2015,Arcuri2011}. Each program was seeded by various types of faults to produce a failure region. Table~\ref{TAB:Program} lists detailed information of all six programs. In our experiments, once a failure is detected, IFR methods were employed to identify the $N$ failure-causing boundary test cases. Finally, we recorded the area and identification time of the AFR, together with the number of iterations required for each of these programs.

Similar to the simulation studies, some parameters of FSB and DSB are required to be assigned in advance. For this purpose, we chose the same assigned values as in the simulation studies. For example, the threshold $\lambda$ is set to 20 for stopping Algorithm~\ref{ALG:find_boundary}; while the extension length $L$ is set to the length of the input domain.

To provide a failure-causing source test case as the algorithm input for both FSB and DSB, one particular ART algorithm (i.e., FSCS-ART~\cite{Chen2010}) was used in all our simulations and object programs for this purpose. Moreover, considering the randomness of FSB and DSB, we ran each setup of the experiment 3000 times to obtain averages for analysis.

\begin{figure*}[!t]
\centering
    \subfigure[$d=2,\delta= 1$]
    {
        \includegraphics[width=0.315\textwidth]{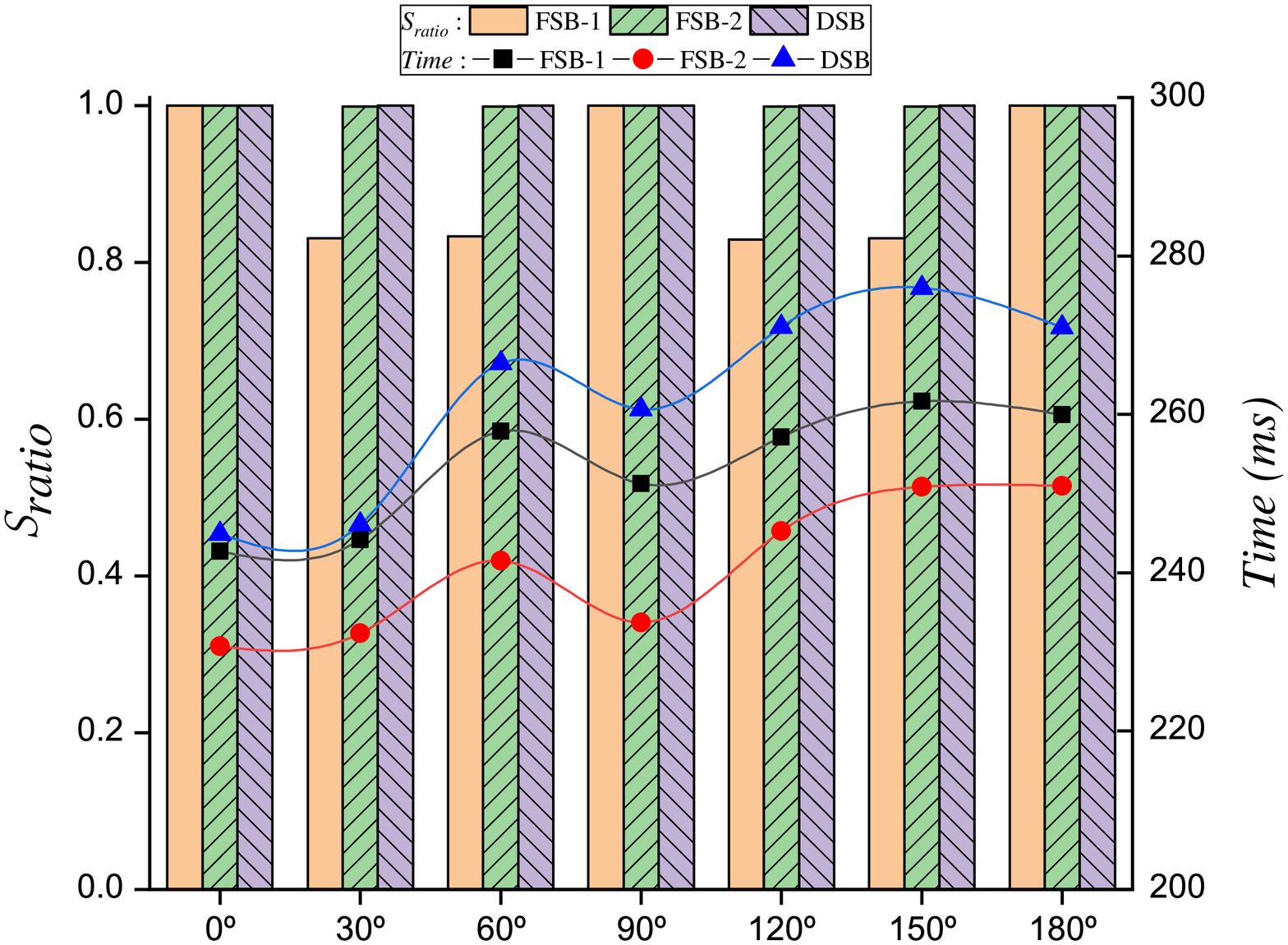}
        \label{r_2d_1}
    }
    \subfigure[$d=3,\delta = 1$]
    {
        \includegraphics[width=0.315\textwidth]{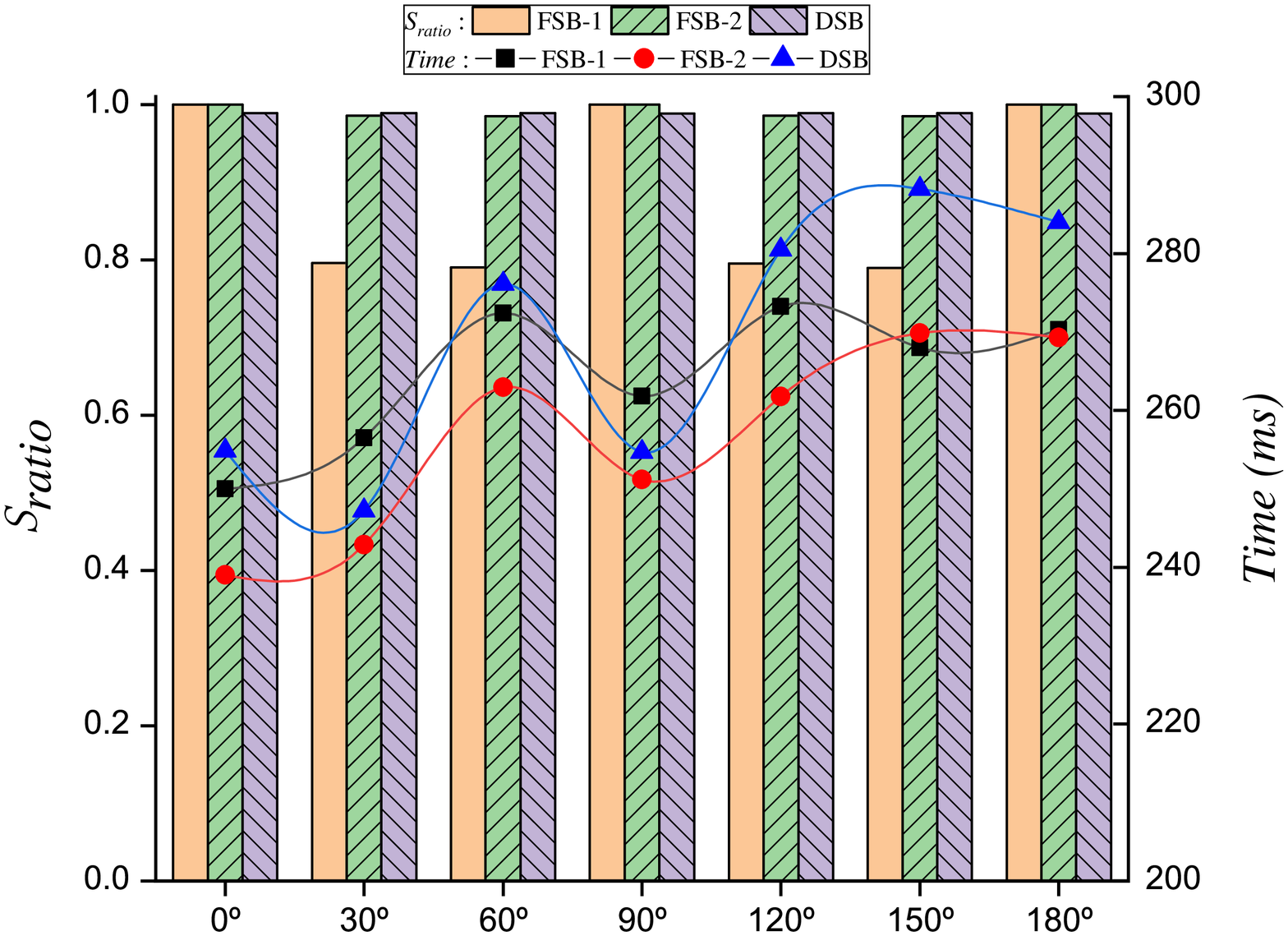}
        \label{r_3d_1}
    }
    \subfigure[$d=4,\delta = 1$]
    {
        \includegraphics[width=0.315\textwidth]{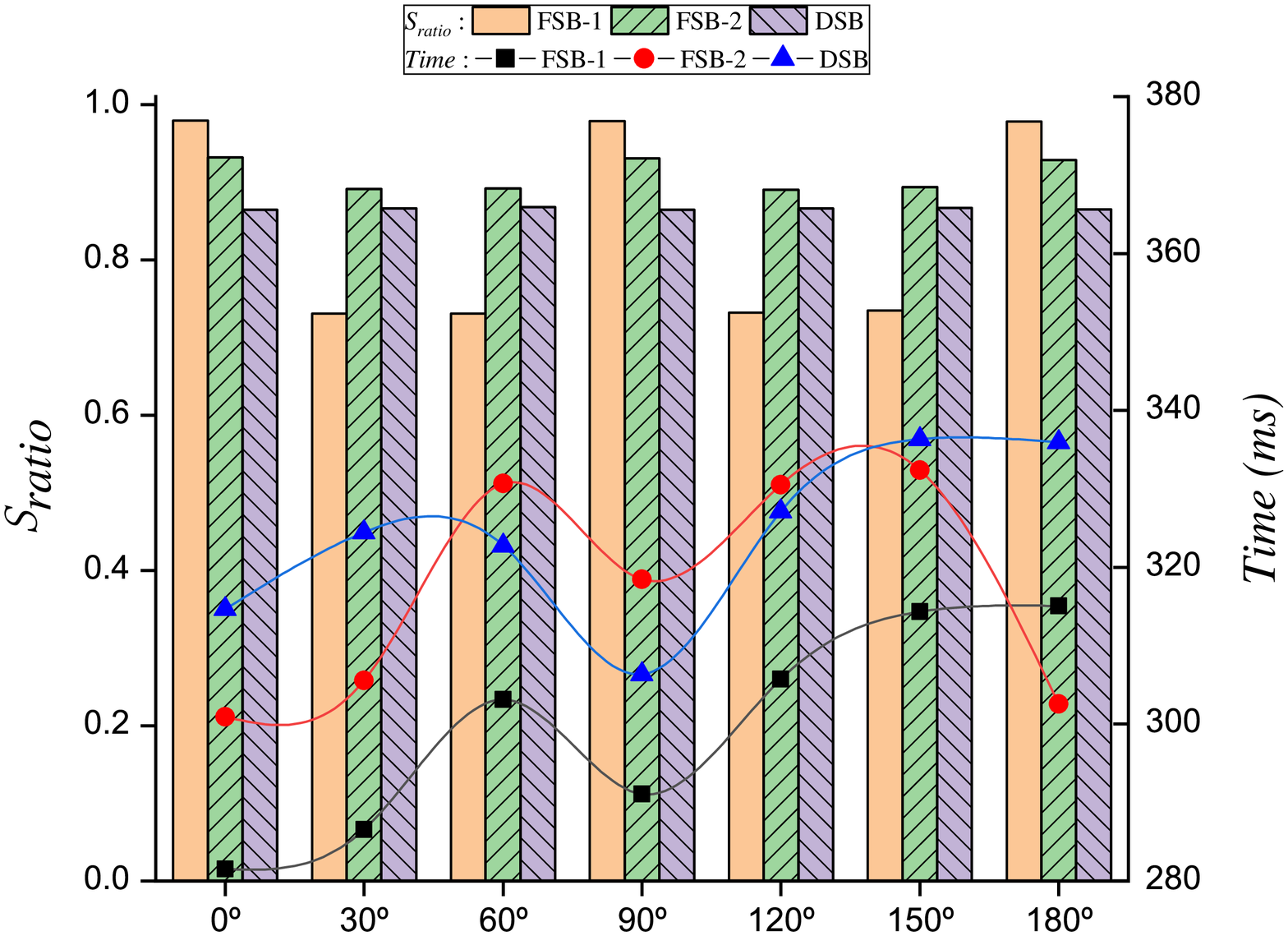}
        \label{r_4d_1}
    }
    \subfigure[$d=2,\delta= 10$]
    {
        \includegraphics[width=0.315\textwidth]{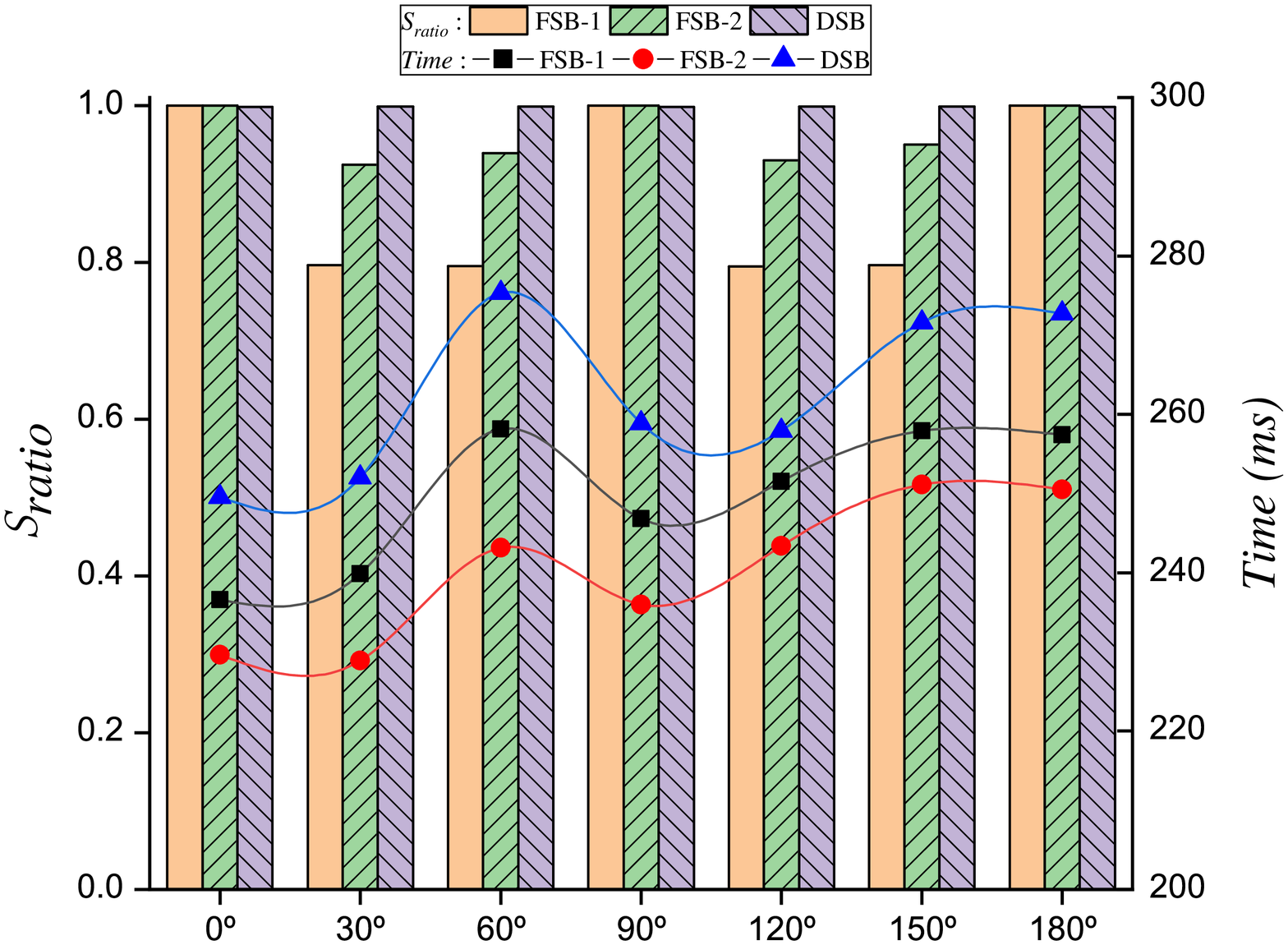}
        \label{r_2d_10}
    }
    \subfigure[$d=3,\delta= 10$]
    {
        \includegraphics[width=0.315\textwidth]{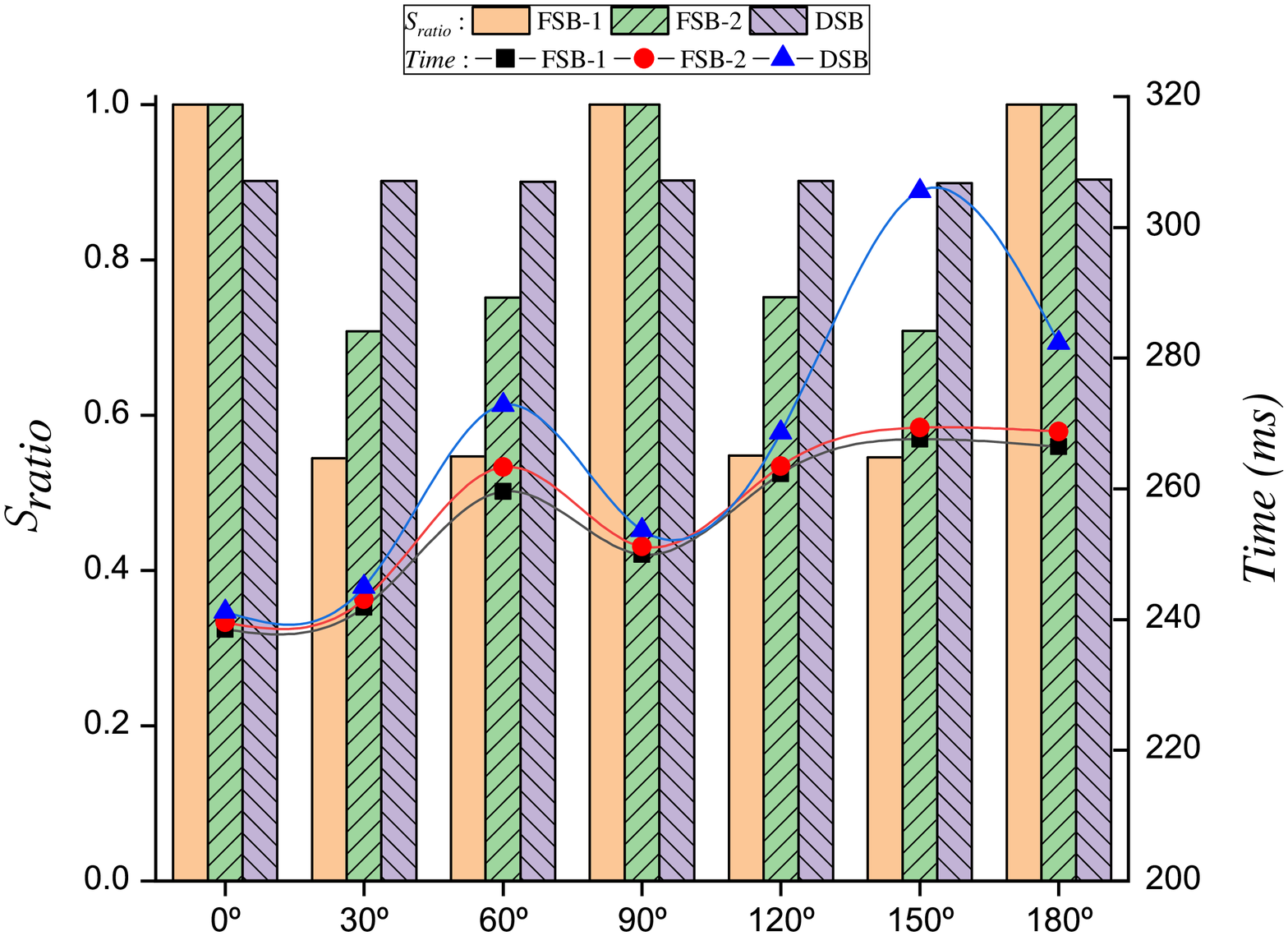}
        \label{r_3d_10}
    }
    \subfigure[$d=4,\delta= 10$]
    {
        \includegraphics[width=0.315\textwidth]{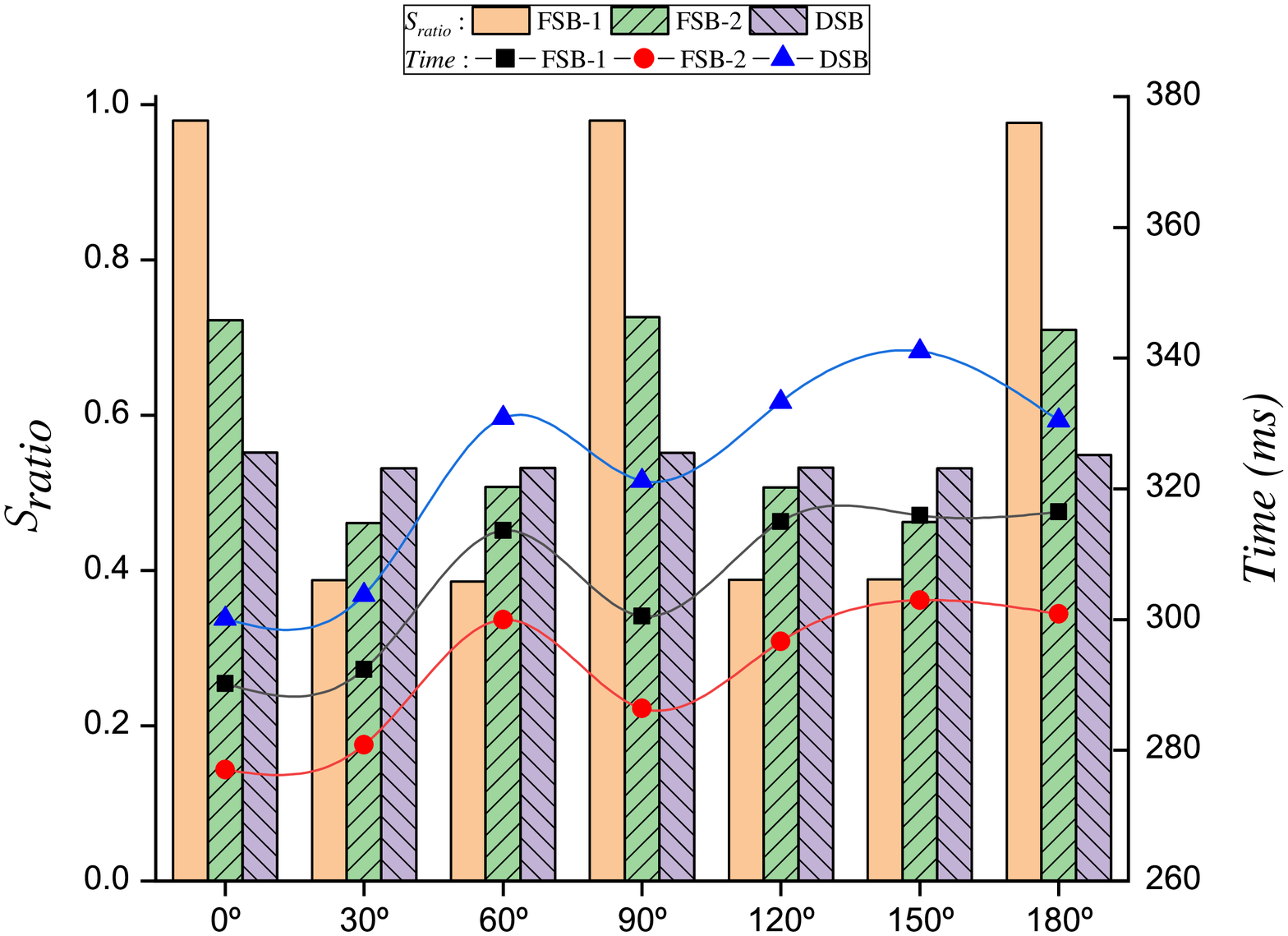}
        \label{r_4d_10}
    }
    \subfigure[$d=2,\delta= 100$]
    {
        \includegraphics[width=0.315\textwidth]{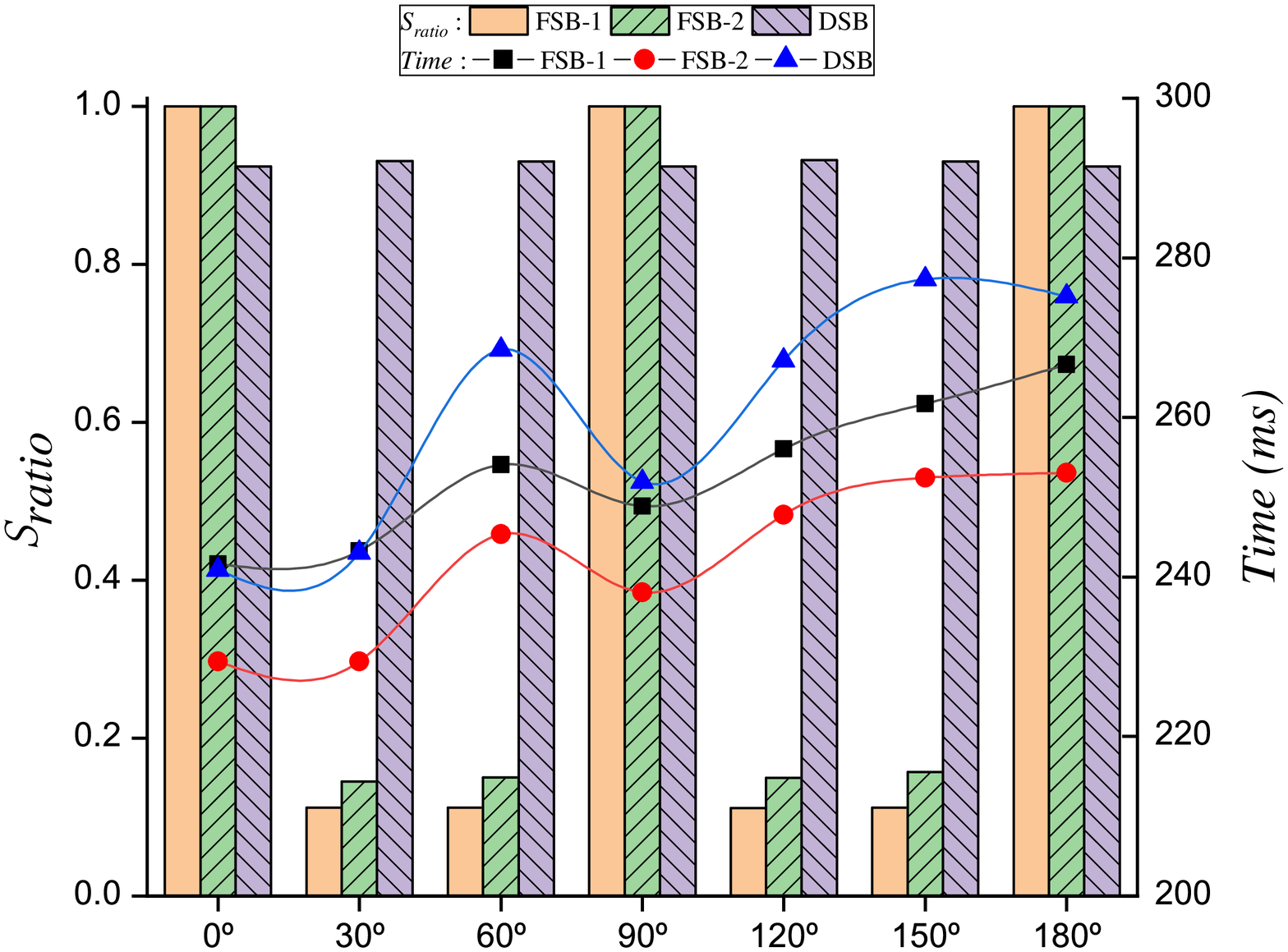}
        \label{r_2d_100}
    }
    \subfigure[$d=3,\delta= 100$ ]
    {
        \includegraphics[width=0.315\textwidth]{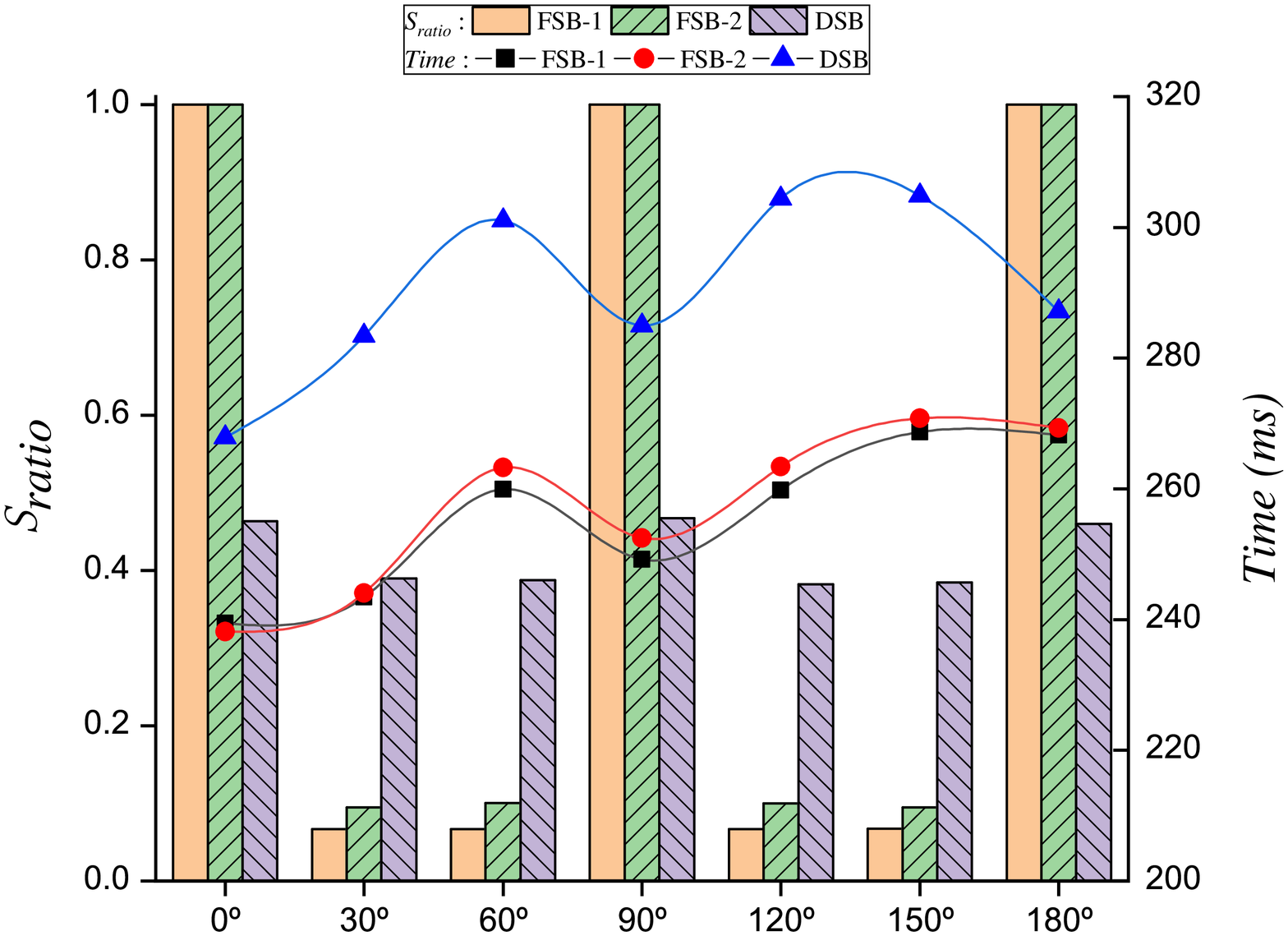}
        \label{r_3d_100}
    }
    \subfigure[$d=4,\delta= 100$ ]
    {
        \includegraphics[width=0.315\textwidth]{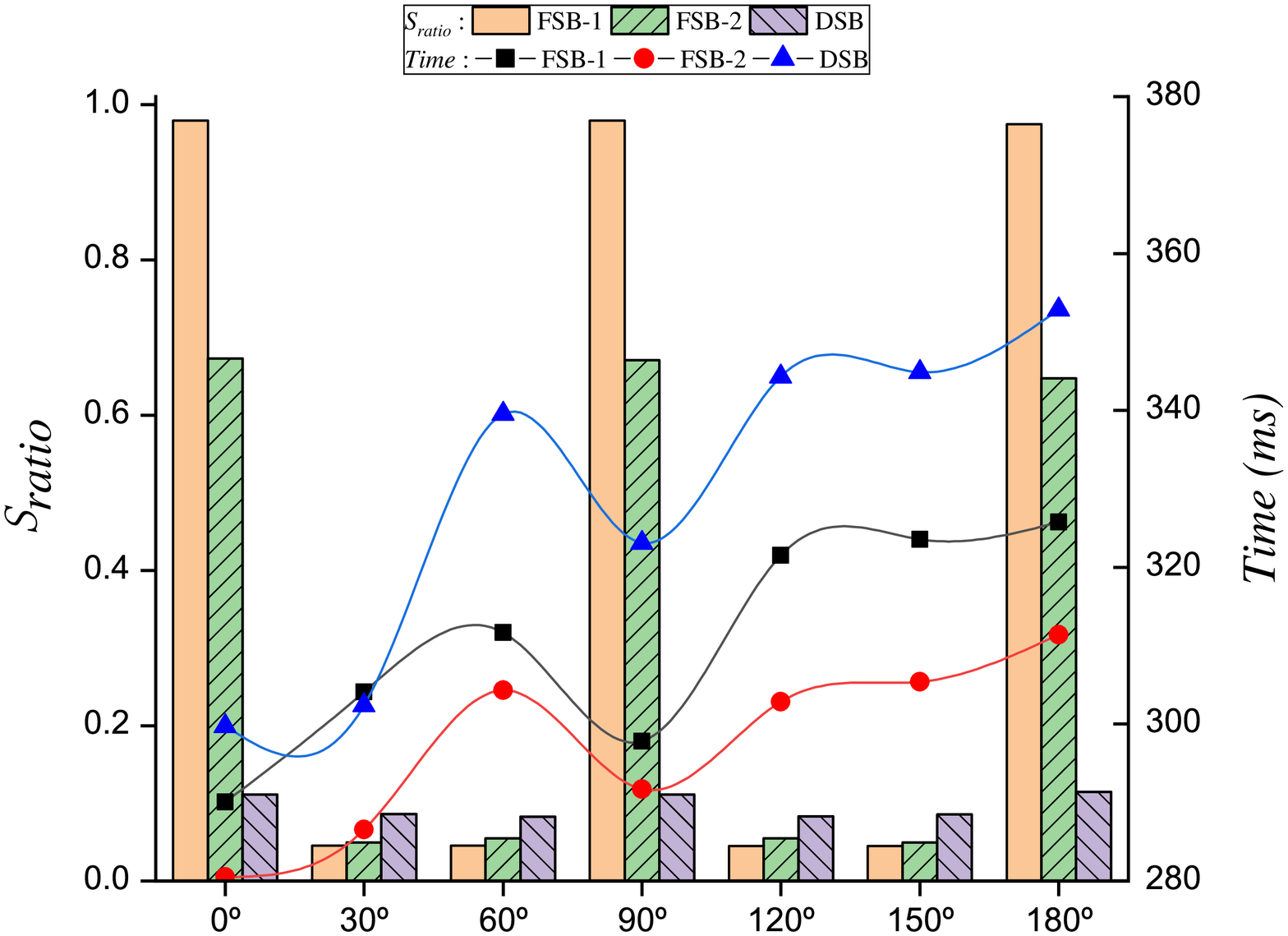}
        \label{r_4d_100}
    }

    \caption{Results for RFRs with the shape being a rectangle or hypercube.}
    \label{rectangle}
\end{figure*}

\subsection{Experiment Environment}

The simulations were conducted on a machine serving Windows 10 Home, equipped with an Intel(R) Core(TM) i5-6200U CPU (2.30 GHz, 4 core) with 8GB of RAM. We adopted Java programming language and conducted all experiments on Eclipse (Neon Release 4.6.0) development platform.

\section{Results}
\label{SEC:results}
In this section, we discuss results from our experiments to address the two research questions listed in Section~\ref{SEC:questions}.

\subsection{Simulations Results}
\label{SEC:simulationResults}
Due to page limitation, only those results for some specific simulation settings were provided in this section. Readers may refer to Appendixes A and B to look up a complete set of results (including all 504 simulations)\footnote{Two appendix files contain all simulation results, which are available at https://github.com/huangrubing/IFR/.}. Figures~\ref{rectangle} and~\ref{eclipse} provide the detailed results of $\mathcal{S}_\textit{ratio}$ and identification time for $\theta= 0.001$ and $N=1000$. In these two figures, the $x$-axis represents the rotation angle $\gamma$ of the RFR; while the left $y$-axis and right $y$-axis represent $\mathcal{S}_\textit{ratio}$ and identification time respectively. The bar chart describes our  $\mathcal{S}_\textit{ratio}$ results; while the line chart shows identification times.

\subsubsection{Answers to RQ1: $\mathcal{S}_\textit{ratio}$ Comparisons}
\label{answers:RQ1}
\begin{figure*}[!t]
\centering
    \subfigure[$d=2,\delta= 1$]
    {
        \includegraphics[width=0.315\textwidth]{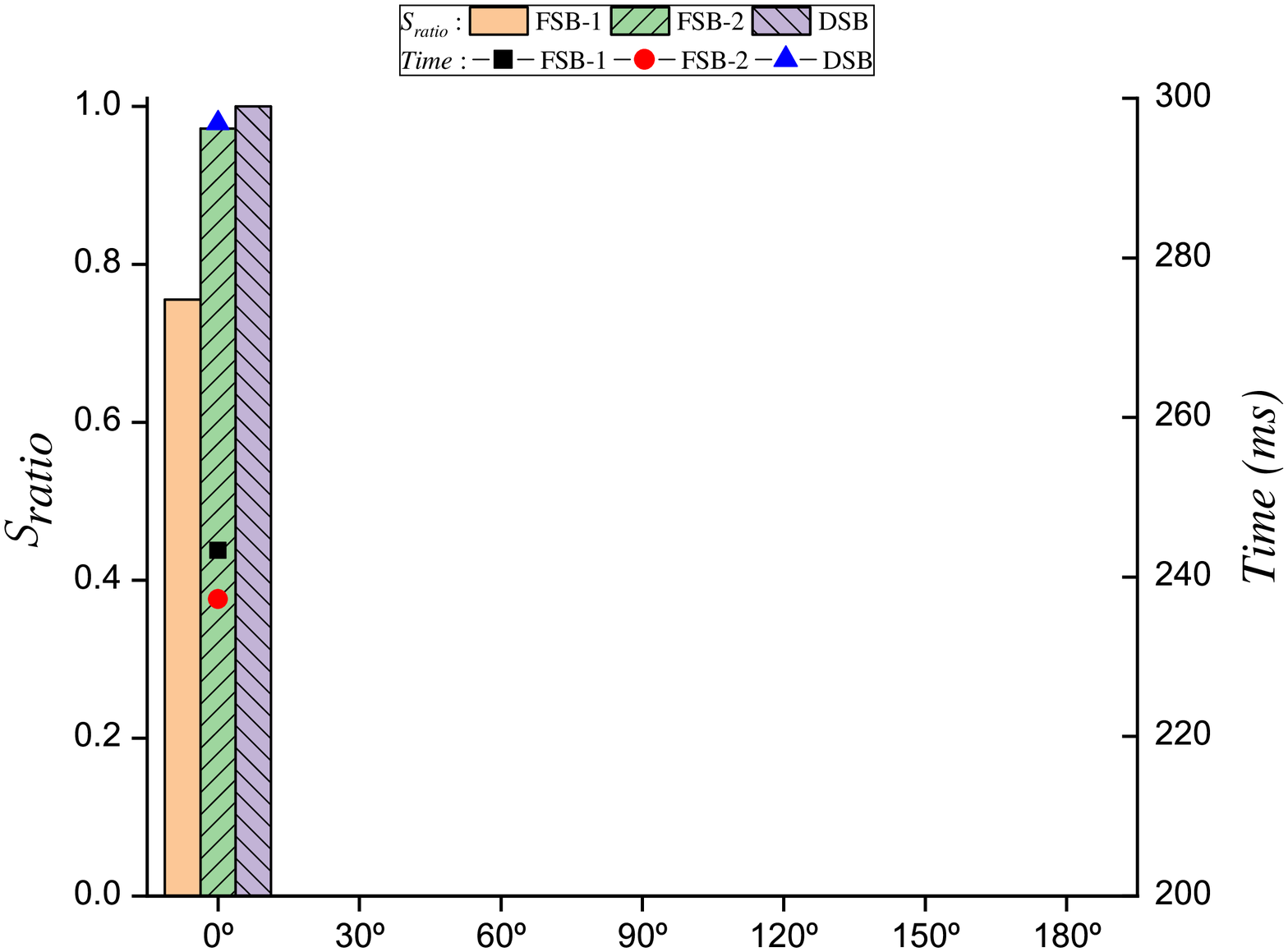}
        \label{c_2d_1}
    }
    \subfigure[$d=3,\delta = 1$]
    {
        \includegraphics[width=0.315\textwidth]{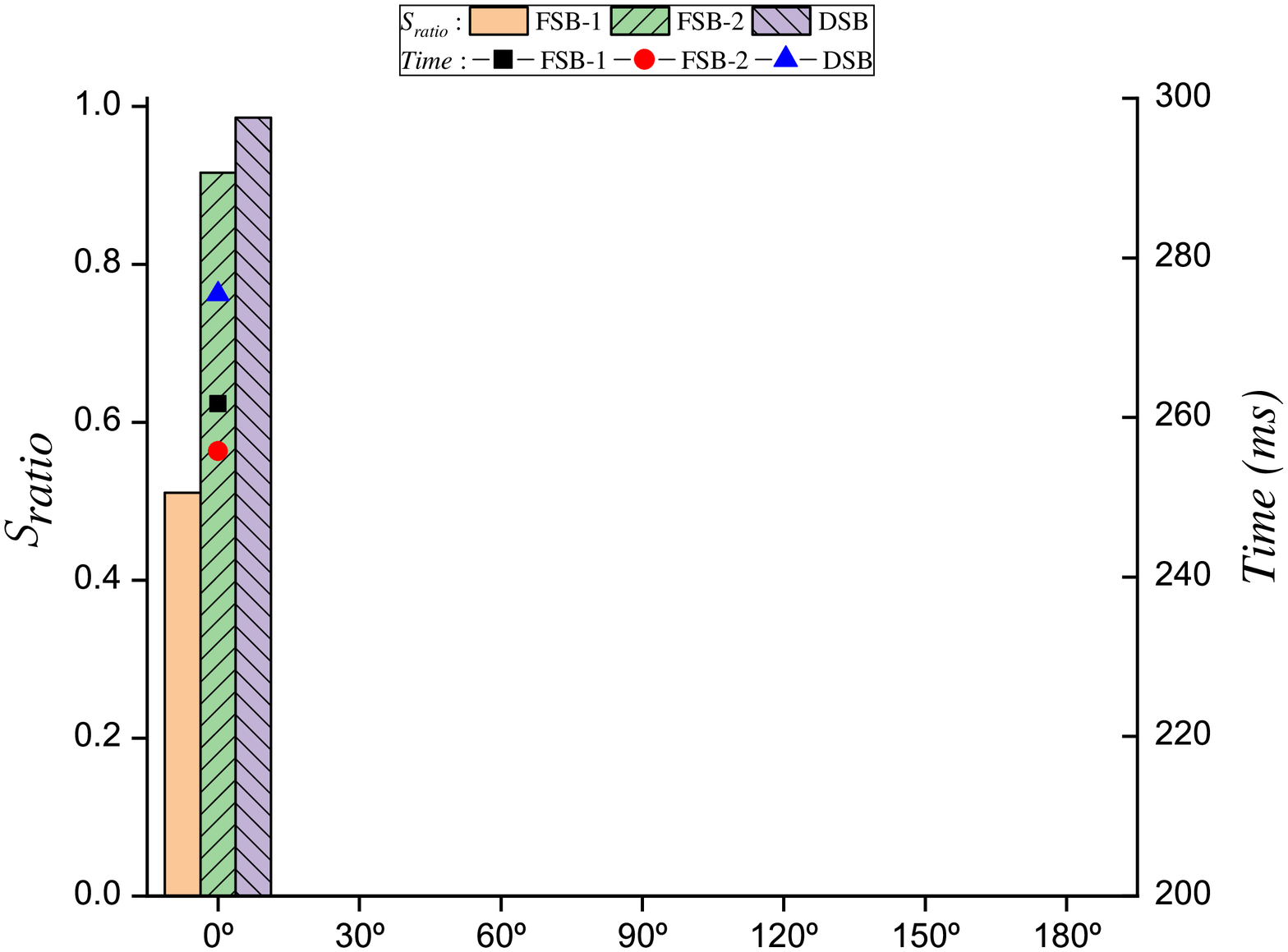}
        \label{c_3d_1}
    }
    \subfigure[$d=4,\delta = 1$]
    {
        \includegraphics[width=0.315\textwidth]{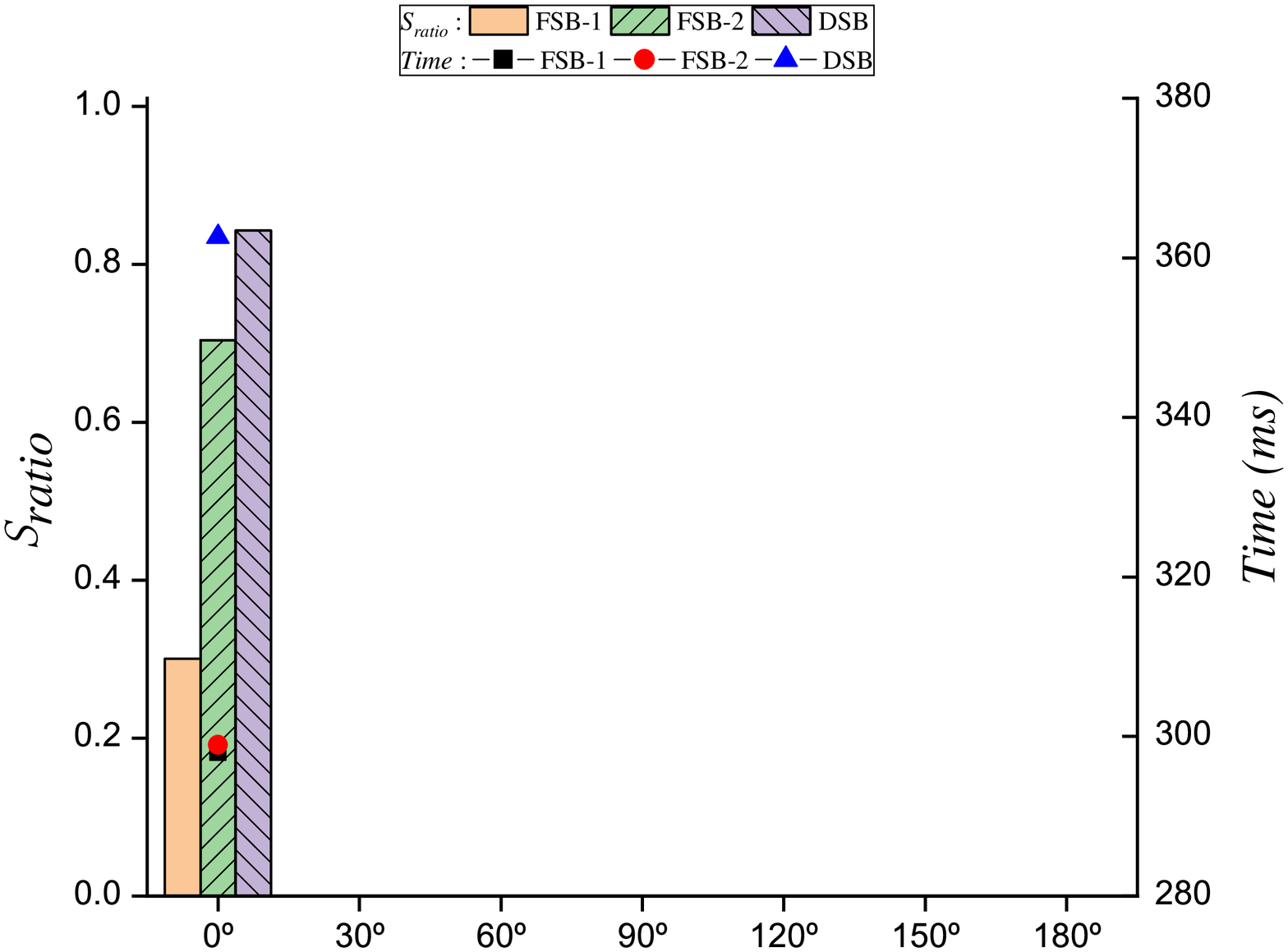}
        \label{c_4d_1}
    }
    \subfigure[$d=2,\delta= 10$]
    {
        \includegraphics[width=0.315\textwidth]{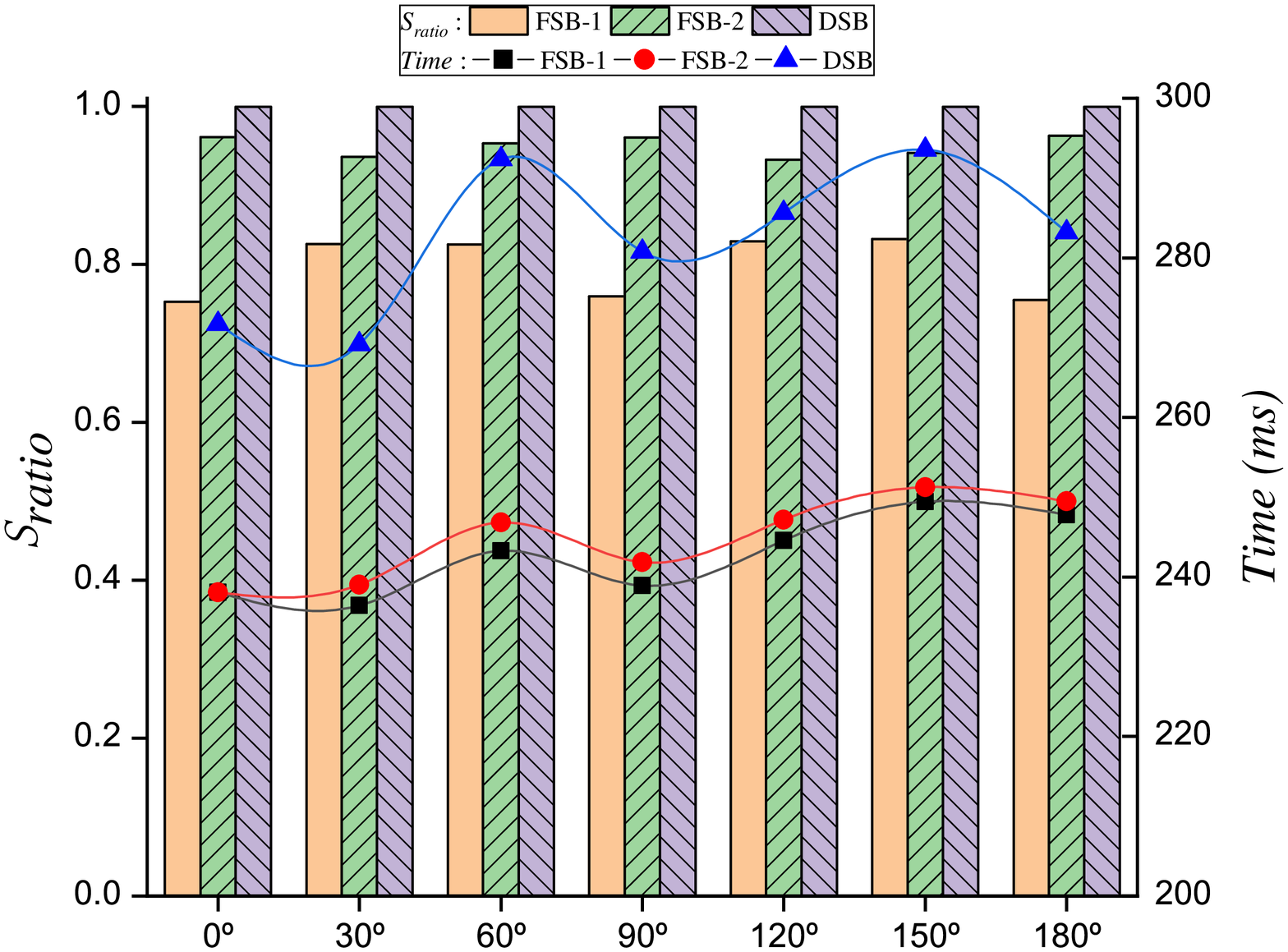}
        \label{c_2d_10}
    }
    \subfigure[$d=3,\delta= 10$]
    {
        \includegraphics[width=0.315\textwidth]{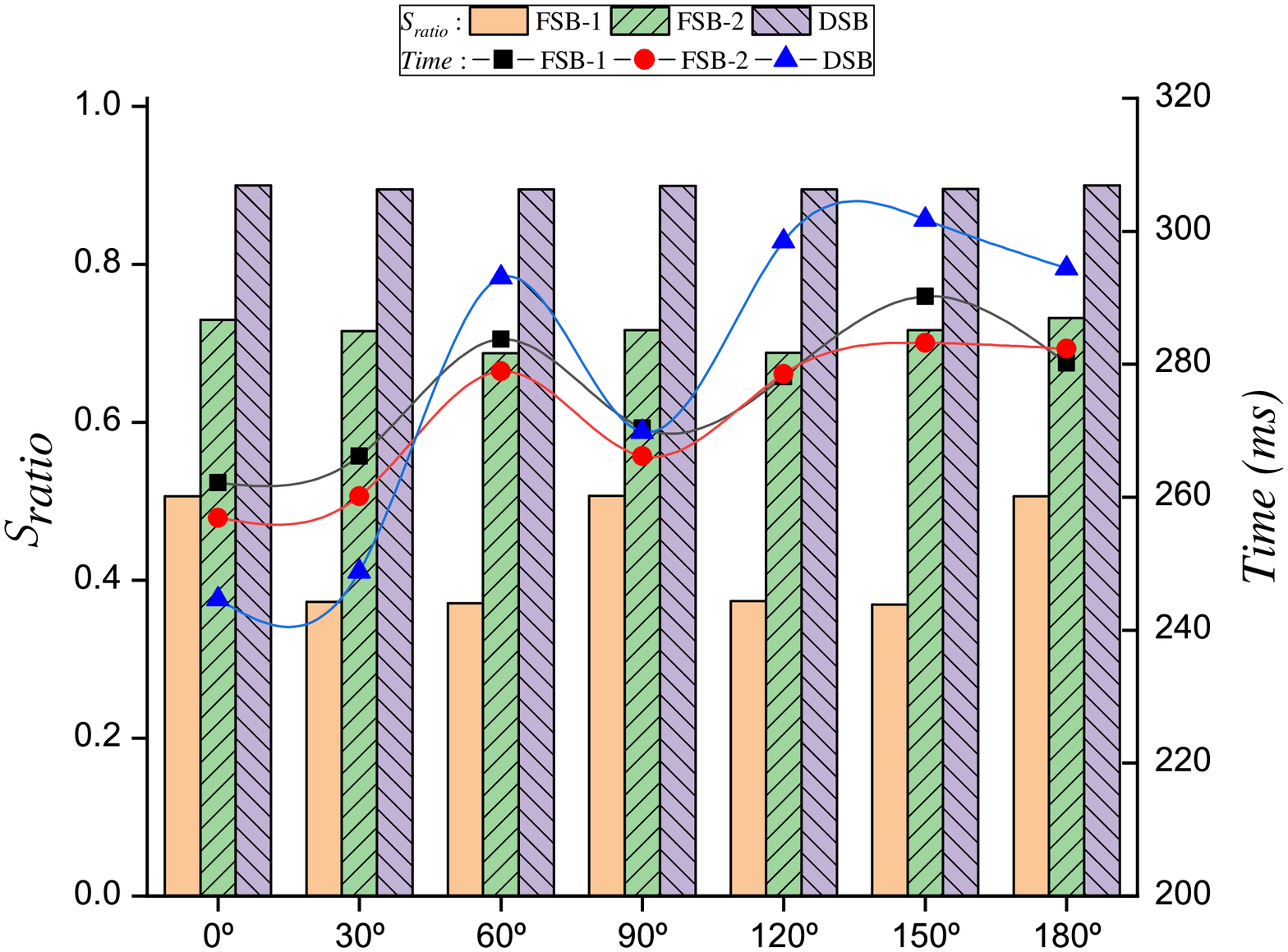}
        \label{c_3d_10}
    }
    \subfigure[$d=4,\delta= 10$]
    {
        \includegraphics[width=0.315\textwidth]{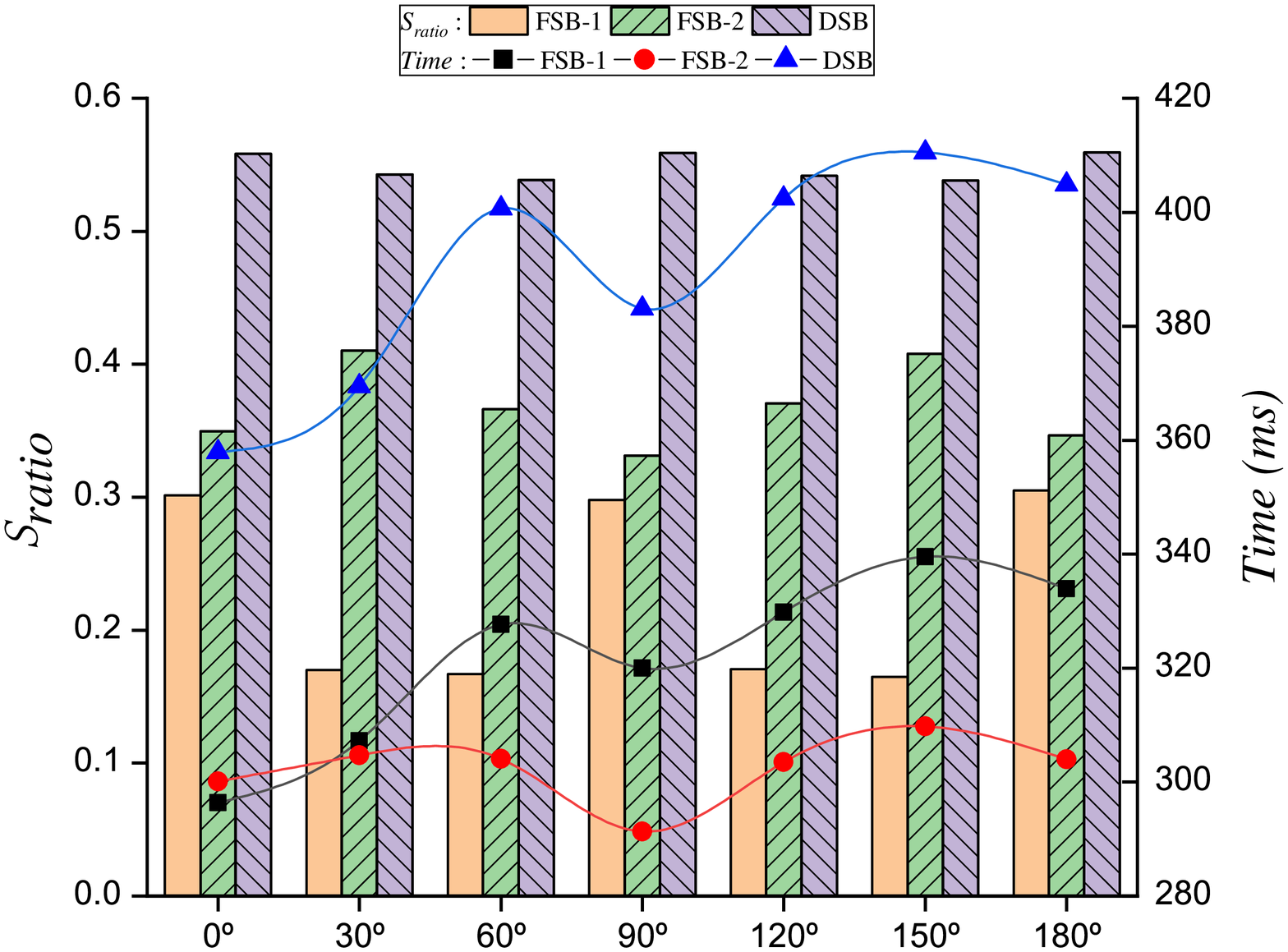}
        \label{c_4d_10}
    }
    \subfigure[$d=2,\delta= 100$]
    {
        \includegraphics[width=0.315\textwidth]{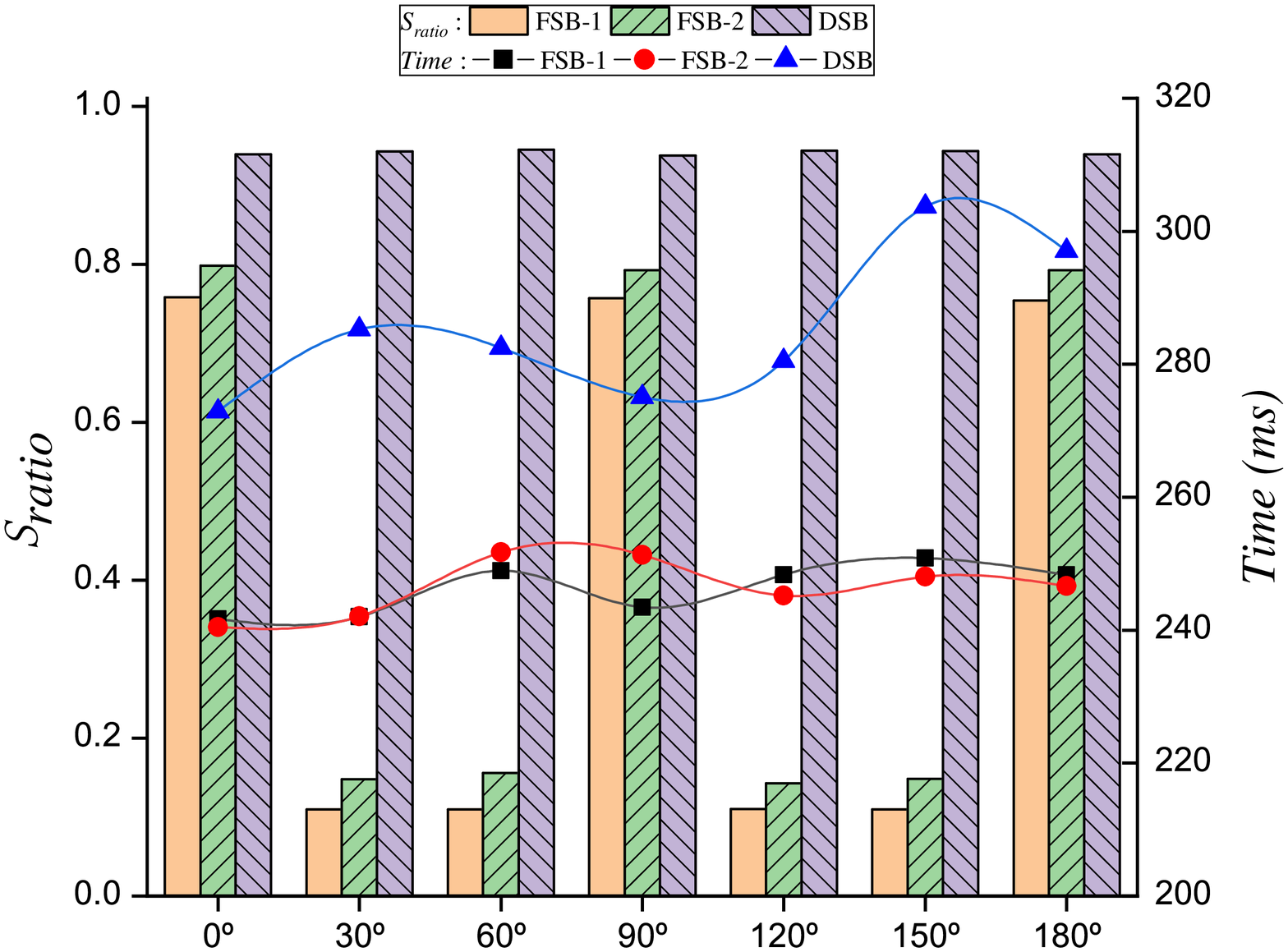}
        \label{c_2d_100}
    }
    \subfigure[$d=3,\delta= 100$]
    {
        \includegraphics[width=0.315\textwidth]{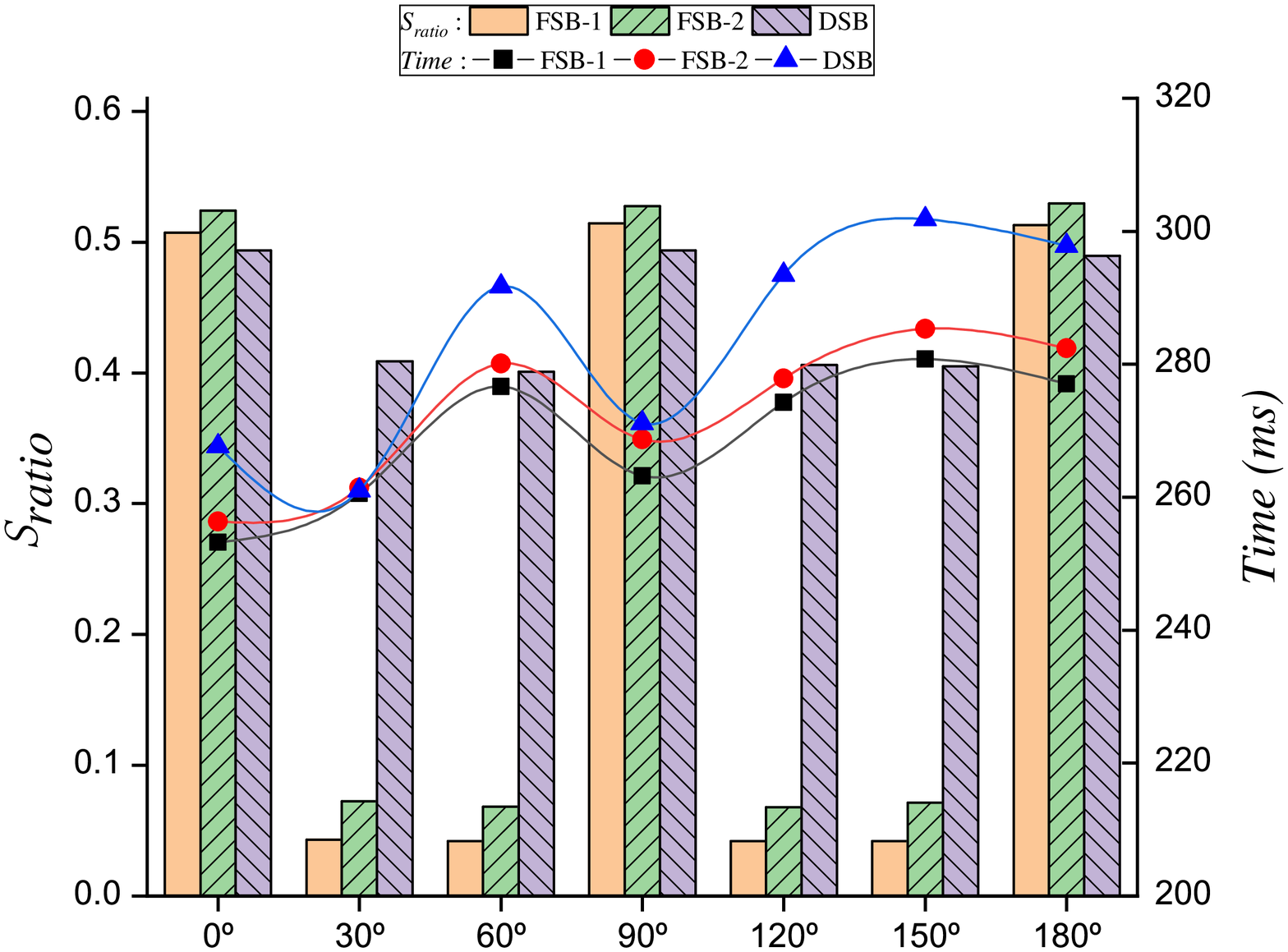}
        \label{c_3d_100}
    }
    \subfigure[$d=4,\delta= 100$]
    {
        \includegraphics[width=0.315\textwidth]{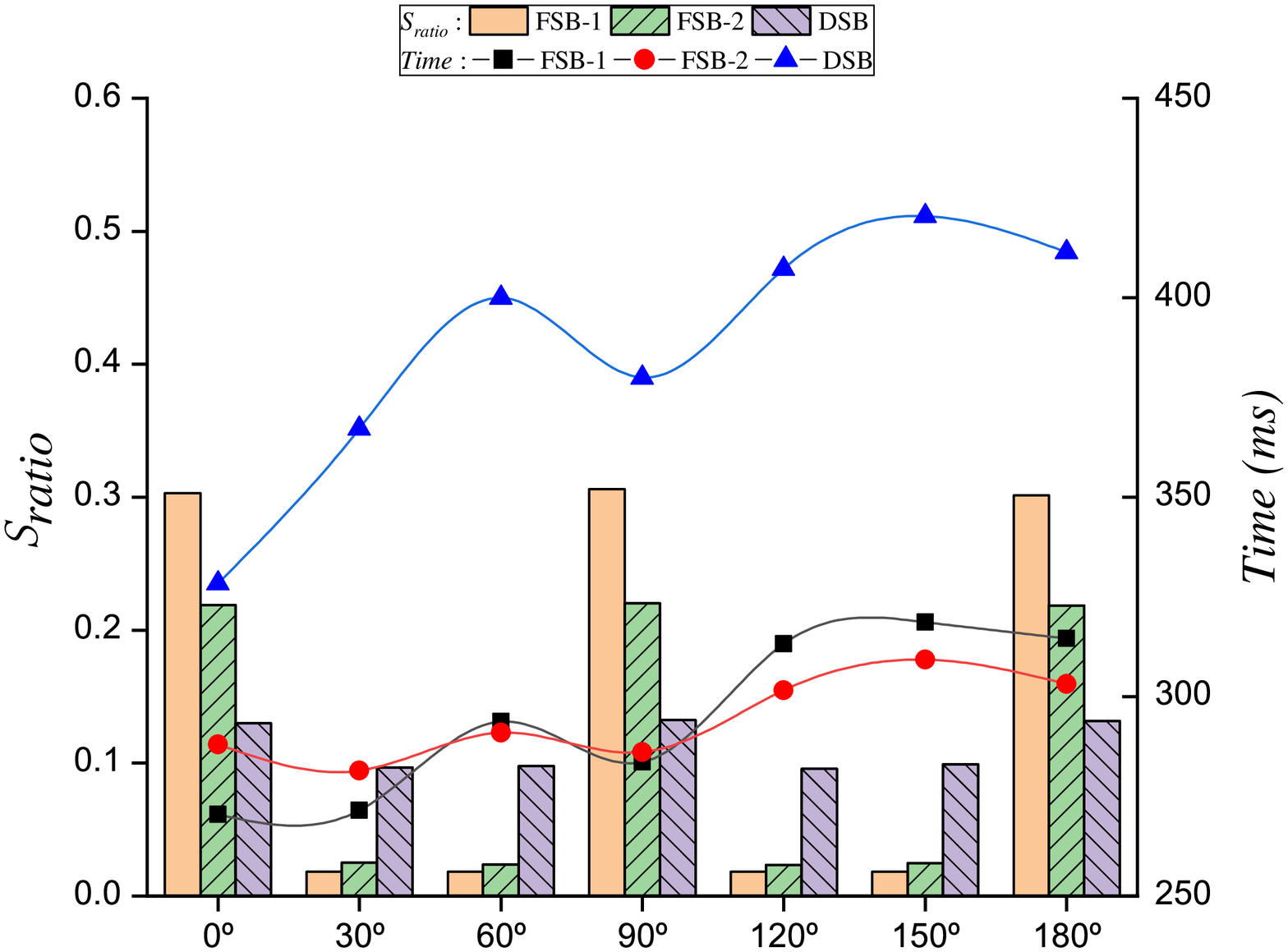}
        \label{c_4d_100}
    }

    \caption{Results for RFRs with the shape being an ellipse or hyperellipsoid.}
    \label{eclipse}
\end{figure*}

Based on the bar charts from Figures~\ref{rectangle} and~\ref{eclipse}, we have the following observations:
\begin{itemize}
\item[1)] With the increase in the dimension $d$ of input domain or the compactness parameter $\delta$ of RFR, the $\mathcal{S}_\textit{ratio}$ performances of each SB method gradually deteriorate, regardless of the shape and rotation angle $\gamma$ of the RFR.

\item[2)] With the increase in the rotation angle $\gamma$ of RFR, DSB achieves very similar $\mathcal{S}_\textit{ratio}$ performances, indicating that the rotation angle provides little impact on the effectiveness of DSB. However, FSB (including FSB-1 and FSB-2) may have very different performances, especially when $\delta$ is larger (i.e., the RFR is less compact). For example, when $\delta$ equals 100, the highest $\mathcal{S}_\textit{ratio}$ values of FSB reach 1.0; while the lowest values are less than 0.1.

{\item[3)] Comparing to DSB, if $\gamma$ equals $0^\circ$, $90^\circ$, or $180^\circ$ (indicating that the orientations of edges or axes of the RFR coincides with those of the coordinate axes), FSB generally performs similarly or better for the rectangle or hypercube RFRs, especially when $\delta$ is large. However, the case is opposite for the ellipse or hyperellipsoid RFRs (i.e., DSB is similar or better than FSB overall), except in the case with $d$ = 4 and $\delta$ = 100. Otherwise, when $\gamma \notin \{0^\circ, 90^\circ, 180^\circ\}$, DSB performs much better than FSB, especially when $\delta$ is large, regardless of parameter settings.}

\item[4)] As for the comparison between FSB-1 and FSB-2, it is evident that FSB-2 is similar or slightly better than FSB-1, irrespective of $d$, $\delta$, $\gamma$, and the shape of the RFR.
\end{itemize}

The above observations can be explained as follows. For observation 1): FSB-1, FSB-2, and DSB are known to be able to identify the same number of failure-causing boundary inputs in each orthant from a given failure-causing source input. Intuitively, a higher dimension $d$ will lead to more orthants (as the number of orthants is equal to $2^d$). Therefore, after collecting $N$ failure-causing boundary inputs, each orthant may contain more failure-causing boundary inputs for a lower dimension, resulting in a higher $\mathcal{S}_\textit{ratio}$ value. Furthermore, if the compactness parameter $\delta$ becomes larger, indicating that the RFR becomes less compact, the RFR would become narrower. Therefore, it is more difficult to identify failure-causing boundary inputs within the narrow regions of the RFR, especially when the RFR contains a rotation angle $\gamma$ other than $0^\circ$, $90^\circ$, $180^\circ$. As a result, the $\mathcal{S}_\textit{ratio}$ value may become lower. As for observation 2): The main reason is that DSB uses diverse orientations instead of fixed orientations (as in FSB), which may diversely distribute failure-causing boundary inputs around the RFR boundary.

Regarding observation 3): Incidentally, FSB adopts special extension orientations, which coincide with the rotation angle $\gamma$ being equal to $0^\circ$, $90^\circ$, or $180^\circ$. Therefore, it may certainly identify failure-causing boundary inputs around the edges of a rectangle (or hypercube) RFR, and then cover actual (or approximately actual) edges. However, for an ellipse  (or hyperellipsoid) RFR, it will be difficult to identify more failure-causing boundary inputs within the narrow RFR regions. For observation 4): Due to all extension orientations adopted by FSB-1 are already subsumed in FSB-2, hence FSB-2 can apply more different extension orientations than FSB-1. When FSB-1 has a good performance under favourable conditions (such as a rectangle or cuboid RFR with $\gamma = 0^\circ$), FSB-2 will have similar performance. However, under unfavourable conditions (such as $\gamma \notin \{0^\circ,90^\circ,180^\circ\}$), FSB-2 may perform better than FSB-1 due to more diverse extension orientations.

With respect to other simulation results of $\mathcal{S}_\textit{ratio}$, listed in Appendix A, we have similar observations overall. As intuitively expected, increasing N will lead to higher $\mathcal{S}_\textit{ratio}$ because more failure-causing boundary inputs would identify more better AFR for SB. In addition, the failure rate $\theta$ of the RFR apparently provides little impact on the performances of FSB and DSB. This is because the same number of failure-causing boundary inputs identified by either FSB or DSB should identify a similar AFR, regardless of the size of the RFR.

\begin{figure*}[!b]
\centering
    \subfigure[\texttt{airy}]
    {
        \includegraphics[width=0.315\textwidth]{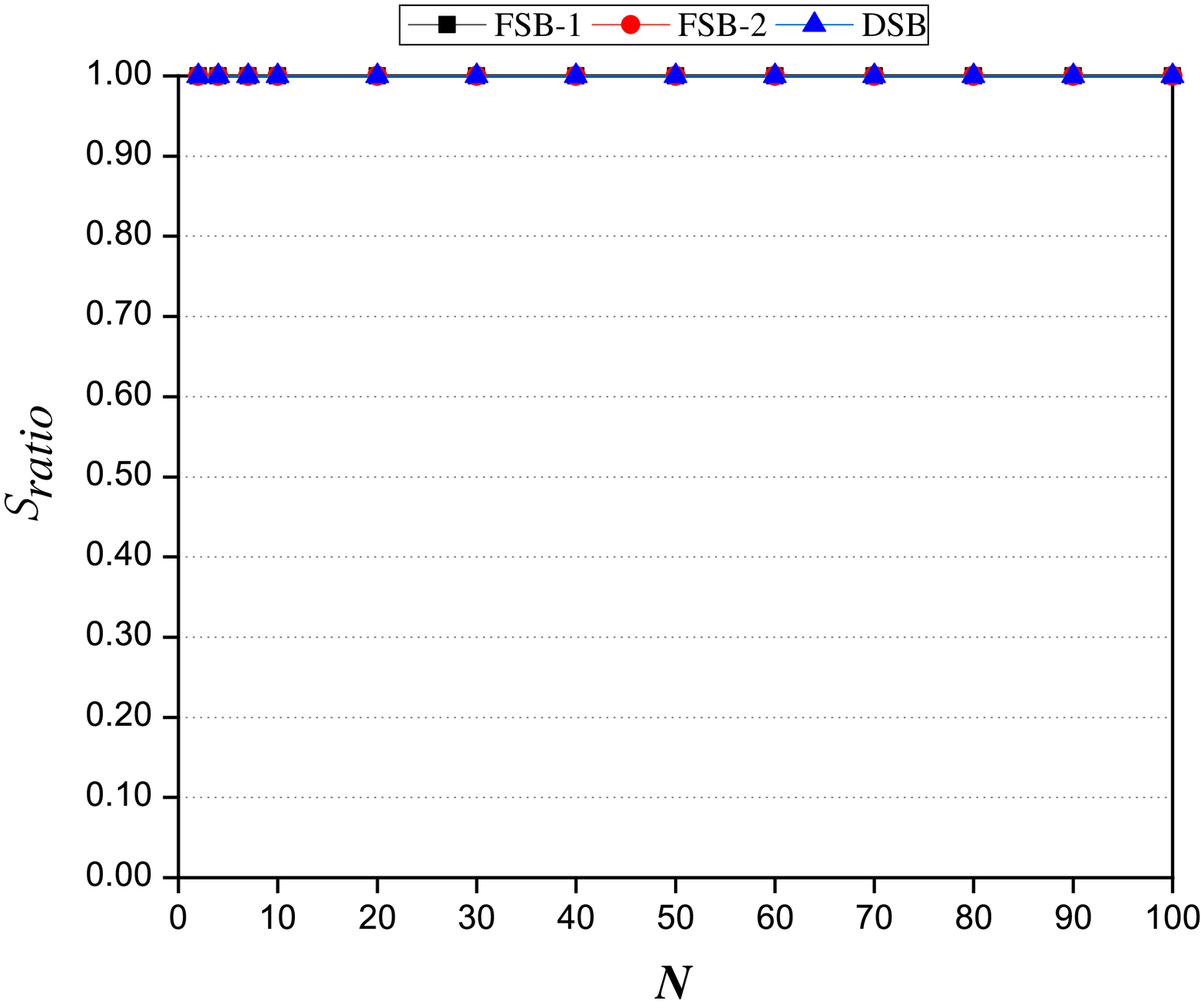}
        \label{airy-ratio}
    }
       \subfigure[\texttt{gammq}]
    {
        \includegraphics[width=0.315\textwidth]{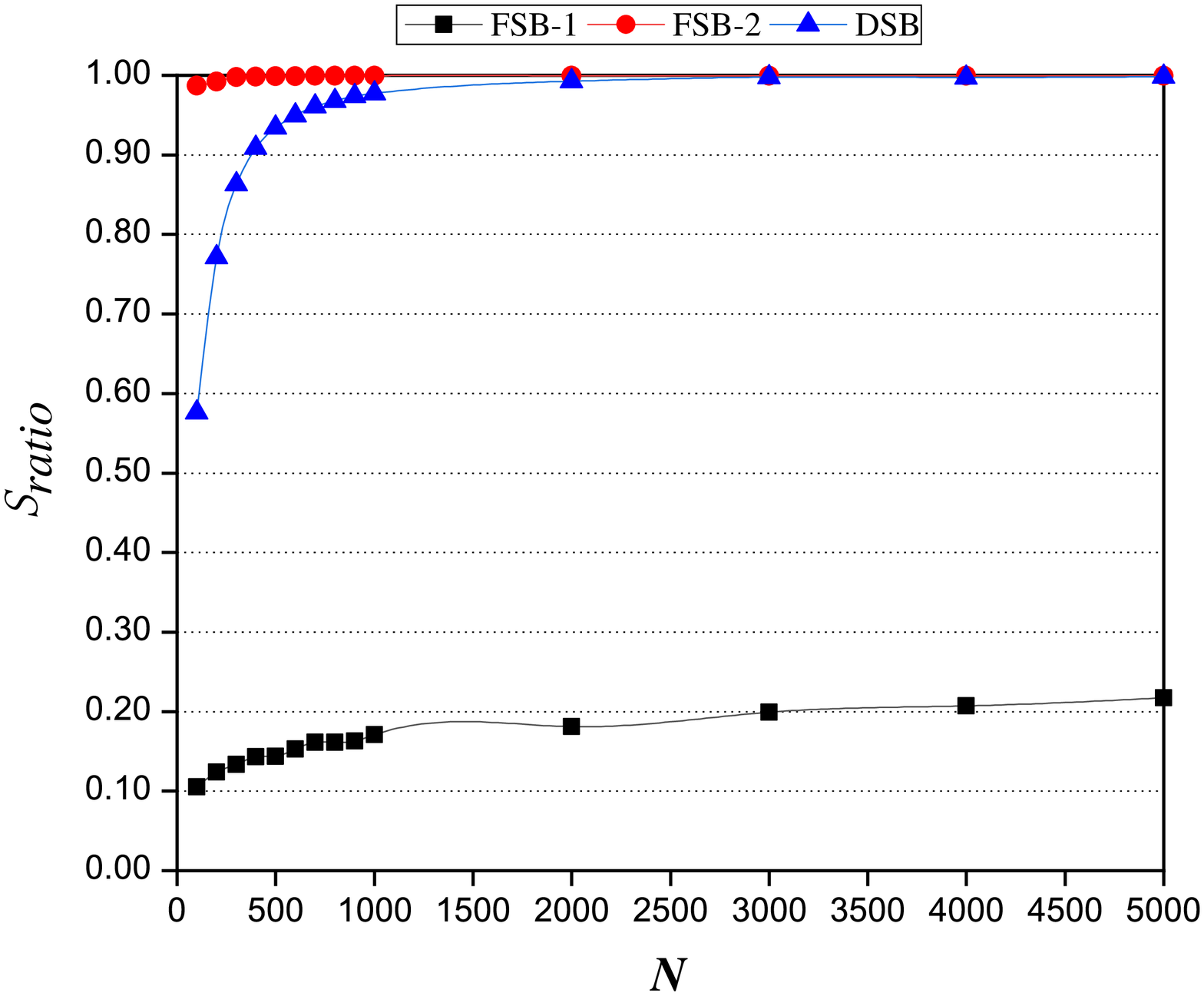}
        \label{gammq-ratio}
    }
        \subfigure[\texttt{Expint}]
    {
        \includegraphics[width=0.315\textwidth]{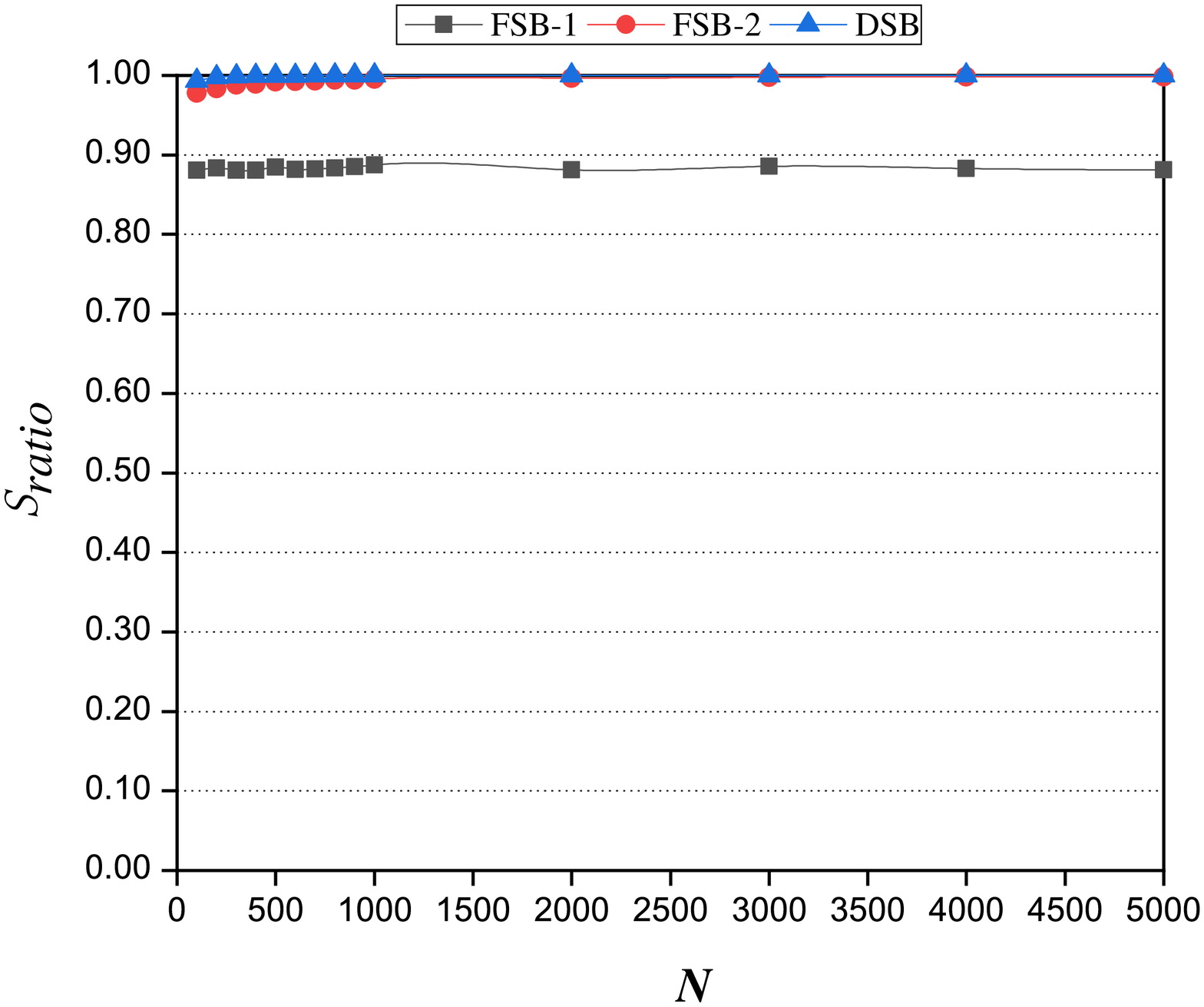}
        \label{Expint-ratio}
    }
        \subfigure[\texttt{bessj}]
    {
        \includegraphics[width=0.315\textwidth]{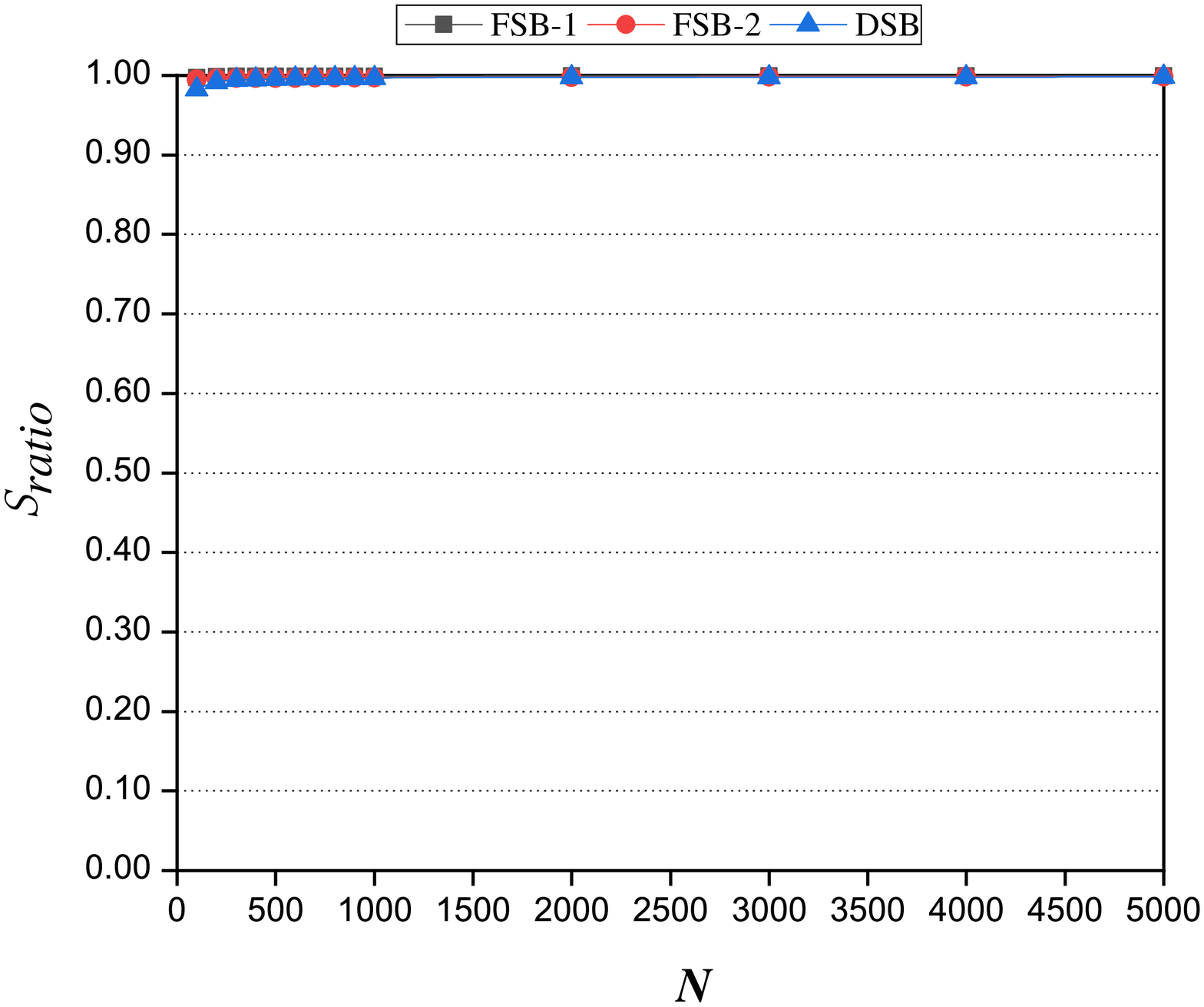}
        \label{bessj-ratio}
    }
        \subfigure[\texttt{triangle}]
    {
        \includegraphics[width=0.315\textwidth]{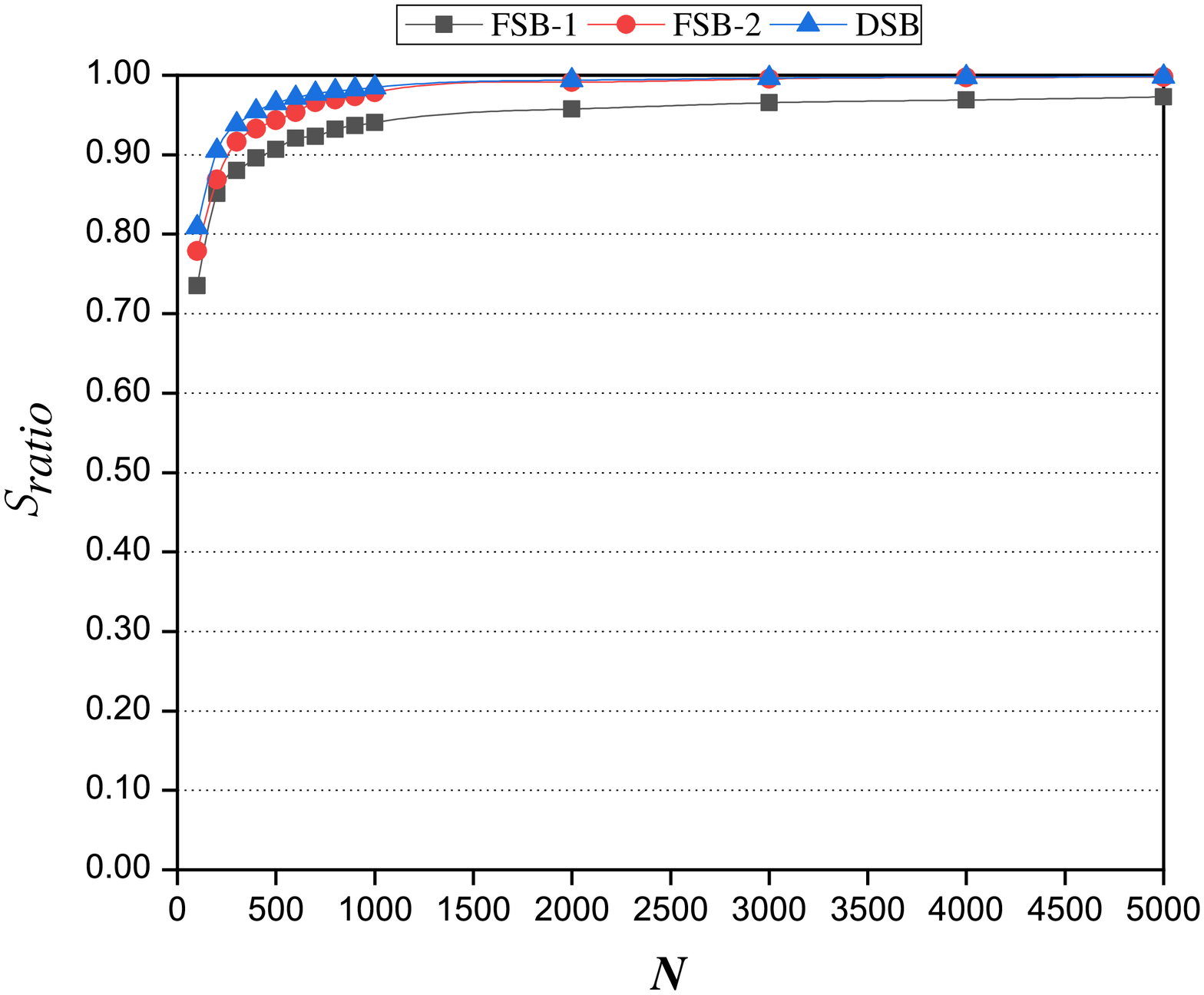}
        \label{triangle-ratio}
    }
          \subfigure[\texttt{cel}]
    {
        \includegraphics[width=0.315\textwidth]{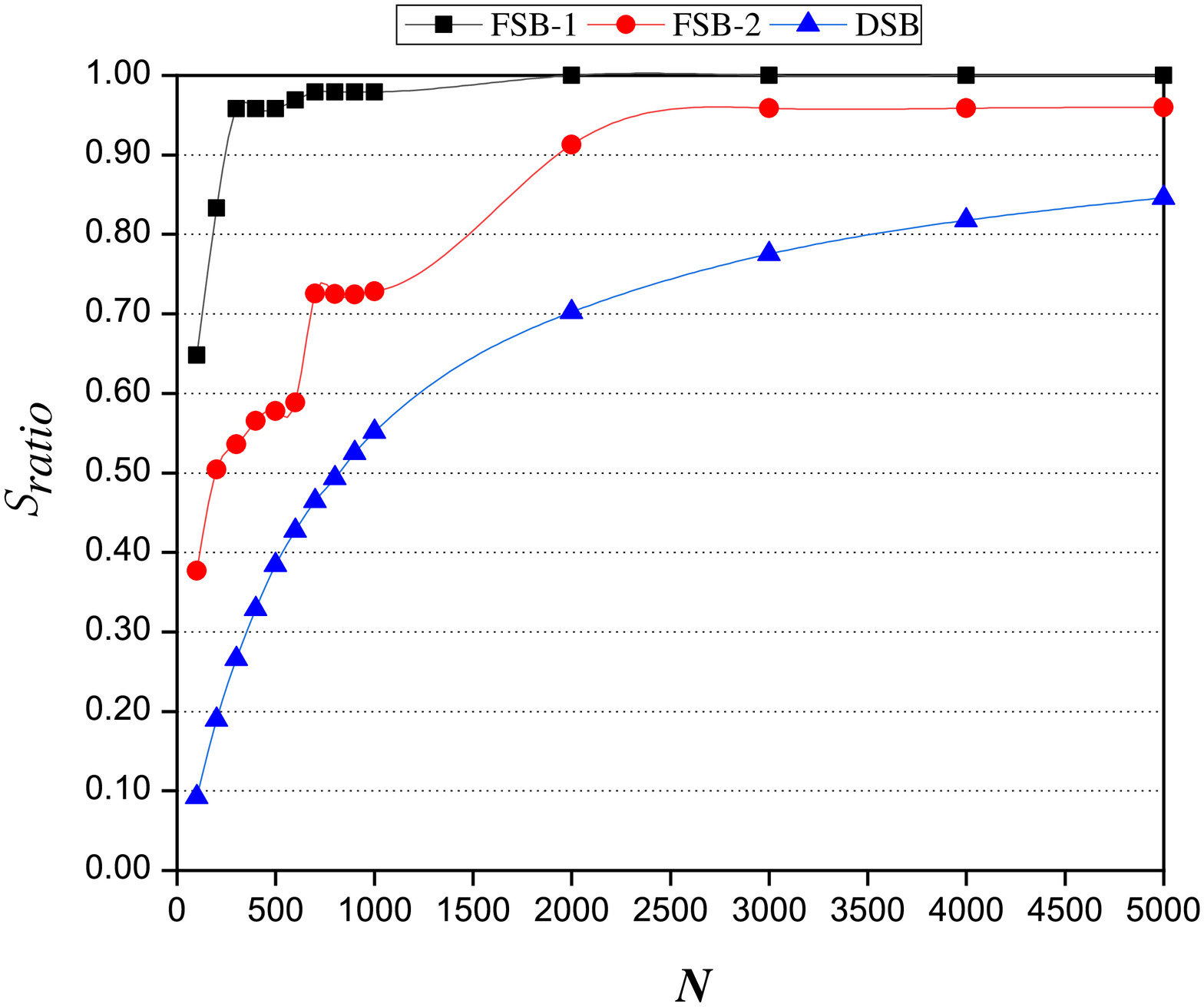}
        \label{cel-ratio}
    }

    \caption{$S_\textit{ratio}$ results for six object programs.}
    \label{FIG:programRatio}
\end{figure*}

\subsubsection{Answers to RQ2: Comparisons of Identification Time}
Based on the line charts from Figures~\ref{rectangle} and \ref{eclipse}, it can be observed that:

\begin{itemize}
\item[1)] FSB-1 has comparable identification time to FSB-2, while DSB generally requires similar or slightly longer identification time than FSB-1 or FSB-2. Referring to the more comprehensive results of the identification time listed in Appendix B, similar observations can be attained. However, the differences of identification time between DSB and FSB (both FSB-1 and FSB-2) appear to be relatively small (the maximum difference is only about 0.1 second).
{ \item[2)] In addition, the identification time shows a slightly increasing trend with the increase of rotation angle.}
\end{itemize}

The above observations can be explained by both FSB and DSB taking similar time in identifying the next failure-causing boundary input using Algorithm~\ref{ALG:find_boundary}. Hence, the difference in final identification times between FSB and DSB is due to different methods used in determining the extension orientation. As discussed in Section~\ref{SEC:algorithm}, DSB adopts diverse extension orientations by choosing the best candidate each time as the next extension orientation, while FSB-1 and FSB-2 make use of fixed orientations. As a result, FSB-1 and FSB-2 have similar identification times, while DSB will take longer than FSB for guiding selection of extension orientations. In addition, identifying failure-causing boundary inputs generally takes up the majority of identification time for both DSB and FSB, hence resulting in marginal differences between them. However, with increasing number of failure-causing boundary inputs, DSB may require significantly longer identification time for selecting the next extension orientation.
{ Regarding the observation 2), as mentioned in Section~\ref{answers:RQ1}, the rotation has an impact on the performance of our methods. In general, when $\gamma \in \{0^\circ, 90^\circ, 180^\circ\}$, they all show good results; however, as the rotation angle increases, it will become more difficult to identify failure-causing boundary inputs within the irregular RFR, especially when $\delta$ is larger (i.e., the RFR is less compact), which means that longer time is required to identify an AFR.}

\textit{\textbf{To conclude: }} With the exception in some special cases of RFR, DSB is mostly more effective than FSB, especially when the number of failure-causing boundary inputs is large. Among the two implementations of FSB, FSB-2 has similar or slightly better effectiveness than FSB-1. Besides, DSB requires more identification time than both FSB-1 and FSB-2, however, their differences are small. Therefore, it can be concluded that DSB appears to be more cost-effective than FSB; while FSB-1 is similar to FSB-2.

\subsection{Empirical Studies Results}
\begin{figure*}[!b]
\centering
    \subfigure[\texttt{airy}]
    {
        \includegraphics[width=0.315\textwidth]{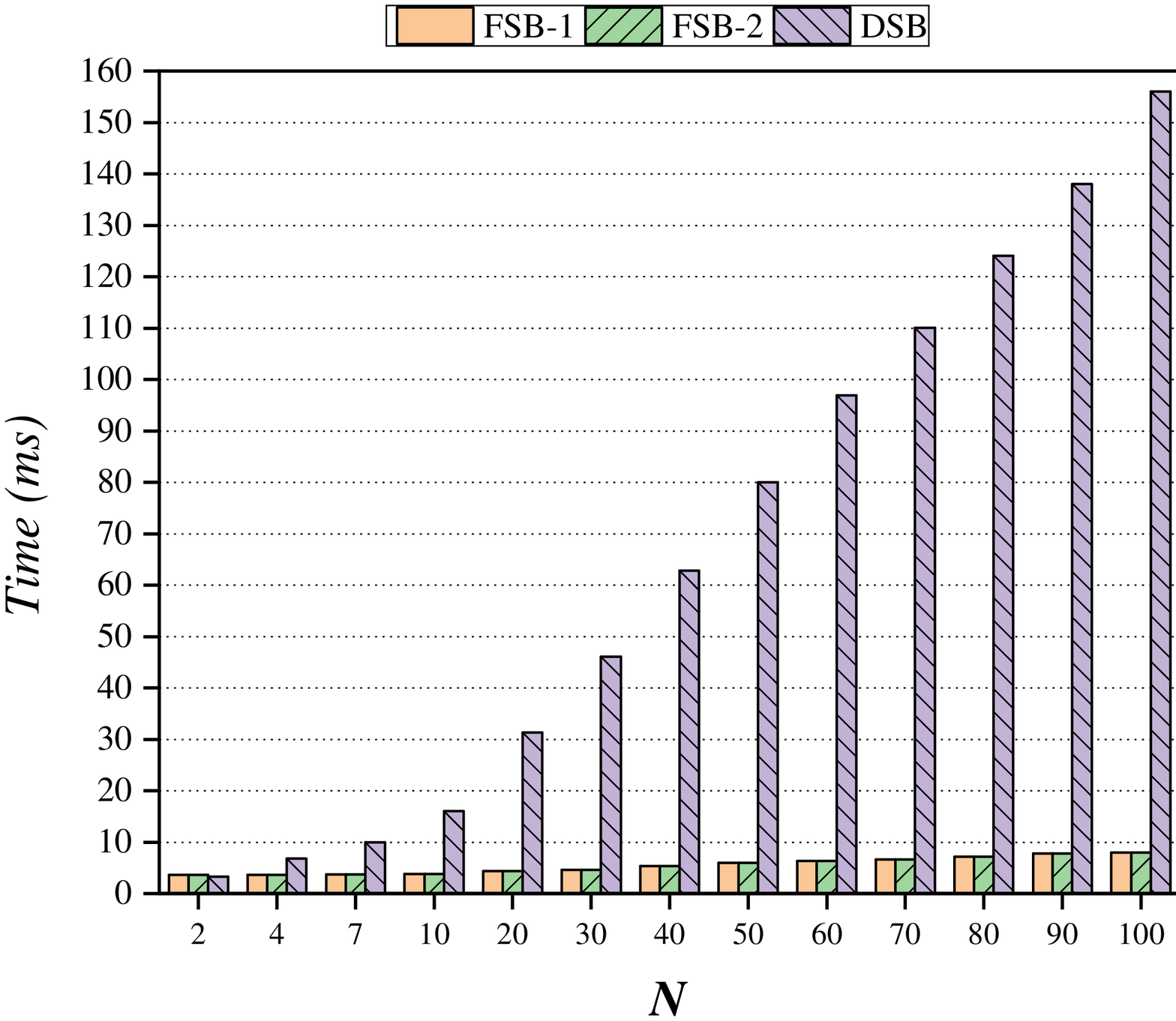}
        \label{airy-time}
    }
       \subfigure[\texttt{gammq}]
    {
        \includegraphics[width=0.315\textwidth]{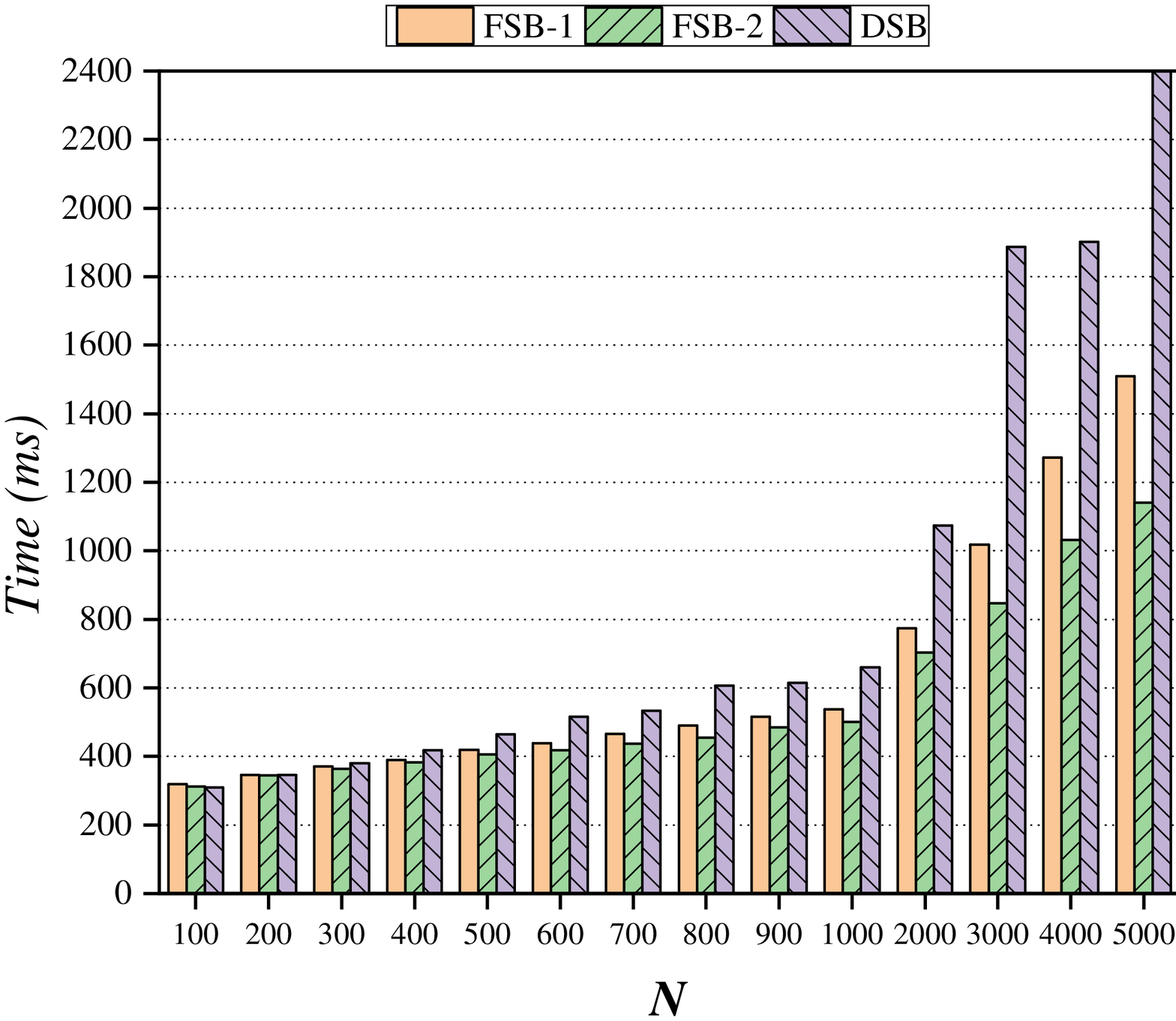}
        \label{gammq-time}
    }
        \subfigure[\texttt{Expint}]
    {
        \includegraphics[width=0.315\textwidth]{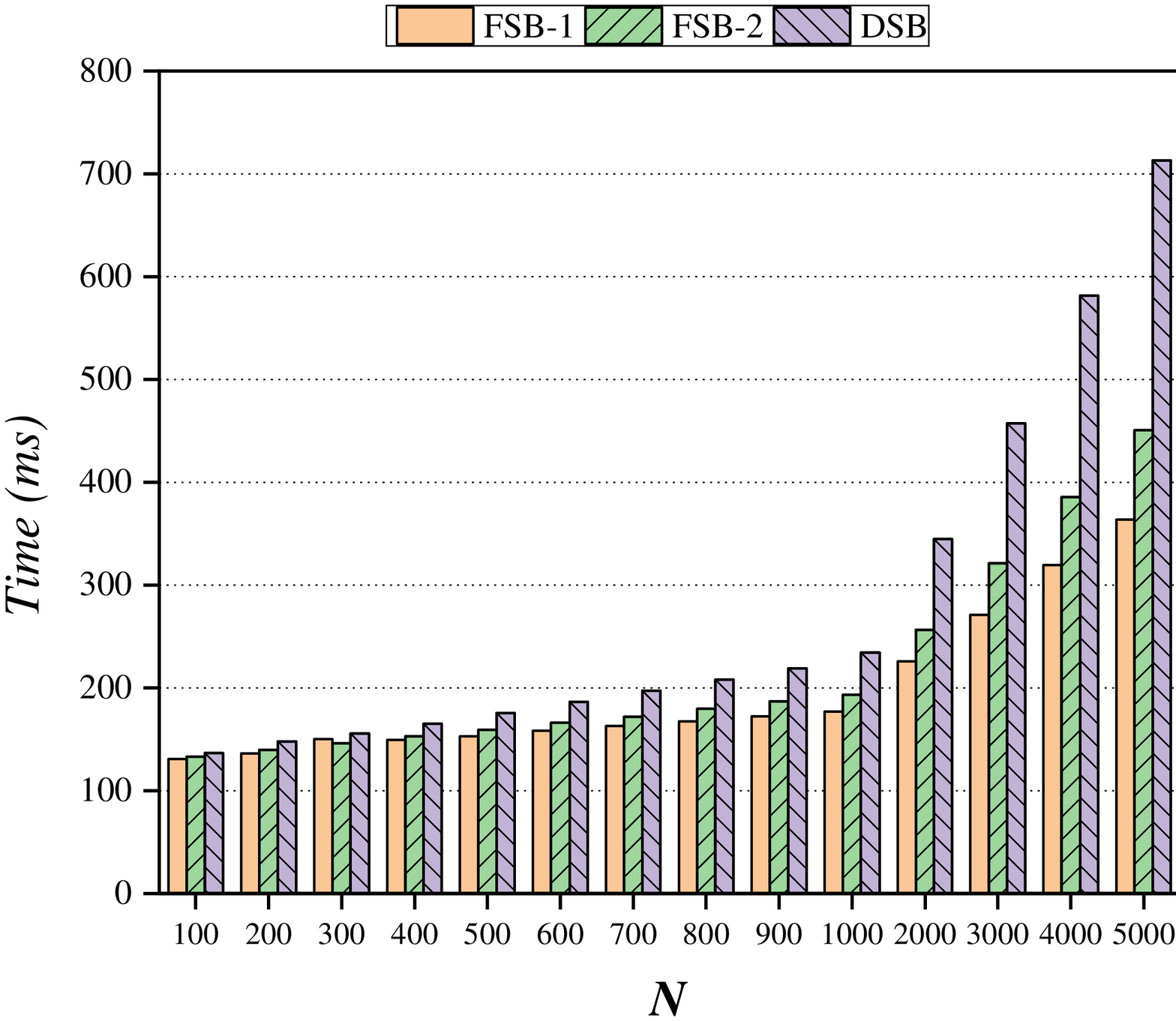}
        \label{Expint-time}
    }
        \subfigure[\texttt{bessj}]
    {
        \includegraphics[width=0.315\textwidth]{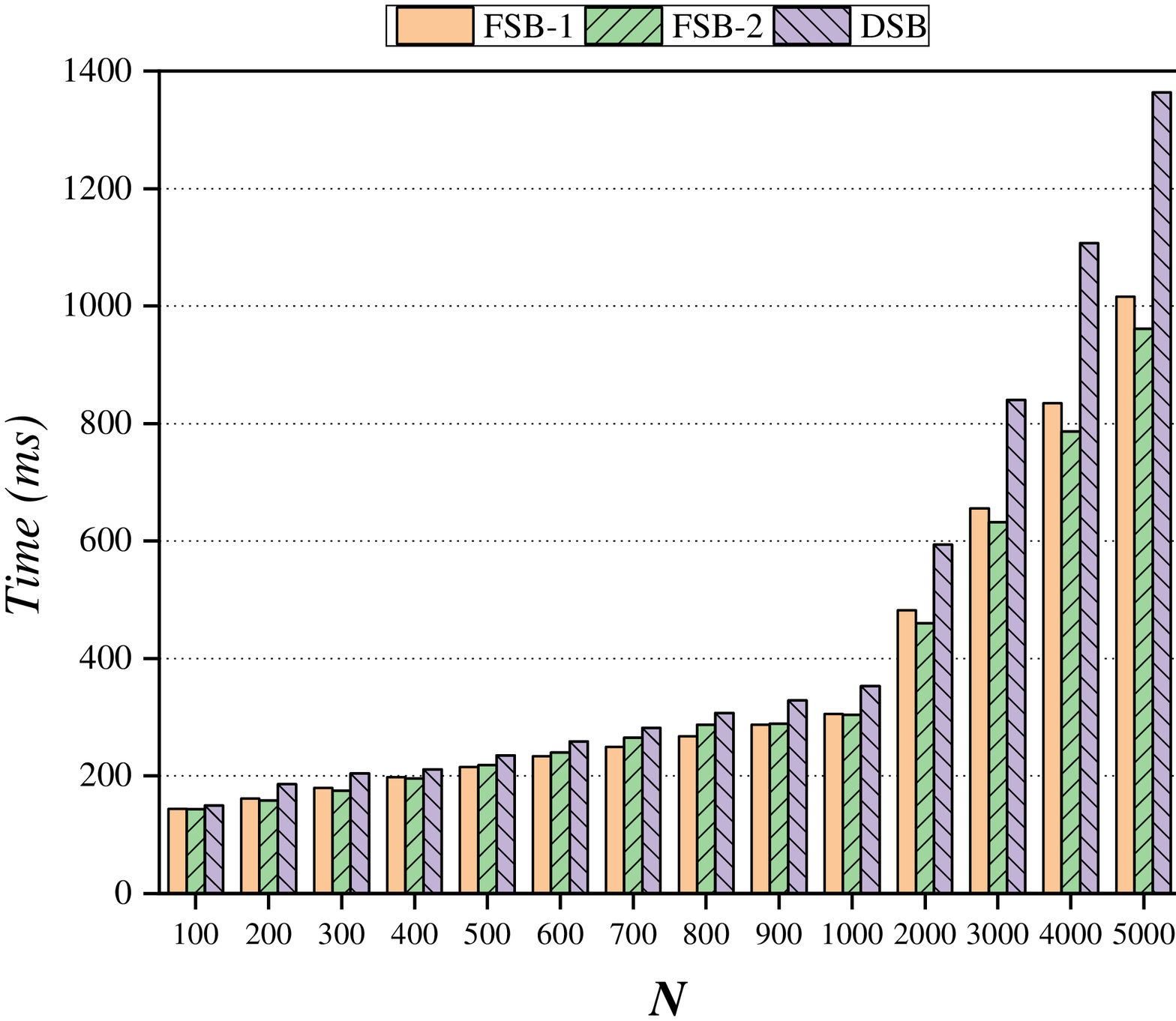}
        \label{bessj-time}
    }
        \subfigure[\texttt{triangle}]
    {
        \includegraphics[width=0.315\textwidth]{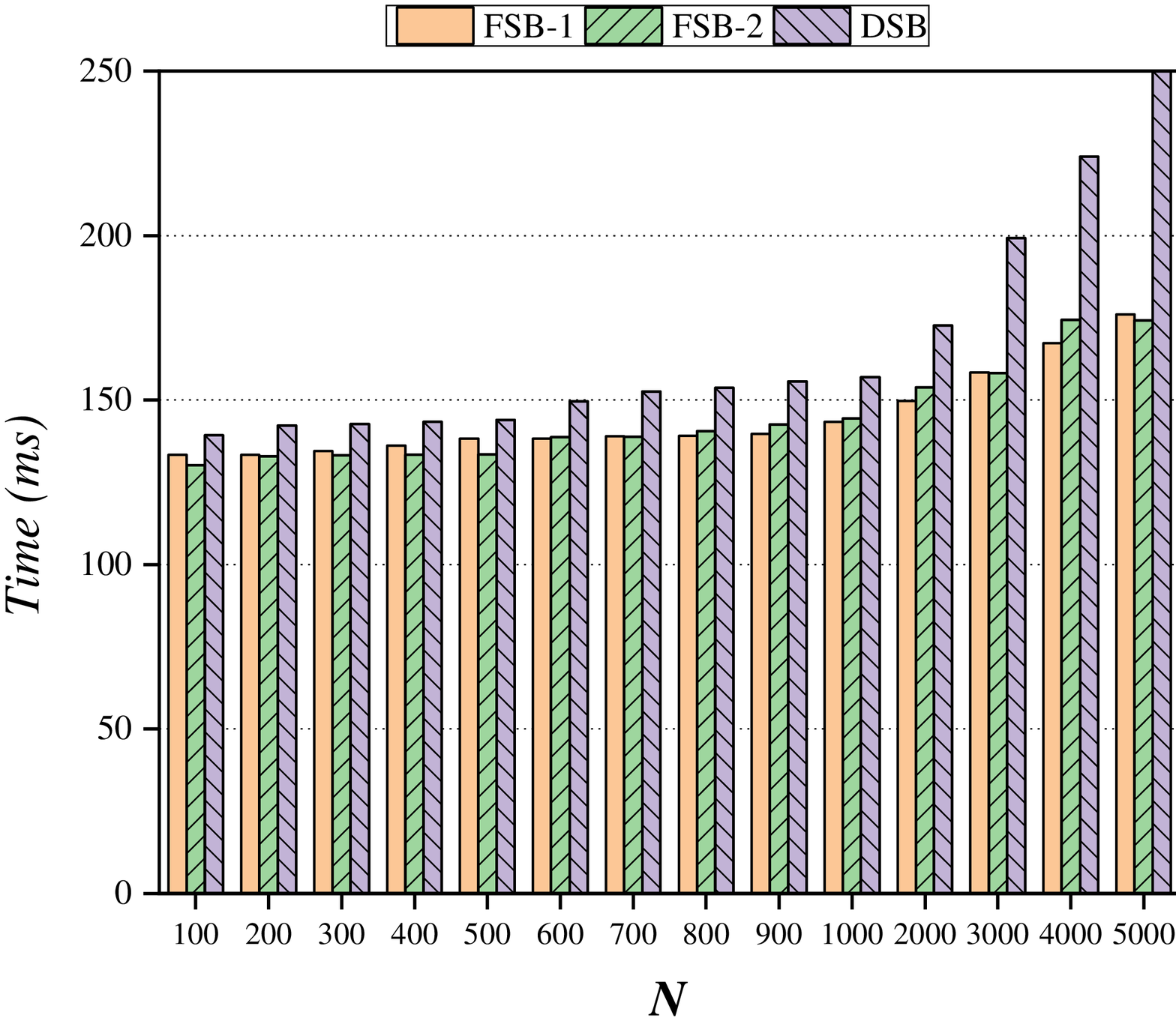}
        \label{triangle-time}
    }
          \subfigure[\texttt{cel}]
    {
        \includegraphics[width=0.315\textwidth]{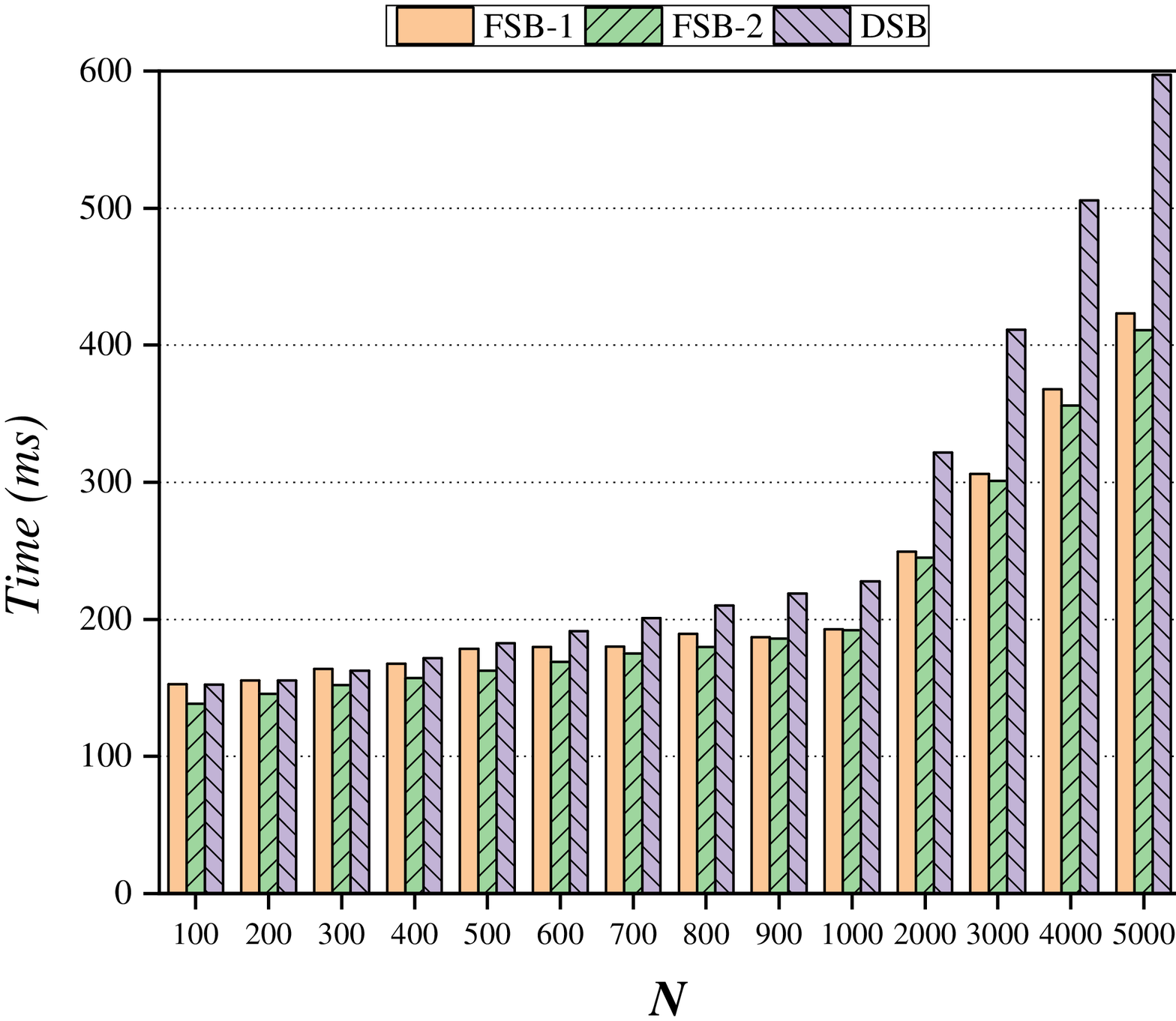}
        \label{cel-time}
    }

    \caption{Results of identification time for six object programs.}
    \label{FIG:programTime}
\end{figure*}

\begin{figure*}[!b]
\centering
    \subfigure[\texttt{airy}]
    {
        \includegraphics[width=0.315\textwidth]{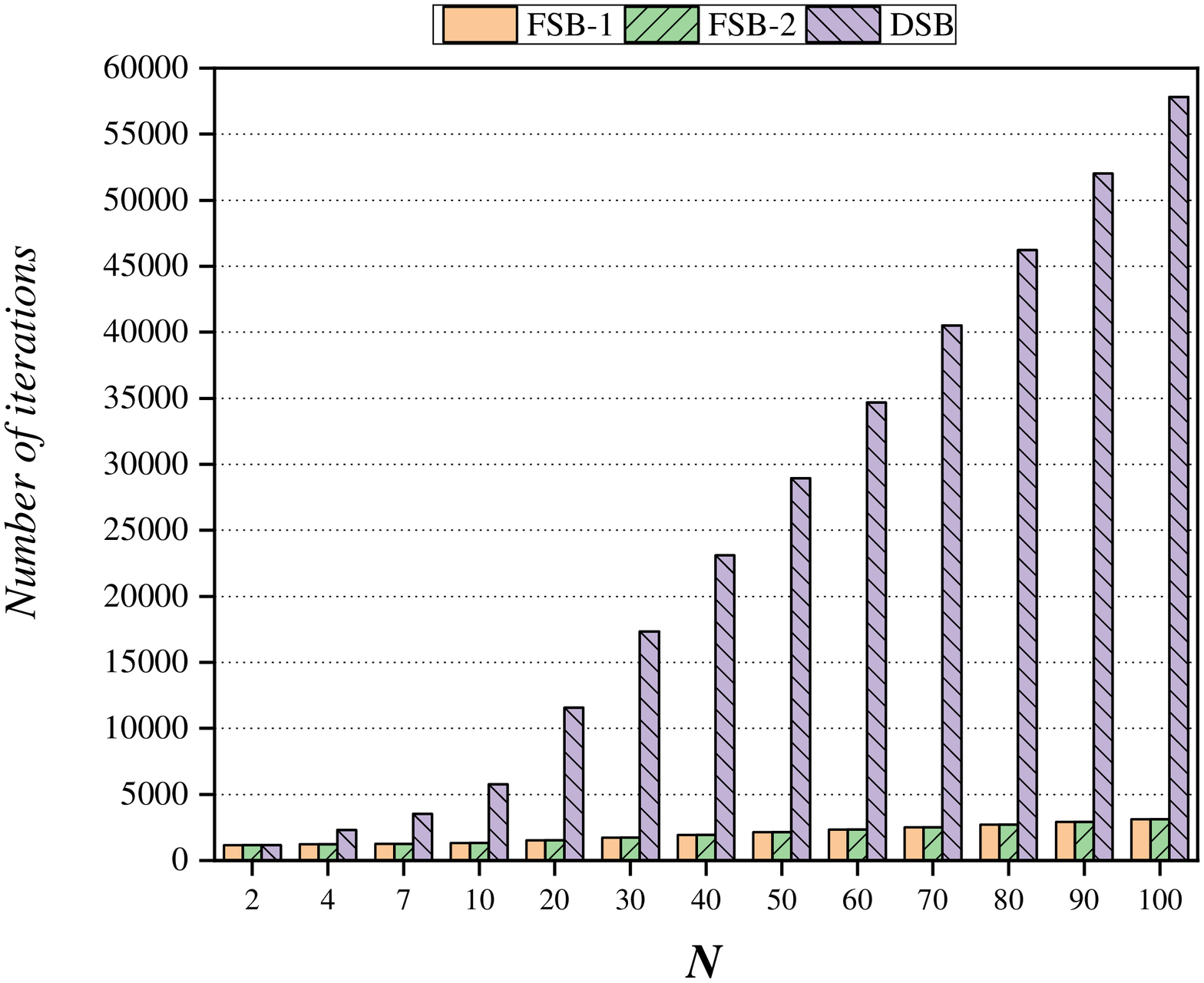}
        \label{airy-ite}
    }
       \subfigure[\texttt{gammq}]
    {
        \includegraphics[width=0.315\textwidth]{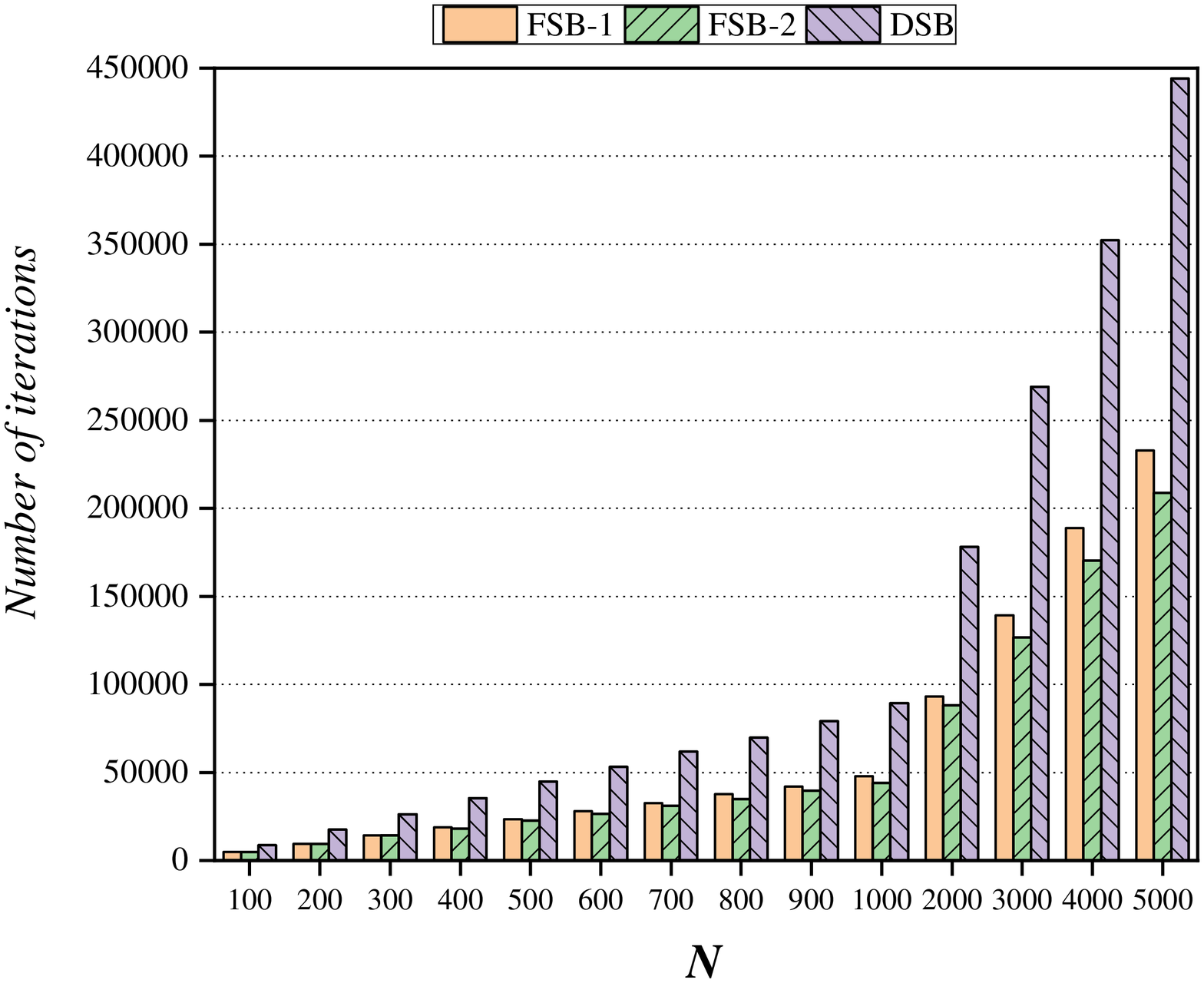}
        \label{gammq-ite}
    }
        \subfigure[\texttt{Expint}]
    {
        \includegraphics[width=0.315\textwidth]{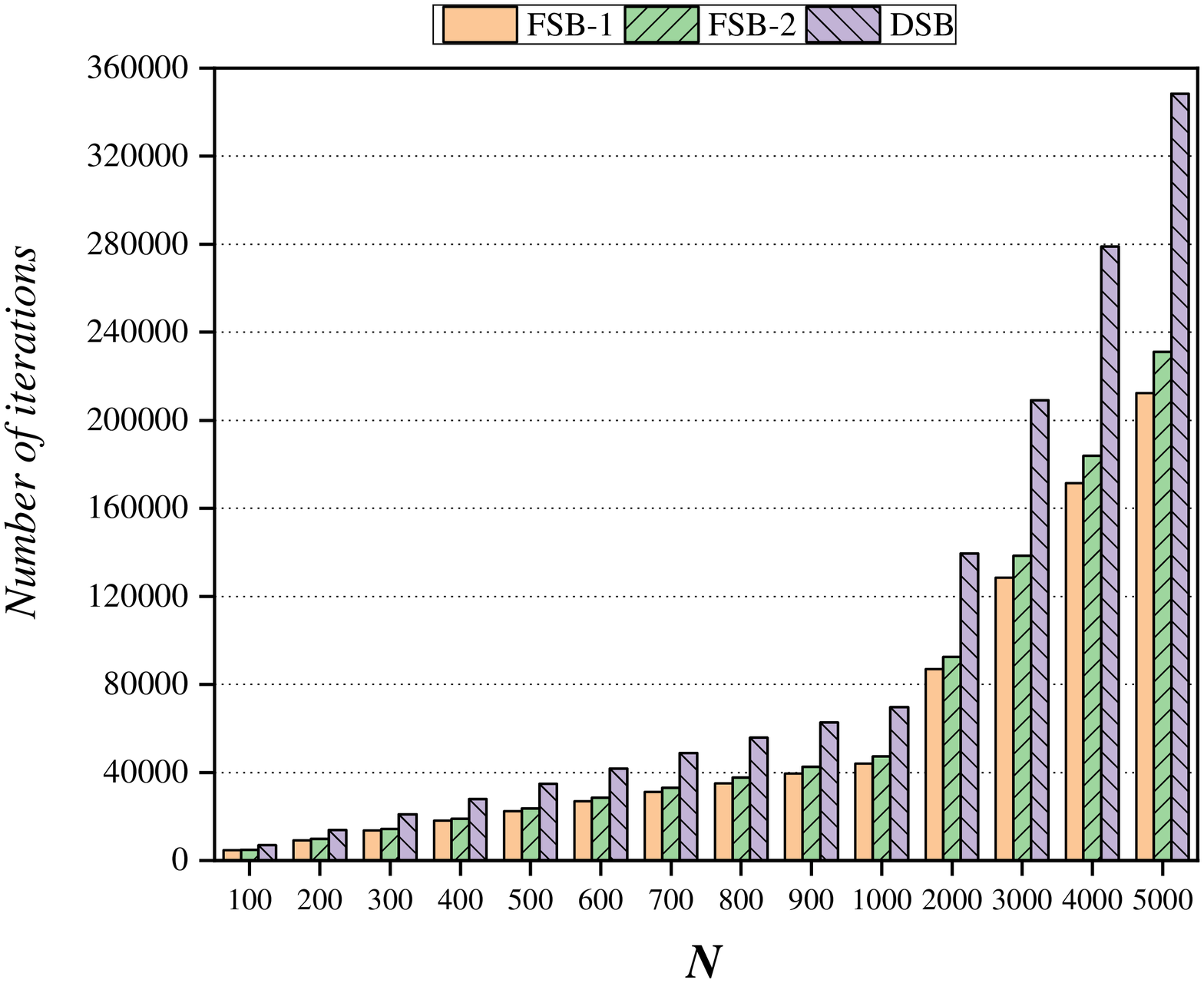}
        \label{expint-ite}
    }
        \subfigure[\texttt{bessj}]
    {
        \includegraphics[width=0.315\textwidth]{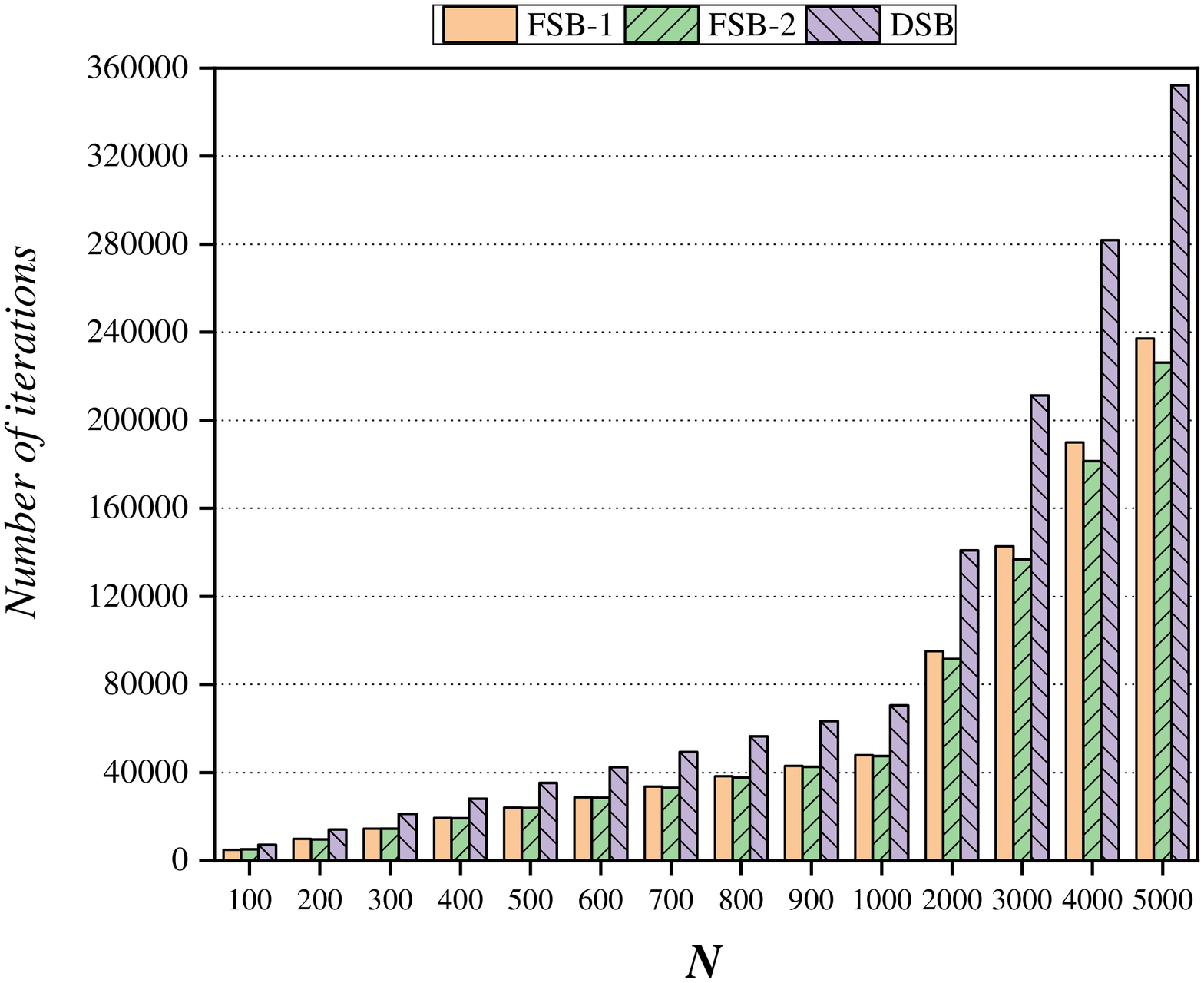}
        \label{bessj-ite}
    }
        \subfigure[\texttt{triangle}]
    {
        \includegraphics[width=0.315\textwidth]{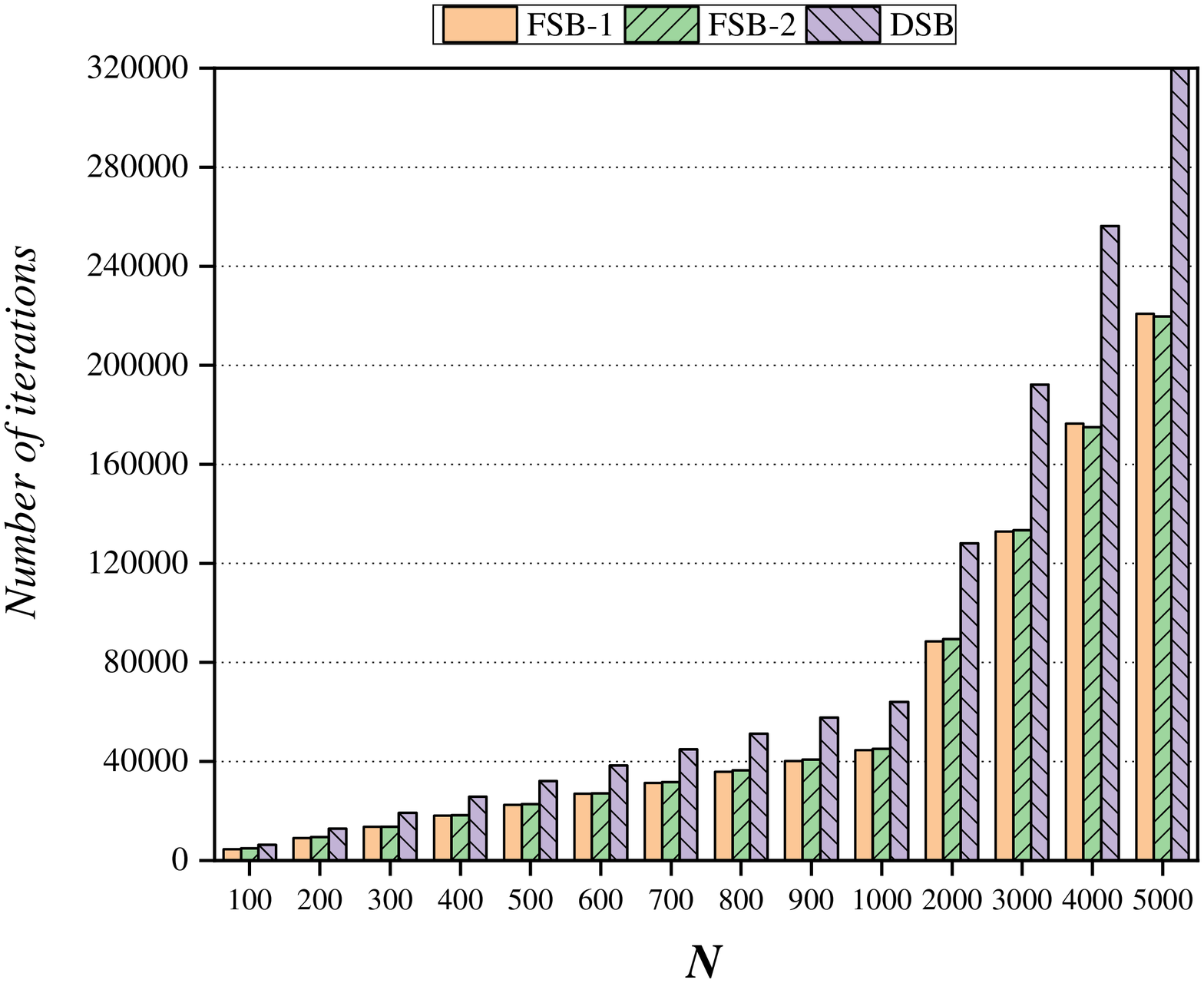}
        \label{triangle-ite}
    }
          \subfigure[\texttt{cel}]
    {
        \includegraphics[width=0.315\textwidth]{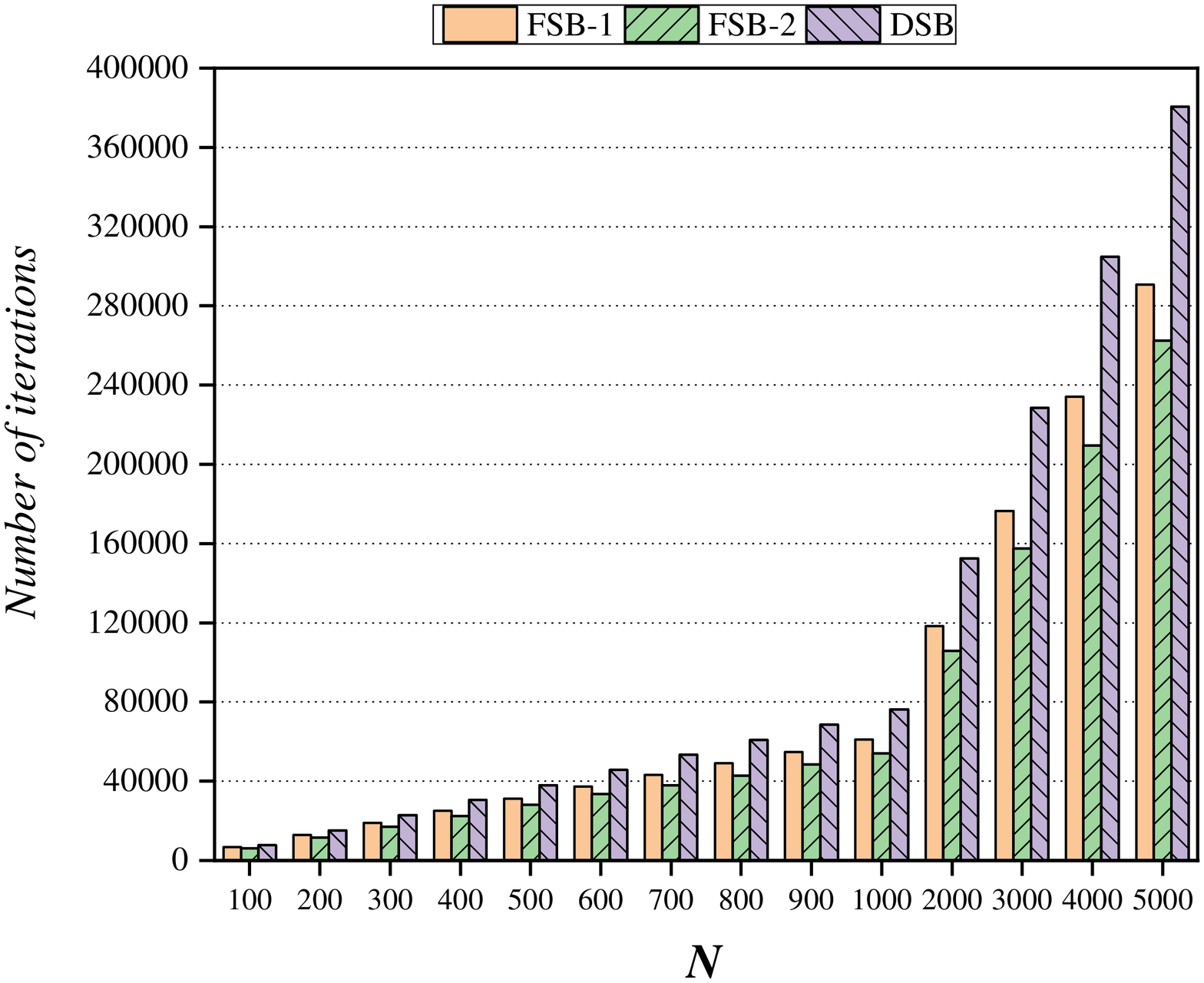}
        \label{cel-ite}
    }

    \caption{Results of the number of iterations for six object programs.}
    \label{FIG:programIterations}
\end{figure*}

Figure~\ref{FIG:programRatio} shows our $\mathcal{S}_\textit{ratio}$ results against $N$, the number of failure-causing boundary inputs. Similarly, Figure~\ref{FIG:programTime} shows the identification time results; while Figure~\ref{FIG:programIterations} provides the results of the number of iterations.

\subsubsection{Answers to RQ1: $\mathcal{S}_\textit{ratio}$ Comparisons}
\label{SEC:RQ1_real}
Based on Figure~\ref{FIG:programRatio}, the followings are observed.
\begin{itemize}
{\item[1)] For the program \texttt{airy}, there are no differences among FSB-1, FSB-2, and DSB, regardless of the number of failure-causing boundary inputs (i.e., $N$). More specifically, when identifying more than 2 failure-causing boundary inputs (i.e., $N \geq 2$), each technique could identify an AFR extremely close to the RFR, because all $\mathcal{S}_\textit{ratio}$ values almost approach 1.0. For the program \texttt{bessj}, when $N$ is small, FSB has slightly better $\mathcal{S}_\textit{ratio}$ performances than DSB. However, with the increase of $N$, the case is very similar to that of the program \texttt{airy}, i.e., the $\mathcal{S}_\textit{ratio}$ values of both FSB and DSE approach 1.0.}

\item[2)] For the program \texttt{gammq}, FSB-1 performs worse than FSB-2 and DSB, irrespective of $N$. As for the comparisons between FSB-2 and DSB, when $N$ is small, FSB-2 is better. However, with the increase of $N$, they have very similar performances, because their $\mathcal{S}_\textit{ratio}$ values both approach 1.0.

\item[3)] For the program \texttt{cel}, FSB-1 performs the best, followed by FSB-2 and DSB. Obviously, when the number of failure-causing boundary inputs is equal to or larger than 2000, only FSB-1 has the $\mathcal{S}_\textit{ratio}$ values approaching 1.0.

{\item[4)] For the programs \texttt{expint} and \texttt{triangle}, FSB-1 performs worse than FSB-2 and DSB, irrespective of $N$. Comparing to DSB, when $N$ is small, FSB-2 performs slightly worse than DSB. However, with an increase of $N$, they have very similar performances because their $\mathcal{S}_\textit{ratio}$ values both approach 1.0.}
\end{itemize}

Here, we briefly provide an analysis to explain the above observations. In fact, these observations in empirical studies are highly consistent with those in our simulation studies discussed in Section \ref{SEC:simulationResults}. As described in Table~\ref{TAB:Program}, the failure pattern of the program \texttt{airy} is a line segment in a 1-dimensional input domain, meaning that there exist only two extension directions. As a result, all SB methods can quickly identify the failure-causing boundary inputs along these two extension directions to identify an AFR (which is very close to the RFR). For the program \texttt{bessj}, its failure region is a right-angled triangle with two of its sides parallel to the coordinate axes. As discussed in Section~\ref{SEC:algorithm}, FSB makes use of the information of coordinate axes as the extension directions, which means that the extension directions of FSB are parallel to the sides of the failure region. Therefore, FSB can quickly search three vertexes of this triangle failure region and successfully identify the RFR (i.e., $\mathcal{S}_\textit{ratio}=1.0$). Compared with FSB, it is more difficult for DSB to identify the three vertexes of this triangle (due to its property of diverse extension orientations), especially when $N$ is small. With an increase of $N$, however, some failure-causing boundary inputs identified by DSB are very close to these three vertexes, i.e., the AFR is very close to the RFR.

The failure pattern of the program \texttt{gammq} is a strip with the rotation angle $45^\circ$ in a two-dimensional input domain. As shown in Figure~\ref{FIG:direction2}, FSB-2 has two extension directions that are parallel to two lines of the failure pattern (which means that it can reach the corresponding boundaries quickly), resulting in higher $\mathcal{S}_\textit{ratio}$ values than the other two techniques (FSB-1 and DSB). In addition, DSB can choose extension directions similar to FSB-2 due to its diversity, especially when $N$ becomes large. However, FSB-1 chooses only four fixed extension directions, which indicates that it is difficult to identify failure-causing boundary inputs that are close to the narrow lines of the strip, resulting in its worst performances.

As for the program \texttt{cel}, its failure pattern is a strip with the rotation angle $0^\circ$ in a four-dimensional input domain. Through an analysis of the failure pattern, we found that this failure region covers the full range of the first, third and fourth parameter but only partial range of the second parameter, meaning that all failure-causing inputs are related to only the second parameter. In other words, if the second parameter is assigned by an appropriate value, regardless of values of the other three parameters, this failure can be triggered. As a result, this case is favorable to FSB-1 and FSB-2 but not DSB, hence resulting in its worst performance. Since
FSB-1 can directly identify each failure-causing boundary input through each coordinate axis (that is considered as an extension direction), it has better performances than FSB-2.

Regarding the programs \texttt{expint} and \texttt{triangle}, we analyzed the failure region of each program, and observed that the shape of the failure region is irregular. However, each failure region still has some edges or planes that are parallel to the coordinate axes. Therefore, FSB could achieve reasonably good performances with its minimum $\mathcal{S}_\textit{ratio}$ value greater than $0.7$. Nevertheless, it is difficult to identify an irregular failure region when only the information of coordinate axes are used. Apart from the information about coordinate axes, FSB-2 adopts the information about orthants, which may have better performance than FSB-1 for identifying the irregular failure region. Compared with FSB (including FSB-1 and FSB-2), DSB may be preferable to support the identification of irregular failure regions because of its ability to search the failure-causing boundary inputs through diverse extension orientations.

\subsubsection{Answers to RQ2: Comparisons of Identification Time}
{As shown in Figure~\ref{FIG:programTime}, we have the following observations:}
\begin{itemize}
\item[1)] For the programs \texttt{airy} and \texttt{triangle}, FSB-1 and FSB-2 have similar or exactly equal time to identify the AFRs. However, DSB needs longer identification time than FSB-1 or FSB-2, especially when $N$ is large.
\item[2)] For the programs \texttt{gammq} and \texttt{bessj}, when $N$ is small (i.e., $N < 400$), the differences of identification time among FSB-1, FSB-2, and DSB are very small. With the increase of $N$, however, FSB-2 has the best performances, followed by FSB-1 and DSB.

\item[3)] FSB-2 generally has the best performances for the program \texttt{cel}, irrespective of $N$. On the other hand, when $N$ is equal to or less than 600, DSB has a comparable performance to FSB-1; however, when $N > 600$, DSB may have worse results than FSB-1. Nevertheless, DSB only requires about 0.6 seconds to identify 5000 failure-causing boundary inputs.

\item[4)] For the program \texttt{expint}, FSB-1 has a similar or slightly better performance than FSB-2. As for the comparisons between FSB and DSB, when $N$ is small (i.e., $N < 300$), they have very similar performances; however, with an increase of N, DSB performs the worst.
\end{itemize}

Figure~\ref{FIG:programIterations} reports the average number of iterations for these six programs, from which we can observe that the results of iterations are highly consistent with those of identification time overall. In effect, more iterations may generally require longer identification time.

Here, we briefly explain the above observations. Since the program \texttt{airy} has a 1-dimensional input domain, FSB-1 and FSB-2 are equivalent, resulting in the same computational time. As we know, each SB technique only needs two failure-causing boundary inputs (i.e., $N=2$) to identify the 1-dimensional AFR. After that, FSB-1 or FSB-2 could identify the remaining failure-causing boundary inputs by making the first two failure-causing boundary inputs closer to the RFR. However, DSB repeats the previous process to identify the next two failure-causing boundary inputs, resulting in more identification time than FSB. From Figure~\ref{FIG:programIterations}, it can be observed that with the same $N$, DSB needs a larger number of iterations to identify the AFR. For the program \texttt{triangle}, FSB-1 and FSB-2 use a similar number of iterations to obtain the failure-causing boundary inputs. Since both FSB methods use little identification time to select extension directions, hence there is only a very small difference between their identification times.

Regarding the second and third observations, FSB-2 satisfies the favorable condition for the program \texttt{gammq} and \texttt{bessj} (as discussed in Section \ref{SEC:RQ1_real}). Therefore, FSB-2 performs the best with program \texttt{gammq} and \texttt{bessj}. For the program \texttt{cel}, FSB-2 needs the least number of iterations to identify the AFR, irrespective of $N$, leading to the best performance. For the program \texttt{expint}, when $N$ is small, the difference in the number of iterations between FSB-1 and FSB-2 is also small. However, with an increase of $N$, FSB-2 needs a larger number of iterations to identify an AFR, resulting in longer identification time. In addition, since DSB needs to choose diverse extension directions, it generally requires more identification time than FSB, especially when $N$ is large.

\textbf{\emph{To conclude:}} None of the SB techniques is always the best for all the programs, which means that each SB technique has its own favorable and unfavorable conditions. Overall, the observations of empirical studies are very consistent with those of simulation studies.

\subsection{Threats to Validity}

For our simulations, there are many factors that affect the failure pattern of RFR, such as failure rate, shape, compactness, orientation, etc., resulting in many possible settings can be selected. However, in reality, we can only cover some but not all representative parts of them. More specifically, for each possible influencing factor, we selected a limited number of representative settings and obtained a total of $504$ simulation settings. The selection of simulation values covers a wide range, e.g., compactness is from $1$ to $100$, orientation is from $0 ^\circ$ to $180 ^\circ$, etc., aiming at investigating the effectiveness of the proposed methods under various possible factors. In the future, it would be better to select more simulation values with different granularities for providing more comprehensive results. Meanwhile, in order to evaluate the effectiveness of our methods, we adopted the convex hull algorithm to obtain the area of the AFR. This process was implemented through an external program \texttt{qhull}, involving some operations of reading and writing files, which may provide some fluctuations in terms of time overhead.

For the empirical studies, we investigated the effectiveness of SB methods for program with failure-unrelated parameters (including \texttt{cel}). Although our proposed methods have good performances (i.e.,when $N=5000$, the $S_{ratio}$ is greater than $80\%$), it is difficult to confirm which failure-related parameters actually cause a software failure, especially for programs with a large number of parameters. As we know, \emph{delta debugging}~\cite{Zeller2002}, as a fault locating method, has been widely adopted to simplify or isolate failure causes. In future study, we will attempt to use delta debugging technique to locate the failure-related parameters prior to adopting our proposed methods for effectively identifying failure regions. In addition, we only used six real-life programs in the empirical studies. Obviously, it would be better to investigate the performance of SB by using more faulty programs in real-life. As discussed in Section \ref{SEC:assumptions}, SB has three assumptions, which may have some impacts on its applicability in practice. Nevertheless, we will discuss how to possibly relax these assumptions in the next section.

\section{About the Assumptions}
\label{SEC:Weaken}

Although we have justified these assumptions mentioned in Section~\ref{SEC:assumptions}, these assumptions do not always hold in practice. In this section, therefore, we discuss how to narrow the gap between the practical and ideal problem framework.

\subsection{Assumption 1}
{In testing, test cases may be generated by a testing technique to execute the system under test (SUT). Once a software failure is triggered by a test case, testers may stop testing, and then adopt our proposed methods to identify the related failure region of this failure-causing test case. In fact, there is no restriction for either FSB or DSB to be applied for identifying failure region with more than one failure-causing inputs.} As discussed in Section~\ref{SEC:ABM}, FSB uses the following failure-causing boundary inputs to repeat the previous iteration. Therefore, when more than one failure-causing input is considered as the input parameter, FSB can choose these failure-causing inputs one by one for each iteration. In addition, DSB does not change the failure-causing source input during the whole process, however, it can select one arbitrarily as the failure-causing source input when encountering multiple failure-causing inputs. The main reason for this is that DSB adopts diverse orientations to identify failure-causing boundary inputs, indicating that different failure-causing source inputs may provide little impact on the identification of failure region, especially when the number of failure-causing boundary inputs is large.

\subsection{Assumption 2}
In reality, multiple failure regions do exist in faulty programs. The identification of multiple failure regions can be seen as the parallelization of a single failure region identification. If we obtain a failure-causing input from each failure region, then the failure region could be identified by SB.

\begin{figure*}[!t]
\centering
    \includegraphics[width=1\textwidth]{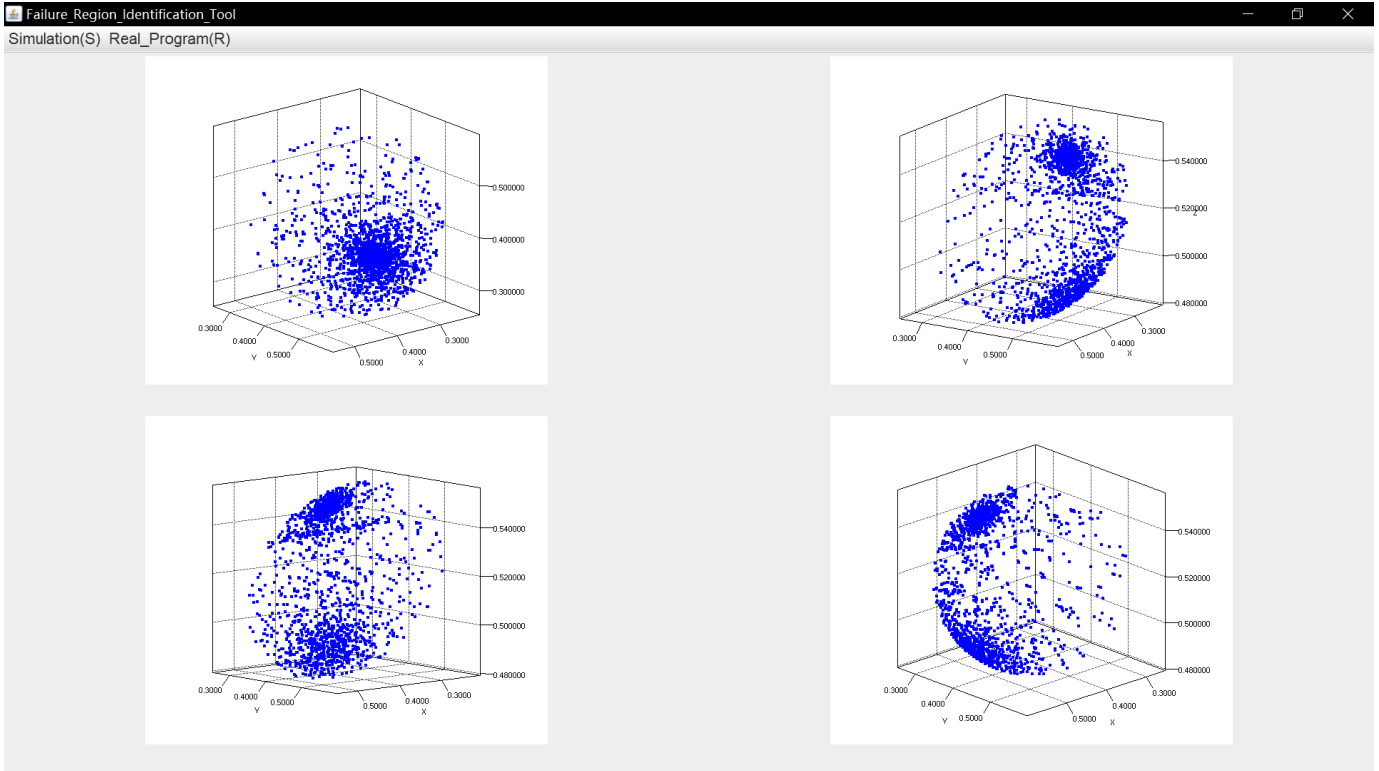}
    \caption{An example of hyperellipsoid AFR visualization with $d=4, \delta=5, \gamma=10$ in three-dimensional spaces.}
    \label{FIG:example}
\end{figure*}

\subsection{Assumption 3}

If this assumption does not hold (that is, the shape of failure region is concave), then the failure region identified by SB may possibly contain some successful inputs, which also means that the actual failure region may not contain the identified failure region. The identified failure region may still provide some insights for further investigations such as fault localization~\cite{Wong2016} and program repair~\cite{Monperrus2018}. Nevertheless, it would be interesting to propose new IFR methods for dealing with concave failure regions in the future.

\section{An Experimentation Platform for IFR}
\label{SEC:Tool}
We proposed two SB methods, FSB and DSB, and then developed an automated experimentation platform to identify and visualize the failure region\footnote{The implementation of our two methods and tool are available at https://github.com/huangrubing/IFR/.}. At this stage, the platform is mainly for simulation experiments, which would be extended to real-life programs.

In the platform, we need to define the features of the input domain such as dimension and scope, and then set the properties of the failure region in advance, including the shape, size, compactness, and orientation. After that, a failure region is randomly placed within the simulated input domain. The first failure-causing input is detected by using a testing method. As soon as the stopping criterion is satisfied, a XML file could be created which stores the experimental settings, the first failure-causing input and the set of failure-causing boundary inputs. Besides, this platform also provides a visual interface to show the AFR using the generated failure-causing boundary inputs. {As far as we understand, any visualization tools (including our platform) can be used for the three-dimensional space. Regarding the visualization of AFR for dimensions higher than 3 (i.e., $d > 3$), our platform can project $d$-dimensional points onto the three-dimensional space by choosing arbitrarily any 3 out of $d$ dimensions. As shown in Figure~\ref{FIG:example}, we provide a three-dimensional visualization example of a hyperellipsoid AFR with the dimension $d=4$.}

\begin{figure*}[!b]
\centering
    \subfigure[]
    {
        \includegraphics[width=0.235\textwidth]{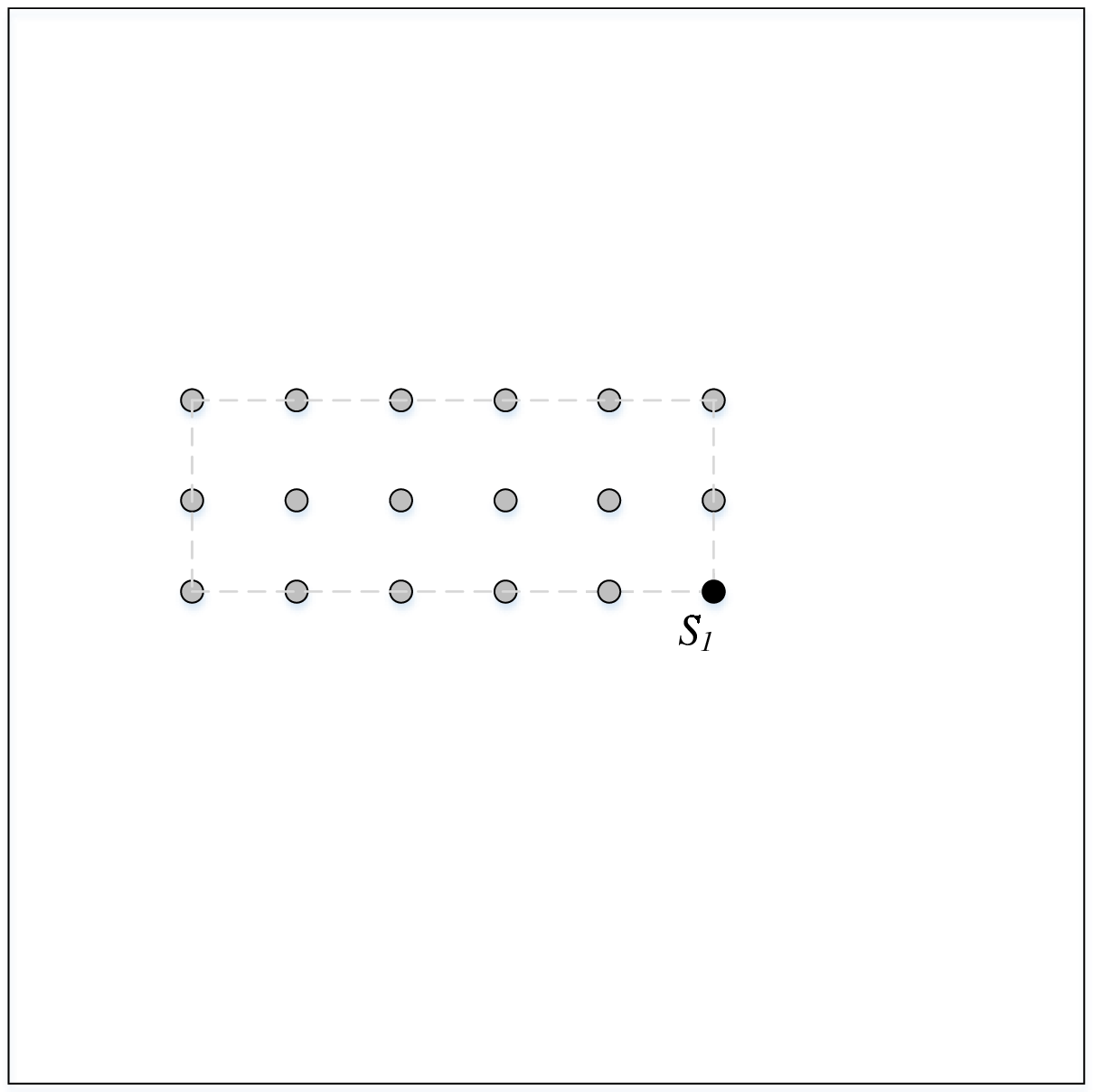}
        \label{ADFD:1}
    }
    \subfigure[]
    {
        \includegraphics[width=0.235\textwidth]{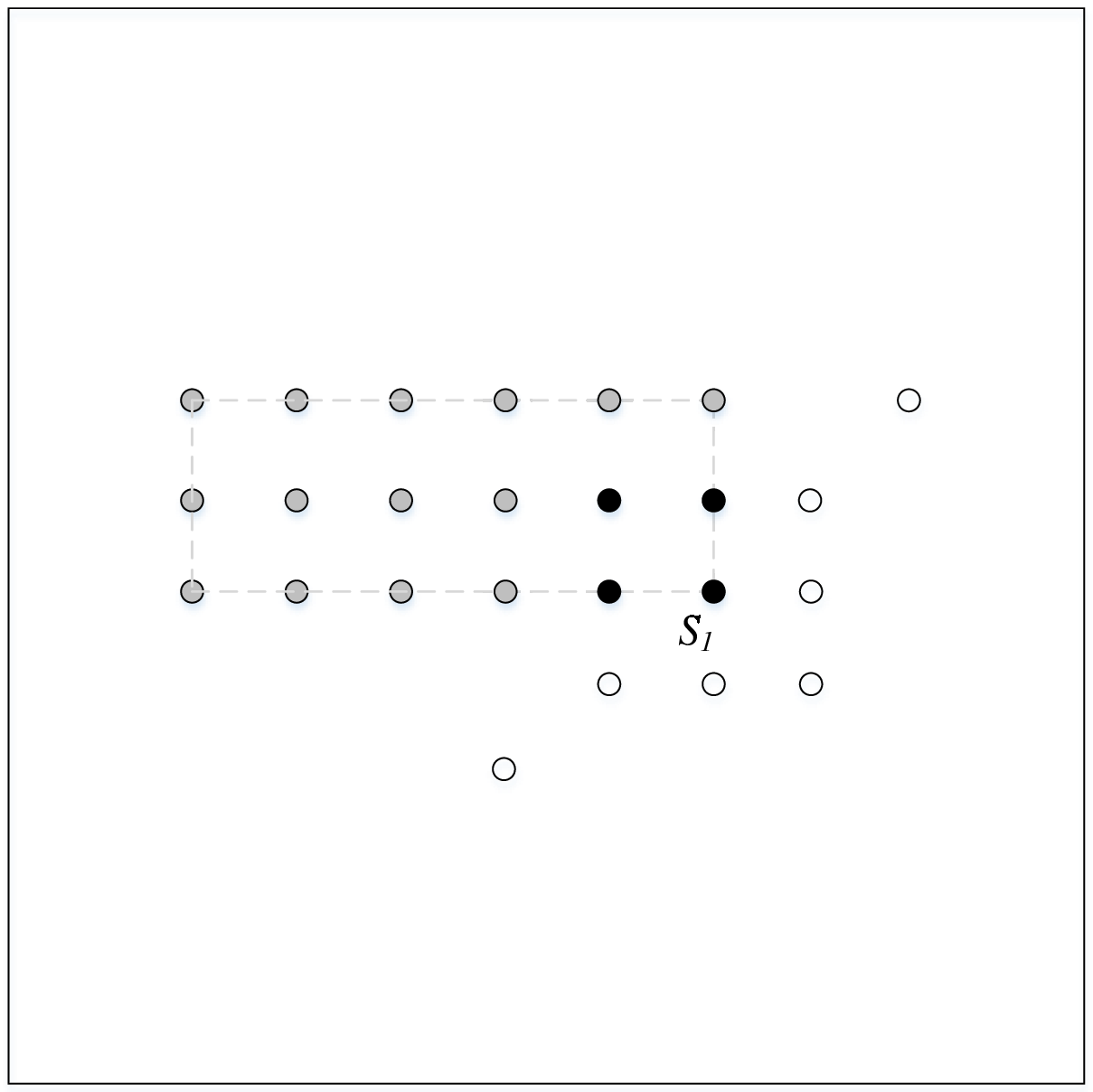}
        \label{ADFD:2}
    }
    \subfigure[]
    {
        \includegraphics[width=0.235\textwidth]{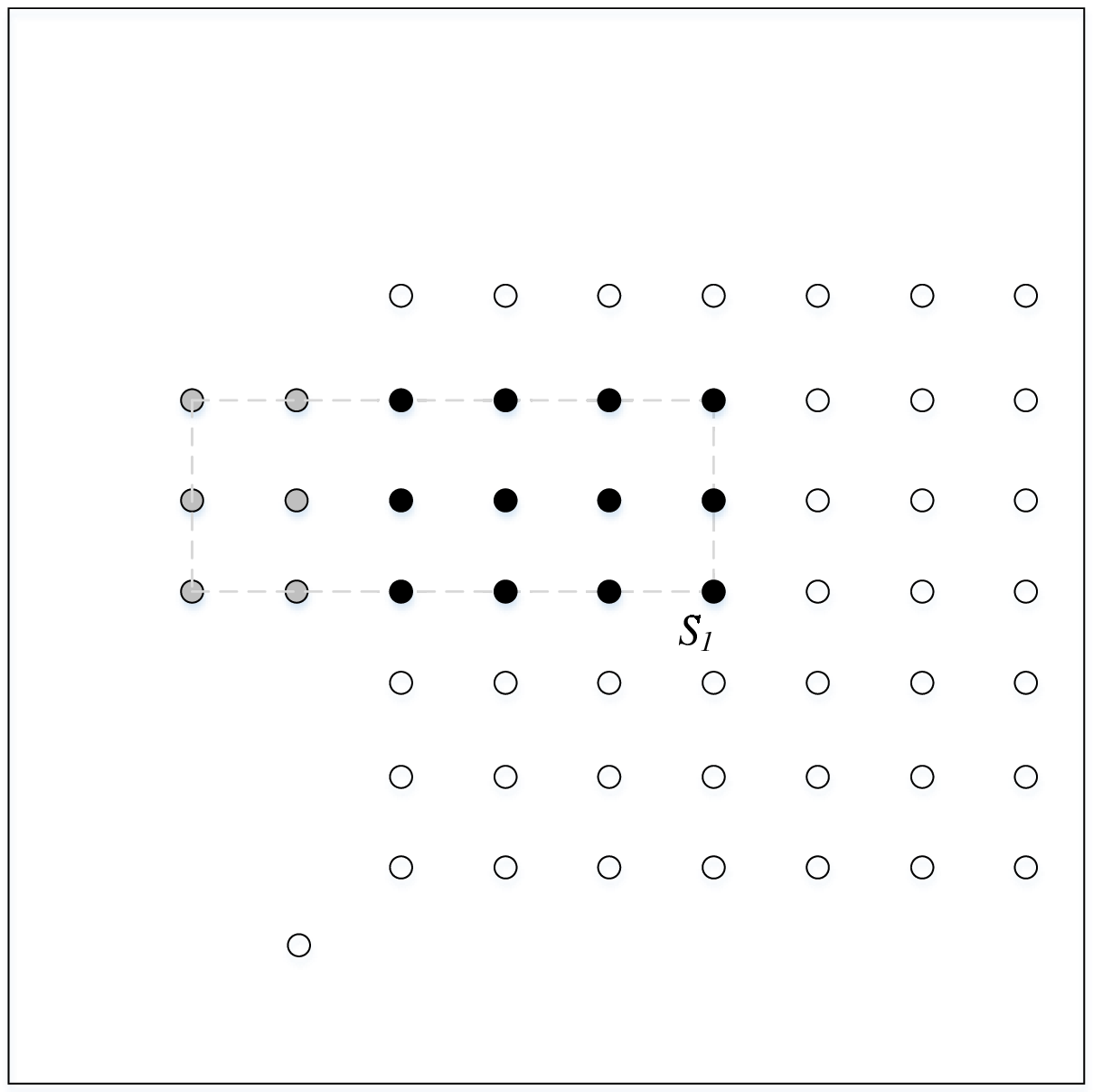}
        \label{ADFD:3}
    }
        \subfigure[]
    {
        \includegraphics[width=0.235\textwidth]{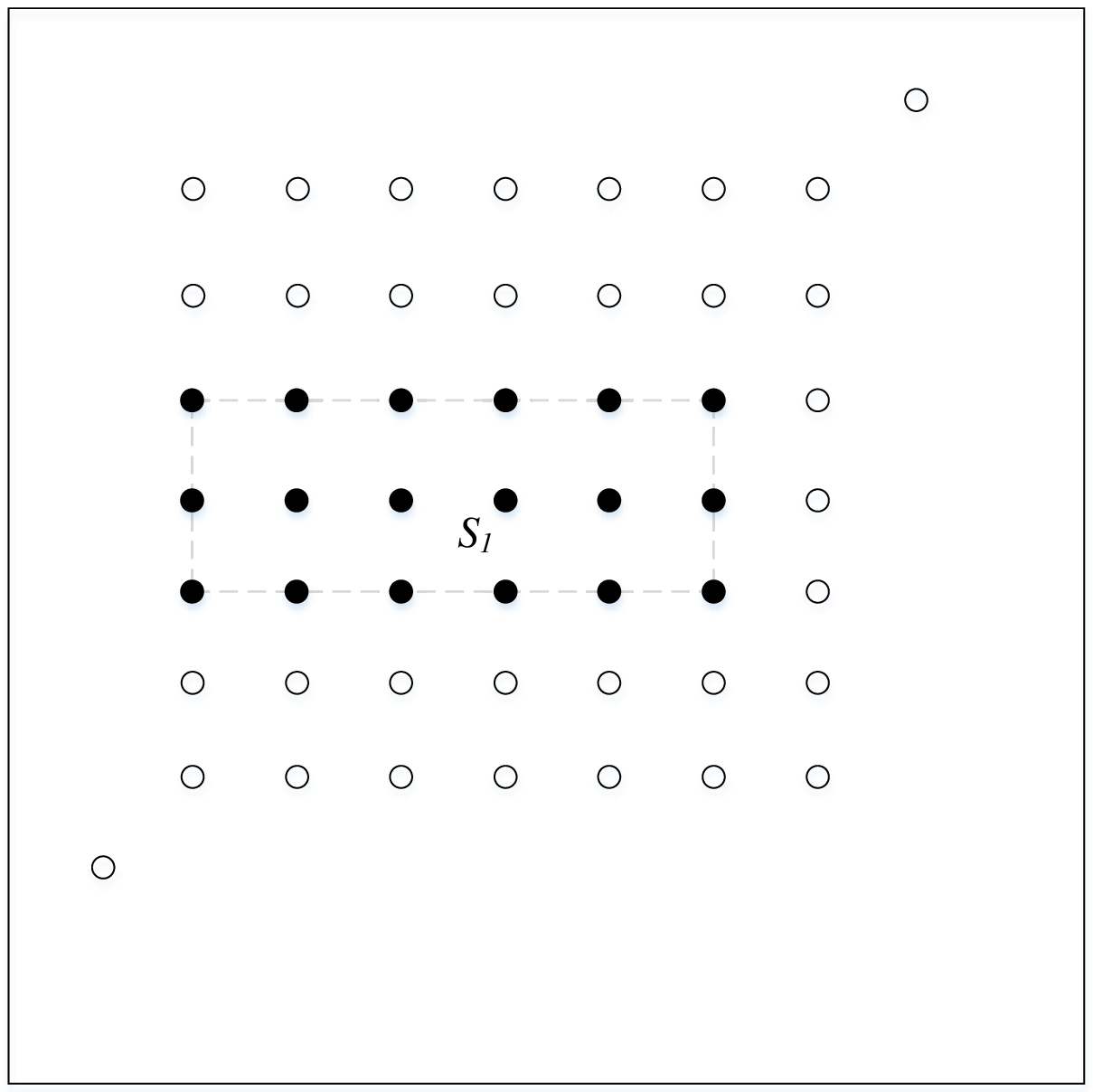}
        \label{ADFD:4}
    }

    \caption{An example of ADFD+ in two-dimensional failure region.}
    \label{ADFD:example}
\end{figure*}

\section{Related work}
\label{SEC:RelatedWork}

In this section, we briefly present some related works about IFR. To the best of our knowledge, only \textit{Automated Discovery of Failure Domain} (ADFD) \cite{Ahmad2013} and its enhancement ADFD+ \cite{Ahmad2014} have attempted to identify the failure regions. In fact, main functions of ADFD are fully covered by ADFD+, which means that ADFD+ is more comprehensive and effective than ADFD \cite{Ahmad2014}. Therefore, we mainly discuss and compare ADFD+ against our proposed methods.

Similar to our proposed methods, ADFD+ needs a failure-causing input $S_1$ as the parameter input. In addition, ADFD+ needs a parameter (i.e., \emph{DomainRange}) to determine the inclusion region, and then scans the whole test inputs within this inclusion region. Compared with our SB methods, ADFD+ suffers from the following two drawbacks: (1) it can only be applied to input domains with integer inputs rather than floating inputs, because it is impossible to exhaustively execute floating inputs; while our proposed methods can be applied to any numeric inputs; and (2) the value of \emph{DomainRange} and the location of $S_1$ have a significant impact on their performances. Figure \ref{ADFD:example} presents an illustrative example of ADFD+ in a two-dimensional input domain. As shown in Figure \ref{ADFD:1}, failure-causing inputs are labelled by the gray circles (to form a block failure region), and $S_{1}$ is the failure-causing source input (labelled by a black circle). When the \emph{DomainRange} is equal to 2, ADFD+ scans the 11 test cases, from which only 4 test cases are failure-causing (as shown in Figure~\ref{ADFD:2}). However, when \emph{DomainRange} is equal to 4, 51 test cases are scaned by ADFD+, from which only 12 test cases are failure-causing inputs (as shown in Figure \ref{ADFD:3}). As shown in Figure \ref{ADFD:4}, if the failure-causing source input $S_1$ is located in the center of the failure region, ADFD+ will identify an exact failure region using $\emph{DomainRange}=4$.

From the above example, it can be observed that the performances of ADFD+ highly depend on the size of the inclusion regions and the location of $S_1$. More specifically, when using the same value of \emph{DomainRange}, ADFD+ with $S_1$ located in the center of the failure region performs better than that with $S_1$ near the boundary of the failure region. In practice, however, the first failure-causing input may be likely to be located in any location within the failure region. Similarly, when the location of $S_1$ is fixed, the larger the \emph{DomainRange}, the better the performance of ADFD+. However, a larger value of \emph{DomainRange} may bring more redundant successful inputs, especially for some narrow failure regions. Unfortunately, the selection of \emph{DomainRange} is a challenge for many testers, and there is no guidance from the previous studies \cite{Ahmad2013,Ahmad2014} about how to choose an appropriate value of it.

As discussed in Section \ref{SEC:method}, our SB methods also need a parameter $L$ during the process of IFR. However, even though a large value of $L$ is selected, our methods can rapidly iterate from a successful input to a failure-causing input, and therefore avoid executing a large amount of successful inputs. In addition, the position of the first failure-causing input $S_1$ has little impact on the performances of our proposed methods, especially when the number of failure-causing boundary inputs is large.

\section{Conclusions and Future Work}
\label{SEC:Conclusions}
To support \emph{Identification of Failure Regions} (IFR), this paper has provided a new technique, namely \textit{Search for Boundary} (SB), to identify an approximate failure region in a numeric input domain. We have proposed two methods, i.e., \textit{Fixed-orientation Search for Boundary} (FSB) and \textit{Diverse-orientation Search for Boundary} (DSB) to support SB. In addition, we developed an automated experimentation platform to support these two methods, and also provided an interface of the visualization process. The simulation results indicate that all SB methods can effectively identify an approximate failure region, and DSB is more cost-effective than FSB overall, while two versions of FSB perform with similar effectiveness and efficiency.

As discussed in the simulation and empirical studies, the proposed methods may encounter unfavorable conditions corresponding to some types of failure patterns. In effect, the key point of IFR is to identify failure-causing boundary inputs as diverse as possible, which can be considered as a search-based task. Since there are many searching techniques developed by the \textit{search-based software testing} community~\cite{Harman2015}, it would be promising to adopt these search-based techniques for supporting IFR work in the future.

%
\ifCLASSOPTIONcompsoc
  \section*{Acknowledgments}
\else
  \section*{Acknowledgment}
\fi
We would like to thank the anonymous reviewers for their many constructive comments.
This work is supported by the National Natural Science Foundation of China under grant nos. 61872167, 61502205, and U1836116, the project funded by China Postdoctoral Science Foundation under grant no. 2019T120396, and the Postgraduate Research \& Practice Innovation Program of Jiangsu Province under grant no.~KYCX19\_1614. This work is also in part supported by the Senior Personnel Scientific Research Foundation of Jiangsu University under grant no.~14JDG039, the Young Backbone Teacher Cultivation Project of Jiangsu University.

\bibliographystyle{IEEEtran}
\bibliography{ABM}

\begin{IEEEbiography}[{\includegraphics[width=1in,height=1.25in,clip,keepaspectratio]{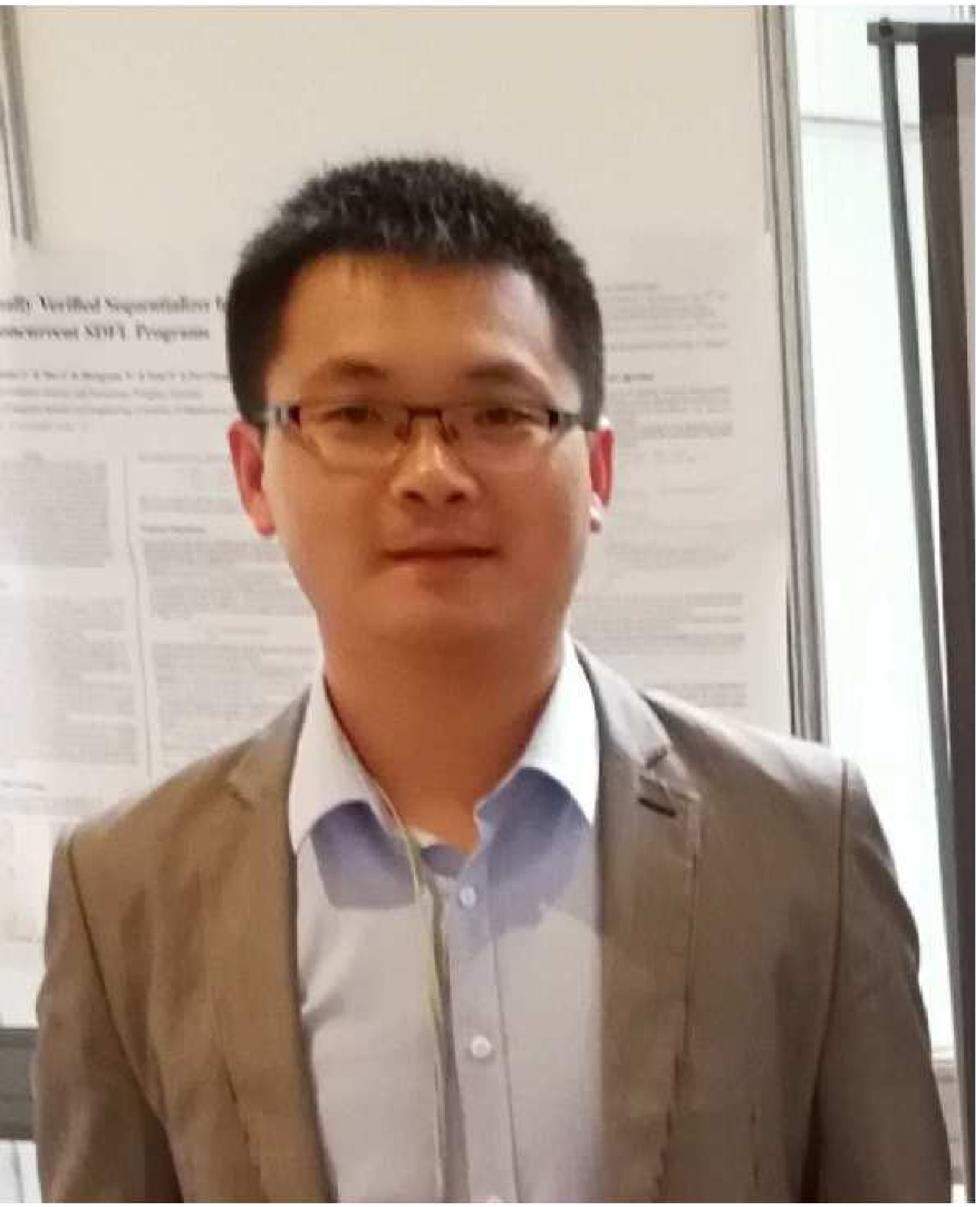}}]{Rubing Huang}
received the Ph.D. degree in computer science and technology from the Huazhong University of Science and Technology, Wuhan, China, in 2013. From 2016 to 2018, he was a visiting scholar at Swinburne University of Technology and at Monash University, Australia. He is an associate professor in the Department of Software Engineering, School of Computer Science and Communication Engineering, Jiangsu University, Zhenjiang, China. His current research interests include software testing (including adaptive random testing, random testing, failure-based testing, combinatorial testing, and regression testing), debugging, and maintenance. He has more than 50 publications in journals and proceedings, including in IEEE Transactions on Software Engineering, IEEE Transactions on Reliability, IEEE Transactions on Emerging Topics in Computational Intelligence, Journal of Systems and Software, Information and Software Technology, IET Software, The Computer Journal, International Journal of Software Engineering and Knowledge Engineering, ICSE, ICST, COMPSAC, QRS, SEKE, and SAC. He is a senior member of the IEEE and the China Computer Federation, and a member of the ACM. More about him and his work is available online at https://huangrubing.github.io/.
\end{IEEEbiography}

\begin{IEEEbiography}[{\includegraphics[width=1in,height=1.25in,clip,keepaspectratio]{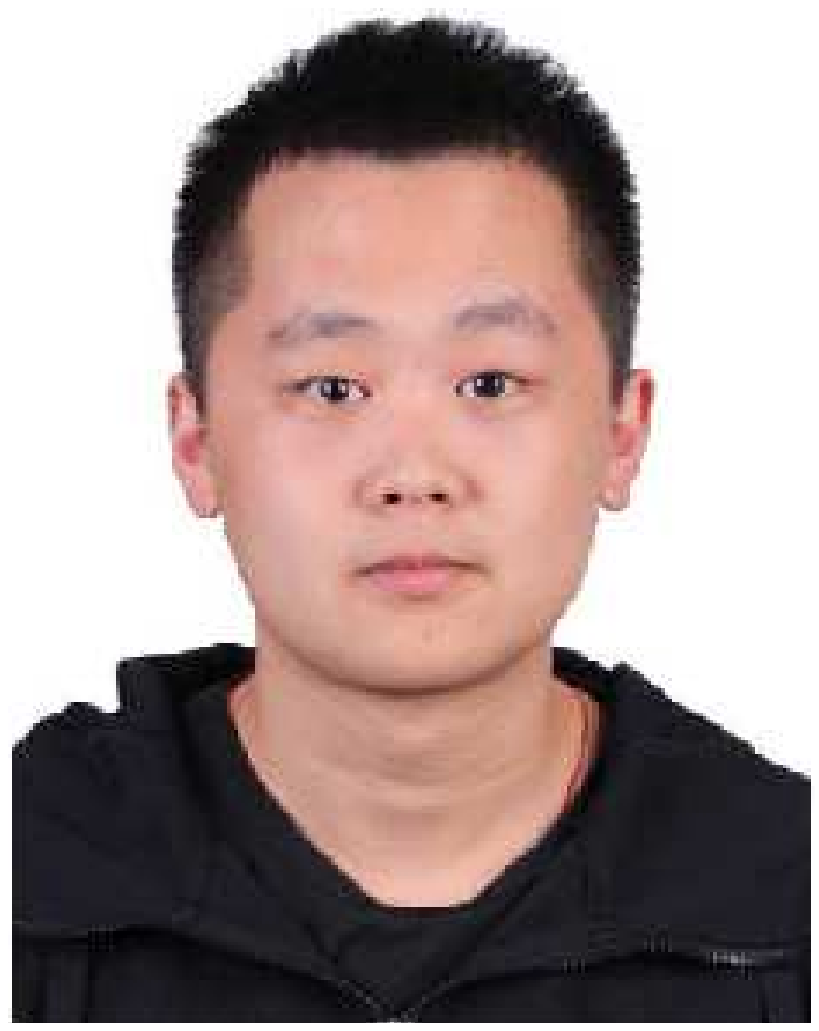}}]{Weifeng Sun}
received the B.Eng. degree in computer science and technology in 2018 from Jiangsu University, Zhenjiang, China, where he is currently working toward the M.Eng. degree with the School of Computer Science and Communication Engineering.
His current research interests include software testing and software debugging. His work has been published in journals and proceedings, including in IEEE Transactions on Software Engineering, IEEE Transactions on Reliability, IEEE Transactions on Emerging Topics in Computational Intelligence, and the IEEE International Conference on Software Testing, Verification and Validation (ICST).
He is a student member of the China Computer Federation and the ACM.
\end{IEEEbiography}

\begin{IEEEbiography}[{\includegraphics[width=1in,height=1.25in,clip,keepaspectratio]{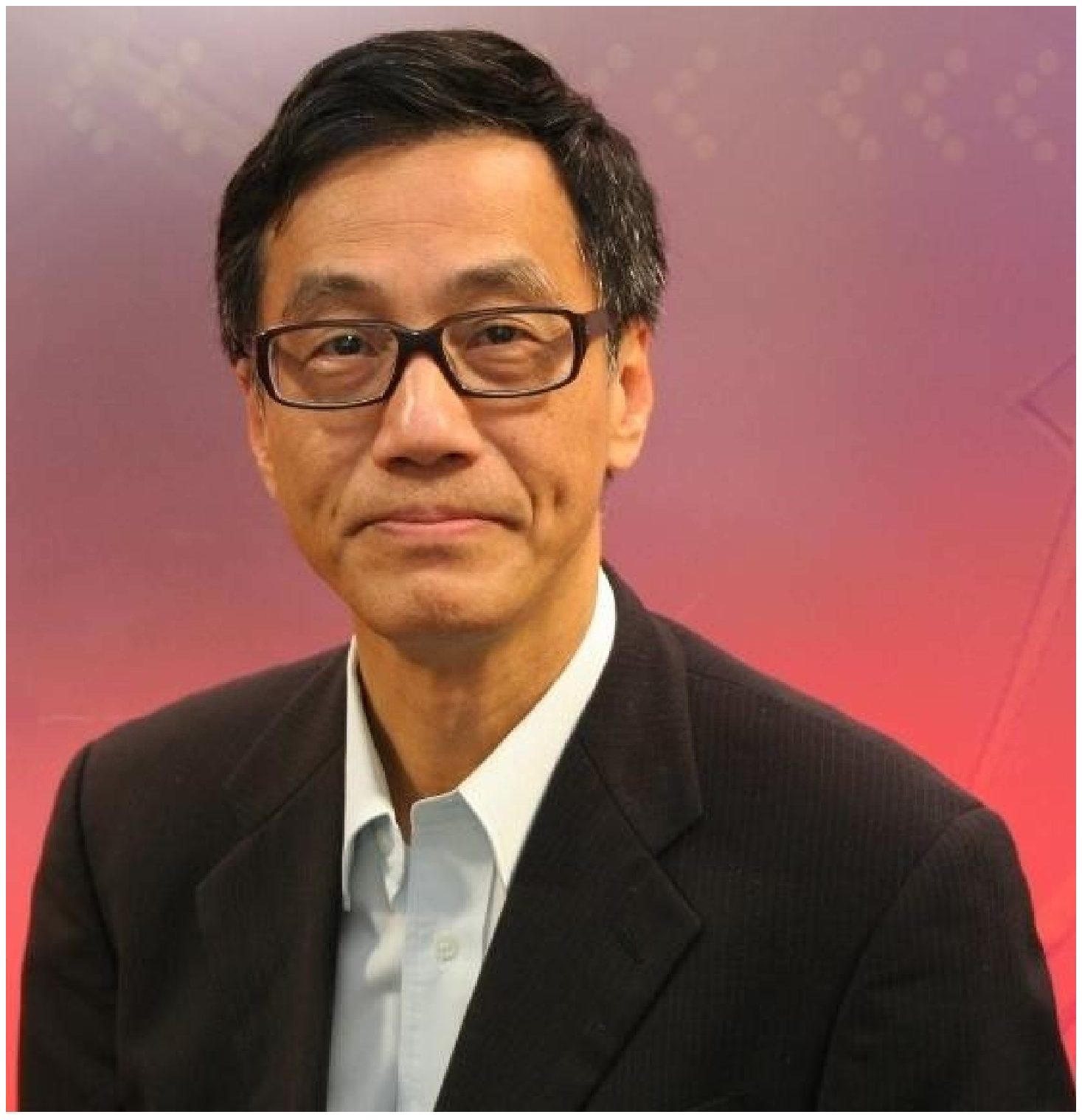}}]{Tsong Yueh Chen}
received his Ph.D. degree from The University of Melbourne. He is currently a Professor of Software Engineering at Swinburne University of Technology, Australia. Prior to joining Swinburne, he taught at The University of Hong Kong and The University of Melbourne. He is the inventor of metamorphic testing and adaptive random testing.
\end{IEEEbiography}

\begin{IEEEbiography}[{\includegraphics[width=1in,height=1.25in,clip,keepaspectratio]{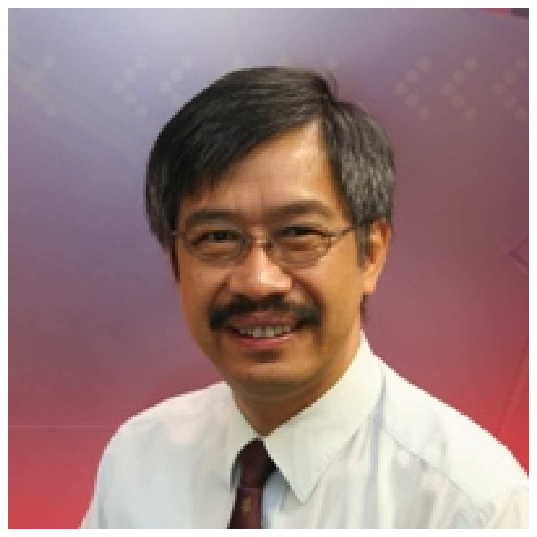}}]{Sebastian Ng}
joined Swinburne University of Technology as an academic staff in 1992, after finishing his doctorate at the University of Hong Kong.
Professor Ng is the current Chair of Computer Science and Software Engineering Department at Swinburne, with Software Engineering Education and Software Testing as his main research interests.  His academic qualifications include PhD, Master, PG Diploma and Bachelor degrees in disciplines ranging from Information Technology and Computing to Science and Education.  He is a senior professional member and Certified Professional of the Australian Computer Society.
\end{IEEEbiography}

\begin{IEEEbiography}[{\includegraphics[width=1in,height=1.25in,clip,keepaspectratio]{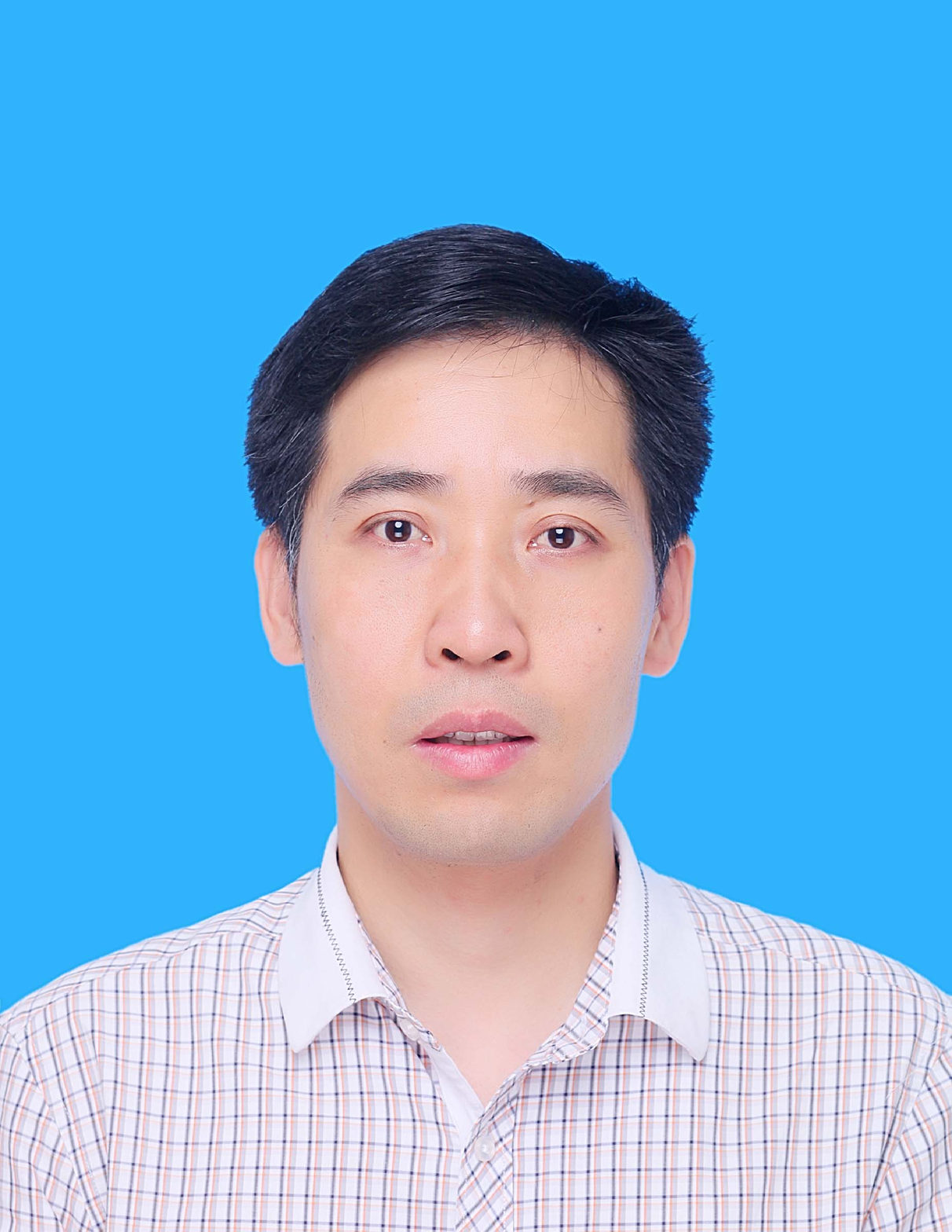}}]{Jinfu Chen}
received the B.Eng. degree in 2004 from Nanchang Hangkong University, Nanchang, China and the Ph.D. degree in 2009 from Huazhong University of Science and Technology, Wuhan, China, both in computer science. He is currently a full professor in the School of Computer Science and Communication Engineering, Jiangsu University, Zhenjiang, China. His major research interests include software testing, software analysis, and trusted software.
\end{IEEEbiography}

\end{document}